\documentclass{aastex62}

\usepackage{graphicx}

\usepackage{natbib}
\accepted{\today}
\submitjournal{ApJ}

\shorttitle{What is the role of stellar radiative feedback in setting the stellar mass spectrum?}
\shortauthors{Hennebelle et al.}

\newcommand{\ad}{{\mathrm{ad}}}

\def\msol{{\rm M}_\odot}

\begin{document}

   \title{What is the role of stellar radiative feedback in setting the stellar mass spectrum?}

\correspondingauthor{Patrick Hennebelle}
\email{patrick.hennebelle@cea.fr}

\author{Patrick Hennebelle}
\affiliation{Laboratoire AIM, Paris-Saclay \\
 CEA/IRFU/SAp -- CNRS -- \\
      Universit\'e Paris Diderot, 91191 Gif-sur-Yvette Cedex, France}

\author{Beno\^it Commer{\c c}on }
\affiliation{
      \'Ecole normale sup\'erieure de Lyon \\
 CRAL, UMR CNRS 5574, 69364, Lyon Cedex 07, France}
\author{Yueh-Ning Lee}
\affiliation{       Department of Earth Sciences, National Taiwan Normal University, \\
 88, Sec. 4, Ting-Chou Road, Taipei 11677, Taiwan
}

\author{Gilles Chabrier}
\affiliation{
      \'Ecole normale sup\'erieure de Lyon \\
 CRAL, UMR CNRS 5574, 69364, Lyon Cedex 07, France}
\affiliation{
      School of Physics \\
 University of Exeter, Exeter, EX4 4QL, UK}

% \abstract{}{}{}{}{}
% 5 {} token are mandatory

\begin{abstract}
%The stellar initial mass function (IMF) is  playing a critical role in
%		the history of our universe. 
%However, 
In spite of decades of theoretical efforts, 
the physical origin of the stellar initial mass function (IMF) is still debated. 
Particularly crucial is the question of what sets the peak 
of the distribution. To investigate this issue we perform high resolution 
numerical simulations  with  radiative feedback exploring in particular the role of 
the stellar and accretion luminosities. 
 We also perform simulations with a simple effective equation 
of state (eos) and we investigate 1000 solar mass clumps
having respectively 0.1 and 0.4 pc of initial radii. 
We found that most runs, both with radiative transfer or an eos, present similar mass
spectra with a peak broadly located around 0.3-0.5 M$_\odot$ and a
	powerlaw-like mass distribution
at higher masses. However, when accretion luminosity is accounted for, 
%due to the temperature of few hundreds of Kelvin that develops in the bulk of the cloud,
the resulting mass spectrum of the most 
compact clump tends to be moderately 
top-heavy. The effect remains limited for  the less compact one, which overall remains  colder.
%Here the net effect of accretion luminosity is to broaden the stellar mass distribution.
Our results support the idea that rather than the radiative stellar feedback,  this is the transition 
from the isothermal to the adiabatic  regime, which occurs at a gas density of
	about 10$^{10}$  cm$^{-3}$, 
that is responsible for setting the peak of the initial mass function.
This stems for the fact that $i)$ extremely compact clumps for which the accretion luminosity 
has a significant influence are very rare and $ii)$ because
of the luminosity problem, which indicates that the effective accretion luminosity is 
likely  weaker than expected.
%We also conduct an investigation of the dependence on numerical resolution and  
%sink particle scheme and we concluded that while our runs, which have 
%a spatial resolution that {\bf goes between 4.6 to 1 AU},  are probably 
%approaching convergence, the numerical results should still be handled with care.
\end{abstract}


%   \maketitle

%________________________________________________________________

\section{Introduction}

Understanding  the origin of the mass distribution of stars, the initial 
mass function 
\citep[IMF][]{salpeter55,kroupa2001,chabrier2003,bastian2010,offner2014,lee2020} 
is a fundamental issue to unravel the history of the Universe.
In particular, the fact that the IMF seems, at first sight, to 
be universal, that is to say weakly varies from one environment
to an other, remains a puzzle
although some recent variations have been claimed \citep[e.g.][]{cappellari2012,chabrier2014,schneider2018}.
Various theories have been proposed to explain the origin of the IMF. This includes 
a correspondence between the core mass function and the initial mass function 
essentially through analytical modeling  \citep{inutsuka2001,padoan1997,HC08,hopkins2013},
numerical simulations of a fragmenting cloud using sink particles to represent
the stars 
 \citep{Girichidis11,Bonnell11,BallesterosParedes15,guszejnov2020,padoan2020}
or analytical statistical description of stellar accretion \citep{basu2004,basu2015}.
In general,  the high mass tail of the IMF is reasonably reproduced in these models although 
the physical reasons invoked are different.  In these models, the universality of the slope 
relies on the invoked scale-free processes, gravity and/or  turbulence 
 \footnote{Strictly speaking turbulence and gravity are not entirely scale-free. However in the fully 
non-linear regime, there is usually a broad range of scales over which a high Reynolds turbulent flow is 
self-similar. The same is true for a self-gravitating fluid which has developed a powerlaw density PDF and for which 
there is not a unique Jeans length.}.
The question of the peak appears however to be more complicated because most theories 
are based on the Jeans mass, which depends on the gas density and temperature and thus 
inferring a characteristic mass, say around 0.3 $M_\odot$, which does not 
vary significantly with the physical conditions is a challenge.  Most proposed
explanations consist in identifying mechanisms which could result in 
a weak dependence of the effective Jeans mass on gas density \citep{leeh2018b,guszejnov2020}.  
For instance \citet{HC08}, \citet{Hennebelle2012} and \citet{lee2016b} proposed that there is 
a compensation between the  density and  Mach 
number variations, \citet{jappsen2005} argued that the change of the effective equation of state 
at a density of about 10$^5$ cm$^{-3}$ makes the corresponding Jeans mass play a dominant role while 
\citet{bate2009,krumholz2016,guszejnov2016} proposed that  radiative feedback 
heats up the gas at very high density \citep{krumholz2007}, i.e. 10$^{8-10}$  cm$^{-3}$ setting up again a Jeans mass 
that weakly depends on density for instance. 

A somewhat different explanation has recently been proposed by 
 \citet{leeh2018b} and \citet{hetal2019} who argue that the peak of the IMF 
is directly linked to the mass, $M_L$, of the 
first hydrostatic Larson core \citep[FHSC][]{larson69,Masunaga98,vaytet2013,Vaytet17,bhandare2018,bhandare2020}, 
which is the hydrostatic core that forms 
when the dust becomes opaque to radiation. 
The mass of FHSC is about $M_L \simeq 0.03 \; M_\odot$ which is about 10 times below the peak 
of the IMF. However, performing high resolution of 
 collapsing 1000 $\msol$ clumps, these authors infer that the peak of the IMF is about 5-10 $M_L$.
This is due to further accretion from the envelope onto the FHSC, which is eventually 
halted when new fragments form \citep{hetal2019}. Because in particular of the stabilizing effect 
of the tidal forces \citep{leeh2018b,colman2019}, the immediate neighbourhood of the FHSC is stable 
against gravitational instability and finding another FHSC requires to go at a distance $L$ such that 
the mass enclosed in the sphere of radius $L$ is about 5-10 $M_L$.
Very importantly, changing the
initial conditions of the initial clumps, initial density, Mach number or
magnetic field \citep{lee2019} by orders of magnitude is found to leave  
the peak of the stellar mass spectrum almost unaffected.
%has been found to be rather insensitive to these variations in a large range of parameters. 

So far the simulations performed to investigate the FHSC based theory have been using an effective barotropic 
equation of state aiming at mimicking the thermal behaviour of the gas at densities above 
10$^{10}$ cm$^{-3}$. While the approach was rather useful to establish and test these ideas, 
 radiative transfer calculations are mandatory for a more realistic treatment. In particular, since the origin
of the  FHSC is the high  optical depth that makes the gas adiabatic, it is important to test the theory in this 
context. Several attempts have been made to study the IMF using  radiative transfer calculations. 
\citet{urban2010} performed SPH calculations and introduce the sink particles at a density of about 
10$^8$ cm$^{-3}$, they include radiative feedback onto the sink particles which includes both the 
stellar and accretion luminosity. They found that radiative transfer calculations are quite different from
the isothermal ones, in particular the stars are much more massive when radiation is considered. 
\citet{bate2009} performed also high resolution  SPH calculations but introduced the sink particles at very high density, i.e. 
$n > 10^{19}$ cm$^{-3}$. This includes the optically thick regime, which 
occurs at a density  $n > 10^{10}$ cm$^{-3}$,  but the simulations do not add any stellar feedback  onto the sink particles. 
By doing this, the IMF  presents a peak at about 0.3 $M_\odot$ and a mass spectrum 
at high mass which is clearly flatter that the Salpeter's exponent of 1.3 (in $dN/d\log m$). 
\citet{krumholz2012} performed adaptive mesh refinement calculations with a resolution of 20-40 AU. The sinks are 
introduced when the Jeans conditions get violated, that is to say when the mesh size 
is larger  than one tenth of the local Jeans length,  and they consider only objects more massive than 0.05 $M_\odot$
as being stars, the smaller ones being allowed to merge. Both intrinsic and accretion luminosity are  added to the sinks. 
By doing so they obtain  mass spectra which are almost flat, that is to say $dN/d \log M \propto M^0$, when winds are not considered while 
in the presence of winds, which allow the radiation to escape, the mass spectra present a peak around 0.3 $M _\odot$
and a powerlaw, $dN/d \log M \propto M^{-\alpha}$, with $\alpha \simeq 0.5-1$. Recently \citet{mathew2020} performed 
simulations with a spatial resolution of 200 AU and compare runs which use either a  polytropic equation 
of state or take into account the stellar heating either assuming spherical symmetry or a polar distribution. They infer 
stellar mass spectra that peaked at about 2 $M _\odot$  and found that when heating is included more massive stars would form.

In the present paper, we want to explore further the influence of the radiative feedback in 
establishing the stellar mass spectrum during the collapse of a massive clump. In particular, we stress
that so far studies that do consider the  accretion luminosity \citep[e.g.][]{urban2010,krumholz2012,mathew2020} 
have been performed at relatively coarse resolution. The lack of resolution can be particularly 
severe in this context because it may result in overestimating the 
mass at which the stellar distribution peaks \citep[e.g.][]{ntormousi2019}. This, in turn, implies that too much mass 
is contained in massive stars and since they exert a strong feedback onto their environment, this may 
lead to  overestimate the importance of radiative feedback. 

To determine the impact of various contributions, we present a set of both barotropic and  radiative transfer calculations 
taking into account the different contributions of the feedback luminosity and for 
different types of initial conditions. By performing these various runs,  we 
can in particular distinguish between the influence of the optically thick and hydrostatic phase (the FHSC)
and the heating of the collapsing envelope by  radiation.
The plan of the paper is as follows.
In  section two, we present the  numerical setup and the 
various assumptions done to perform the two types of  simulations. 
In the third section, we present and discuss our results regarding the temperature distribution 
through the clouds.  The fourth section is devoted to the stellar mass spectra, how they depend 
on the radiative feedback and on the initial conditions. 
The fifth section concludes the paper.

\setlength{\unitlength}{1cm}
\begin{figure}%[h!]
%\centering
\begin{picture} (0,12)
\put(0,6){\includegraphics[width=8cm]{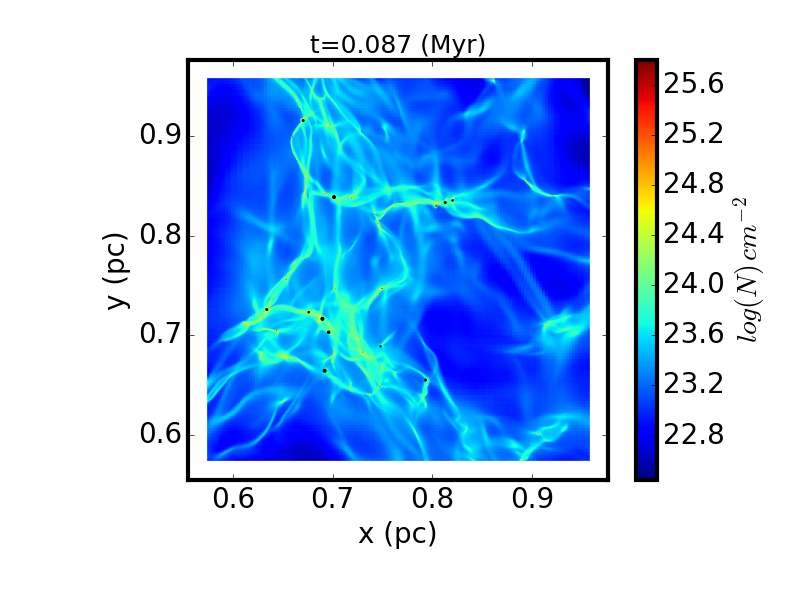}}  
\put(0,0){\includegraphics[width=8cm]{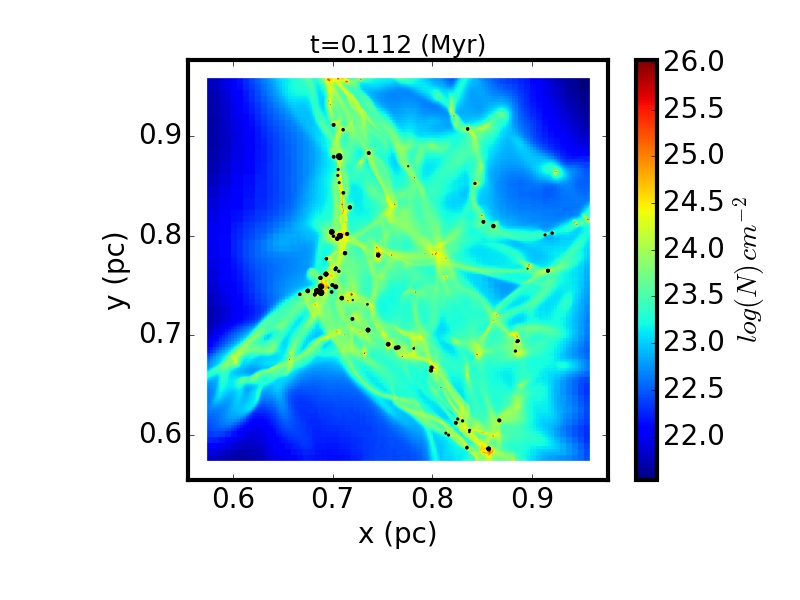}}  
\end{picture}
\caption{Column density at two snapshots for run STAN-ACLUMhrhs. The black dots represent the sink particles.
In the first  and second snapshots about 2 M$_\odot$ and 120 M$_\odot$  of gas have been accreted onto the sink 
particles. }
\label{coldens_diff-aclumhrhs}
\end{figure}

\setlength{\unitlength}{1cm}
\begin{figure*}%[h!]
%\centering
\begin{picture} (0,18)
\put(0,12){\includegraphics[width=8cm]{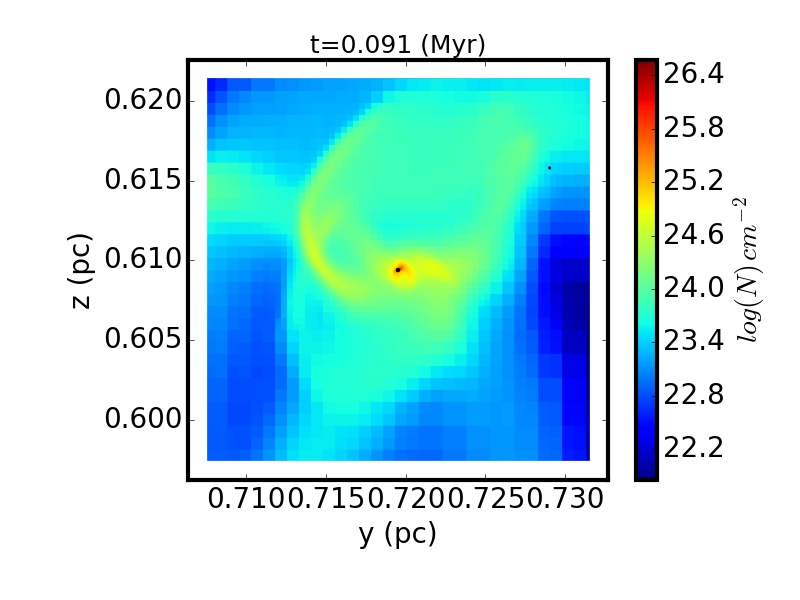}}  
\put(0,6){\includegraphics[width=8cm]{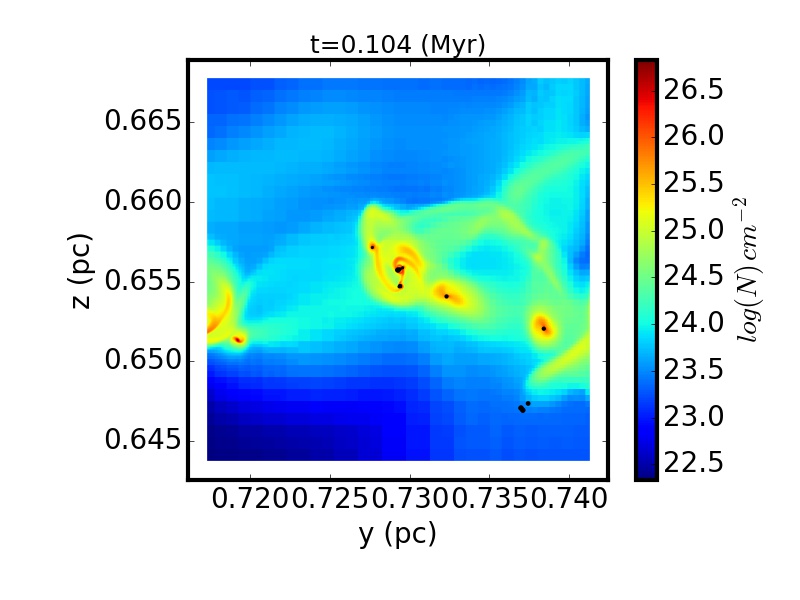}}  
\put(0,0){\includegraphics[width=8cm]{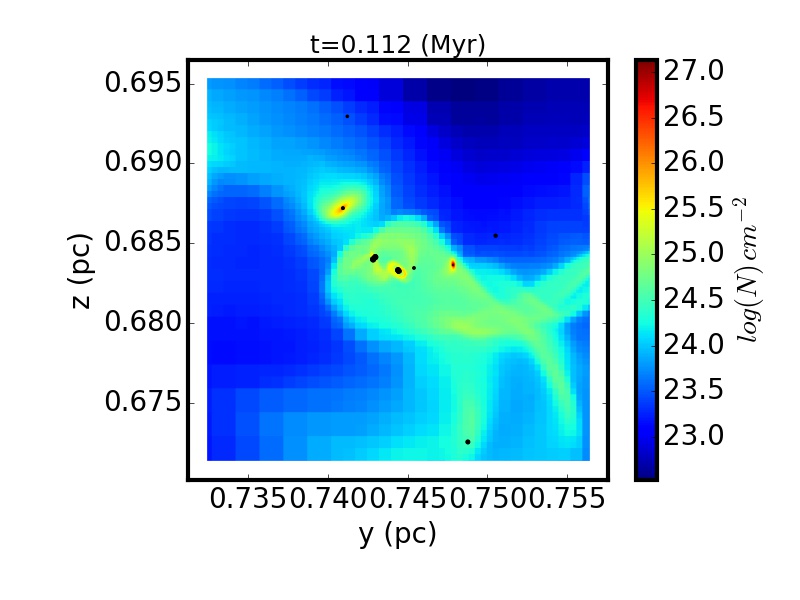}}  
\put(8,12){\includegraphics[width=8cm]{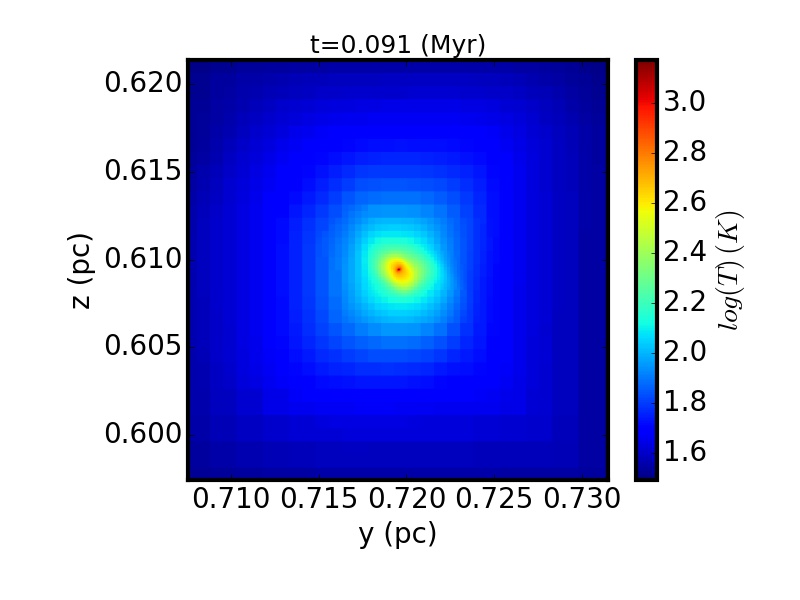}}  
\put(8,6){\includegraphics[width=8cm]{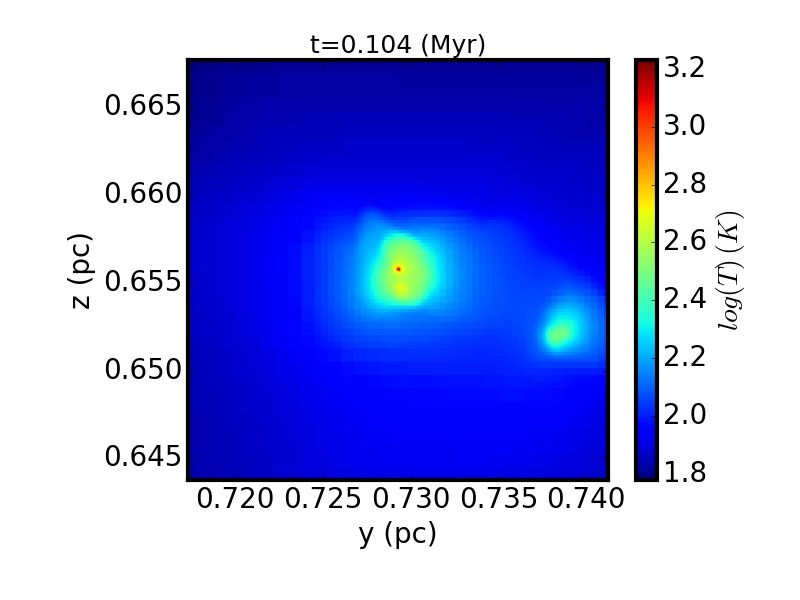}}  
\put(8,0){\includegraphics[width=8cm]{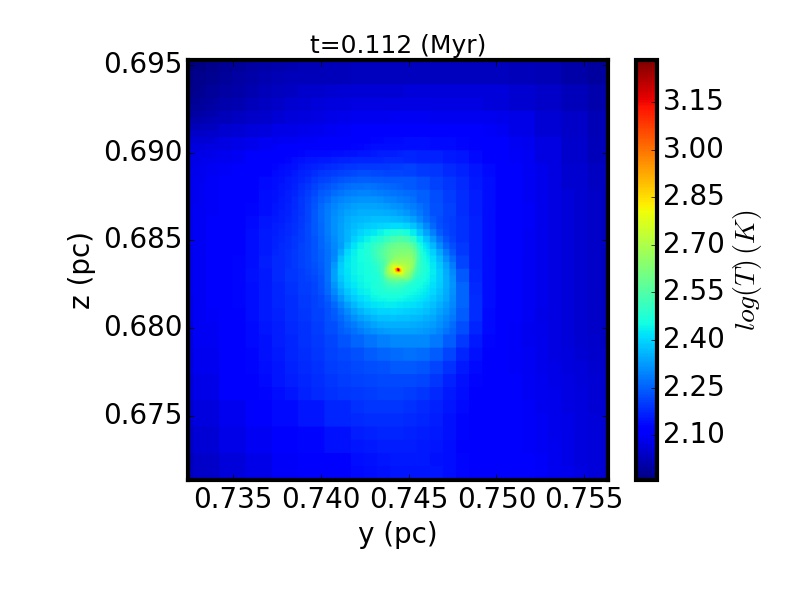}}  
\end{picture}
\caption{Column density (left) and temperature cuts (right) at three snapshots for run STAN-ACLUMhrhs around one of the sink particles.
}
\label{coldens_diff-aclumhrhs_sink}
\end{figure*}

\setlength{\unitlength}{1cm}
\begin{figure*}%[h!]
%\centering
\begin{picture} (0,6)
\put(0,0){\includegraphics[width=8cm]{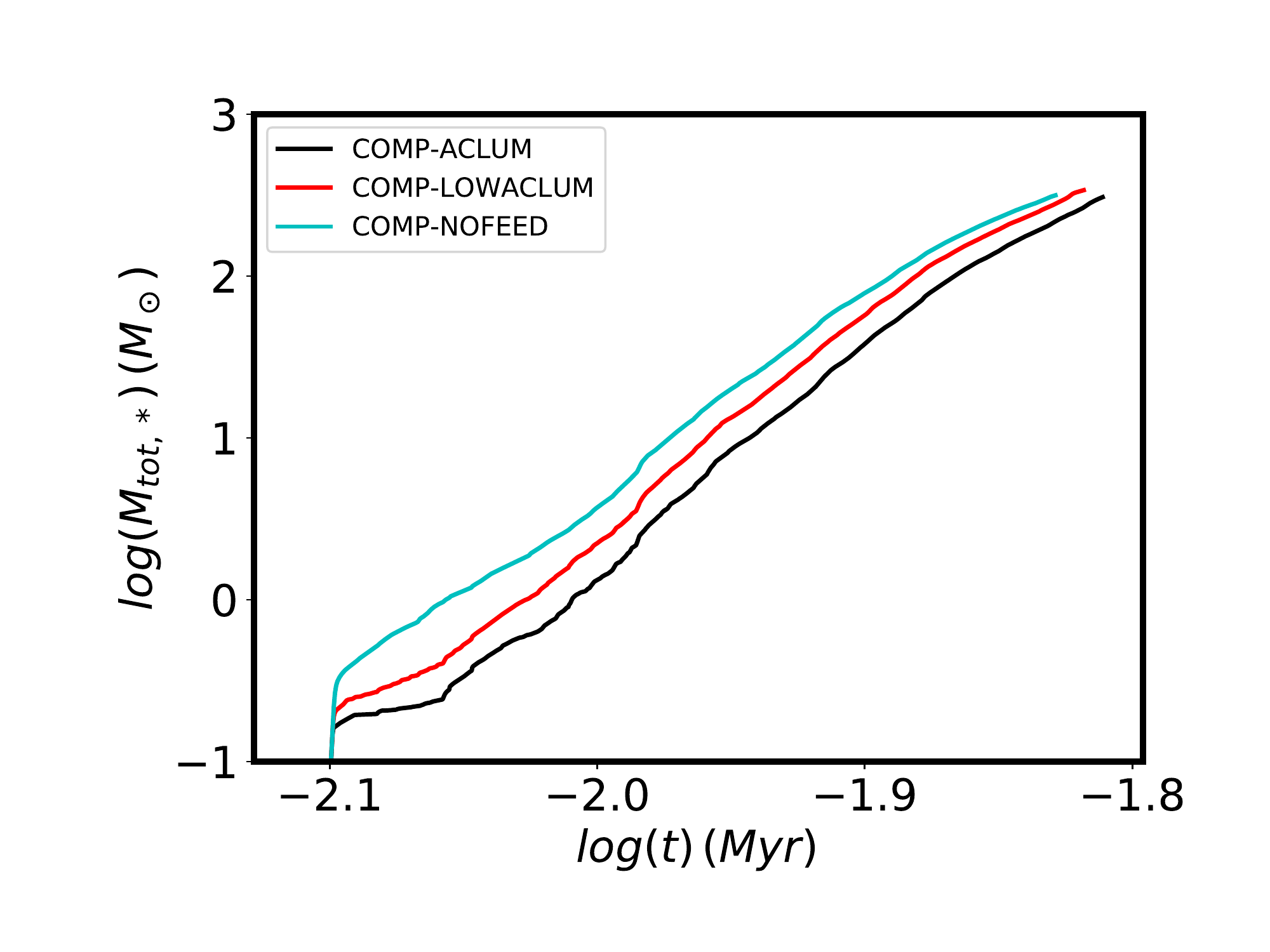}}  
\put(8,0){\includegraphics[width=8cm]{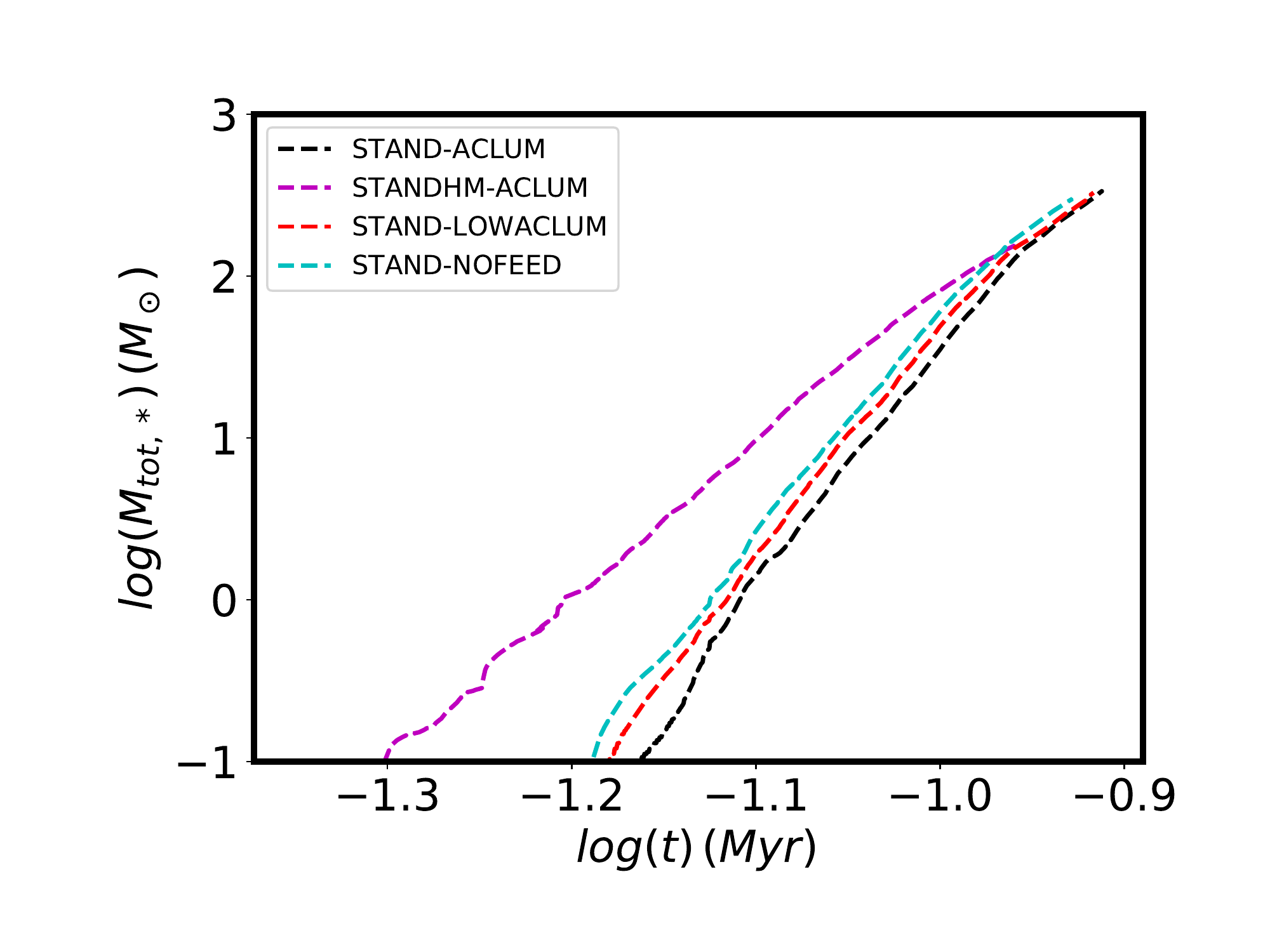}}  
\end{picture}
\caption{Accreted mass as a function of time for the various runs. 
}
\label{time_mass}
\end{figure*}

\setlength{\unitlength}{1cm}
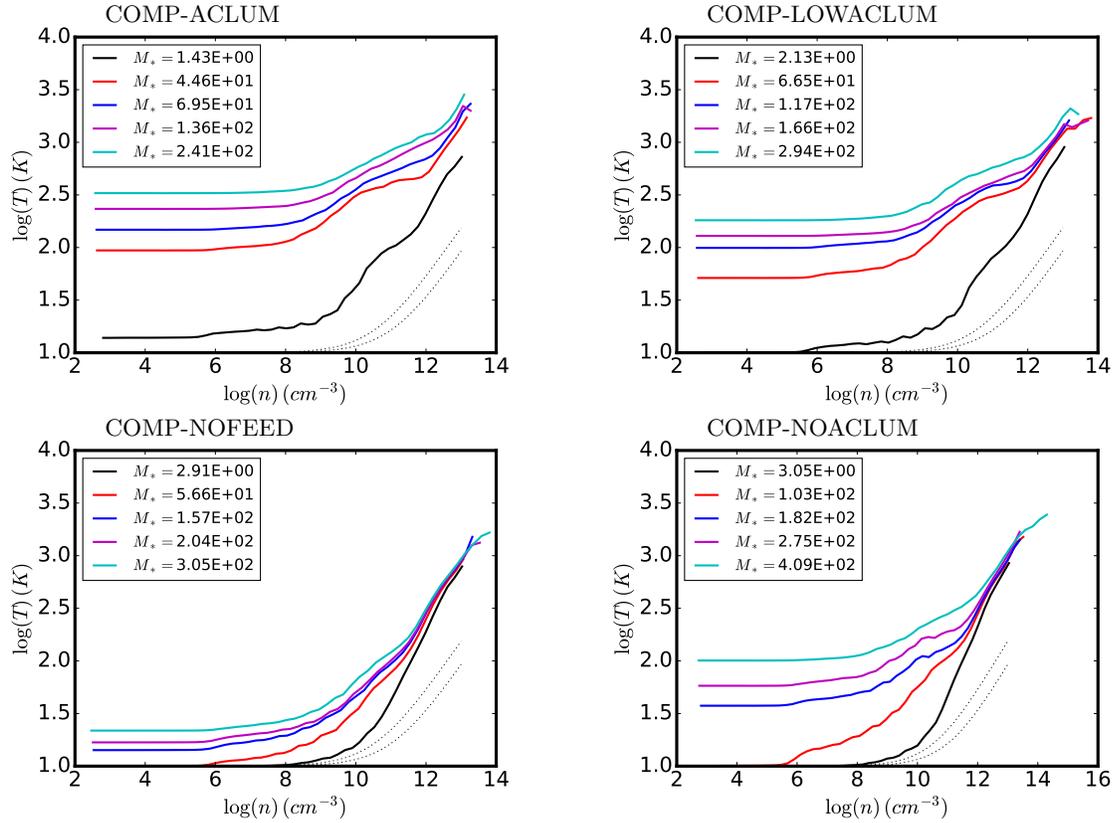
\begin{figure*}%[h!]
%\centering
\begin{picture} (0,12)
\put(8,5.5){\includegraphics[width=8cm]{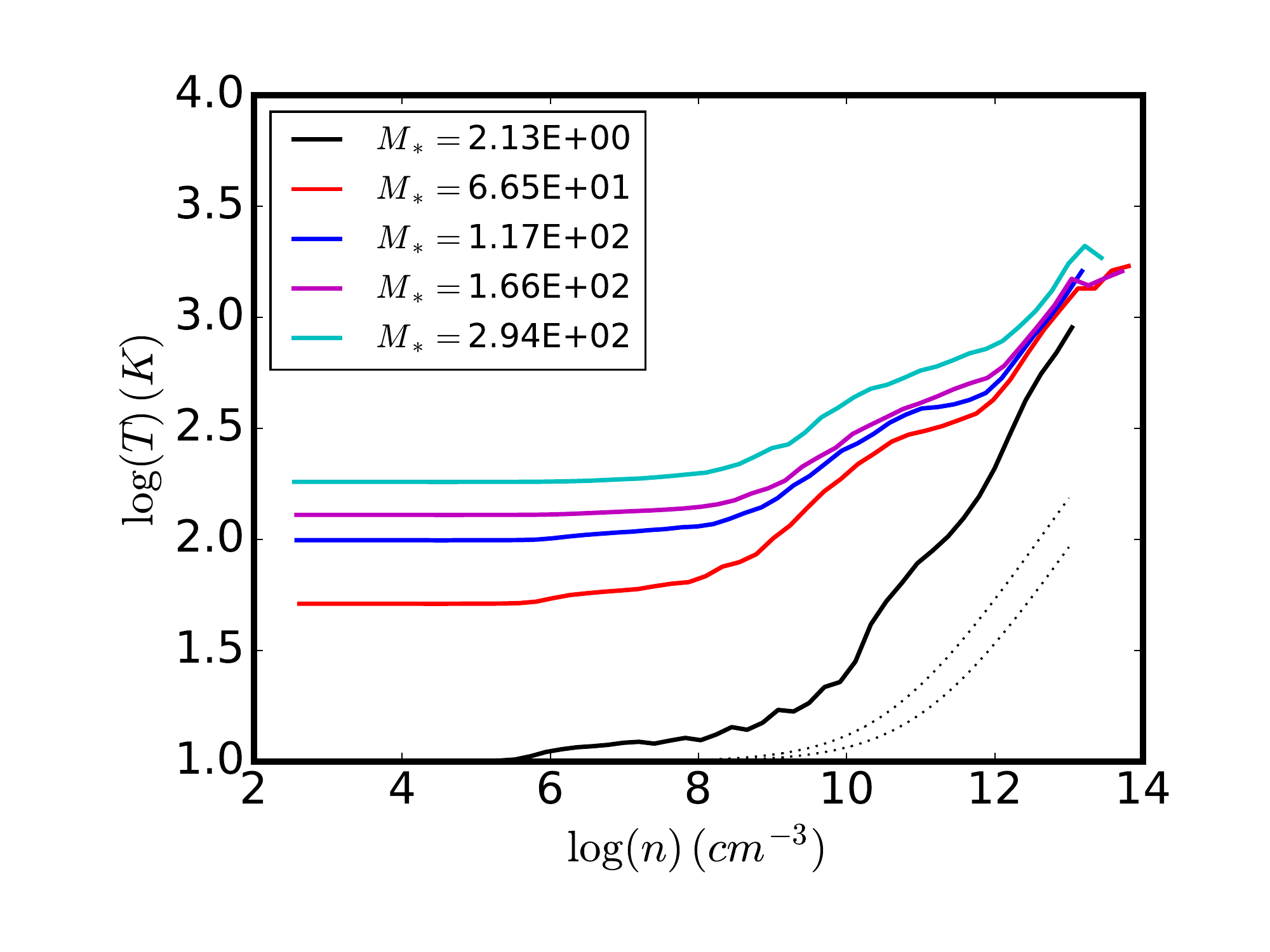}}
\put(0,5.5){\includegraphics[width=8cm]{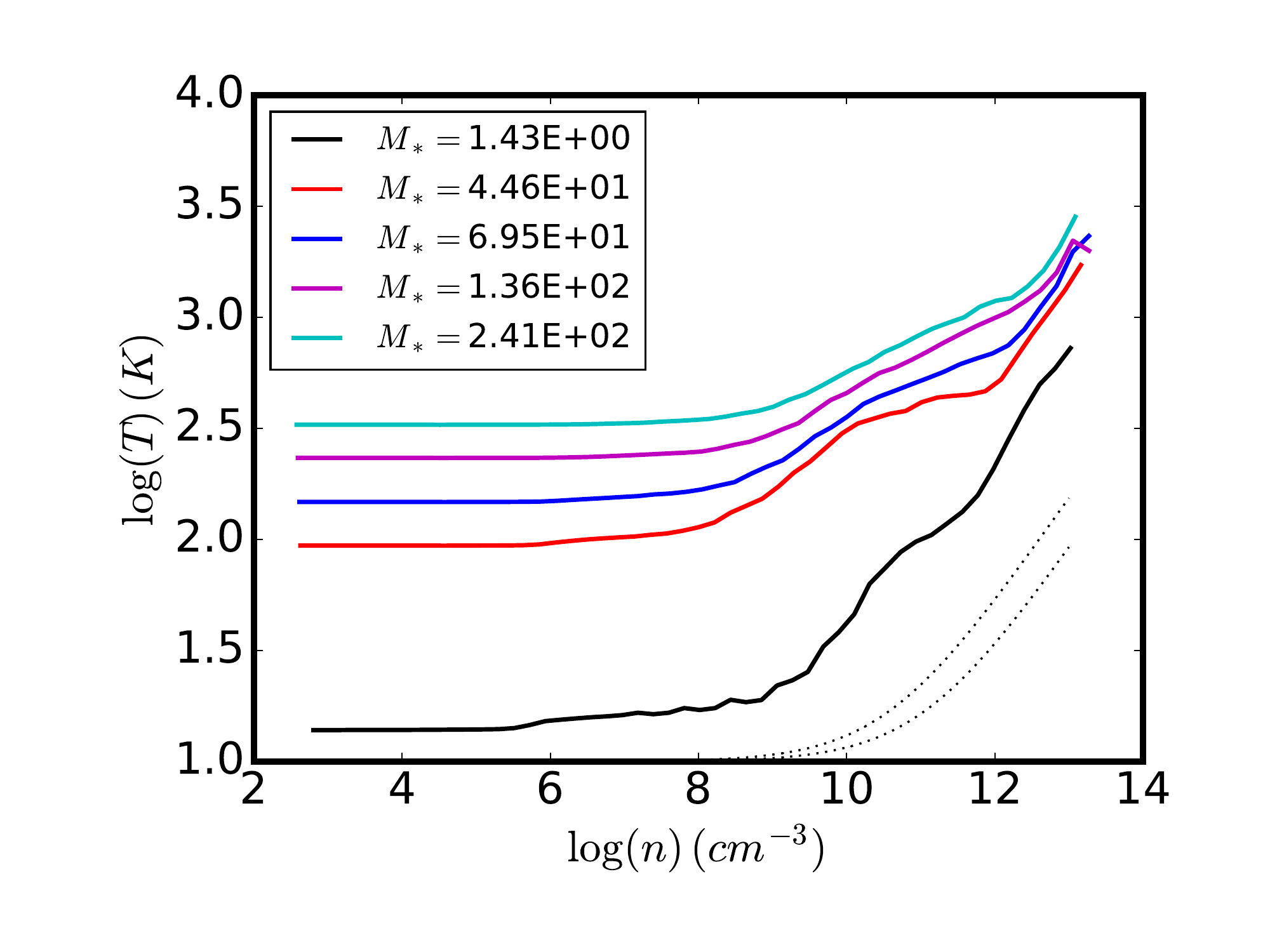}}  
\put(2,11.1){COMP-ACLUM}
\put(10,11.1){COMP-LOWACLUM}
\put(8,0){\includegraphics[width=8cm]{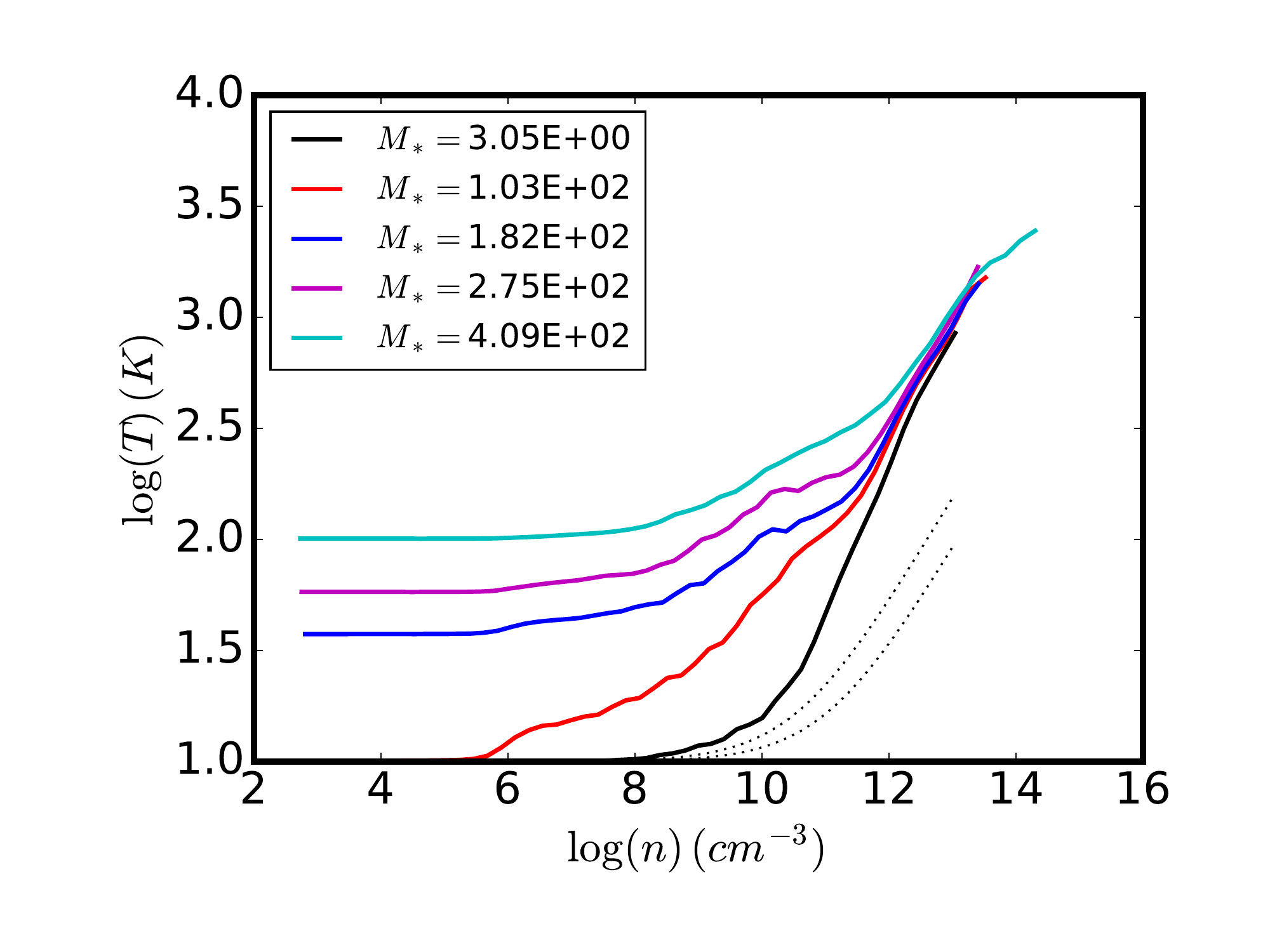}}  
\put(0,0){\includegraphics[width=8cm]{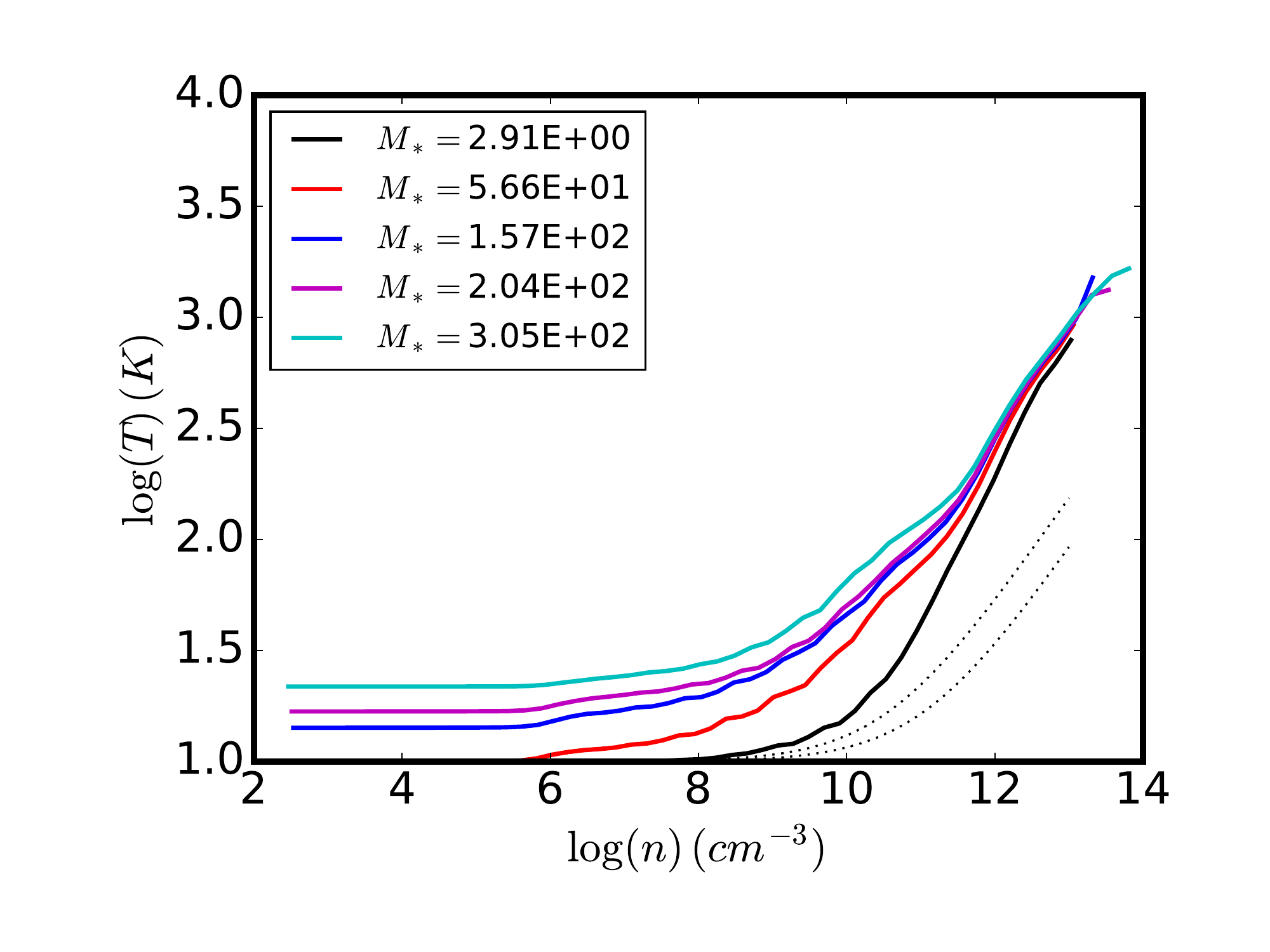}}  
\put(2,5.6){COMP-NOFEED}
\put(10,5.6){COMP-NOACLUM}
\end{picture}
\caption{Mass weighted temperature in density intervals as a function of density at five time steps  (corresponding to five accreted mass, $M_{*,tot}$)
for the four COMP runs. In run COMP-ACLUMN and COMP-LOWACLUMN, respectively 
50$\%$ and 10$\%$ of  the accretion luminosity is taken into account. In run COMP-NOFEED the stellar feedback is not taken into account
while only the stellar luminosity is considered in COMP-NOACLUM. 
}
\label{T_rho_comp}
\end{figure*}

\setlength{\unitlength}{1cm}
\begin{figure*}%[h!]
%\centering
\begin{picture} (0,12)
\put(0,5.5){\includegraphics[width=8cm]{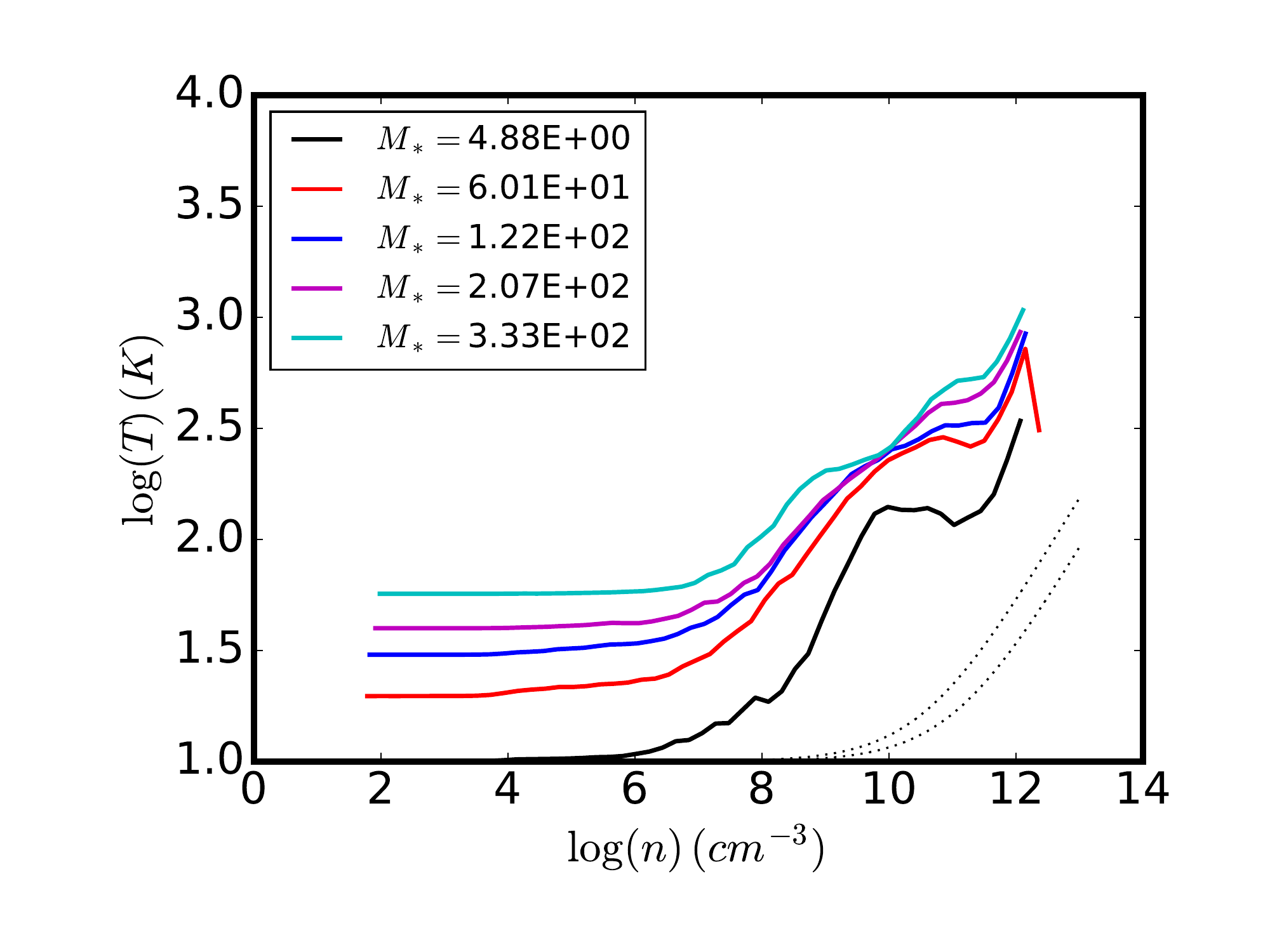}}  
\put(8,5.5){\includegraphics[width=8cm]{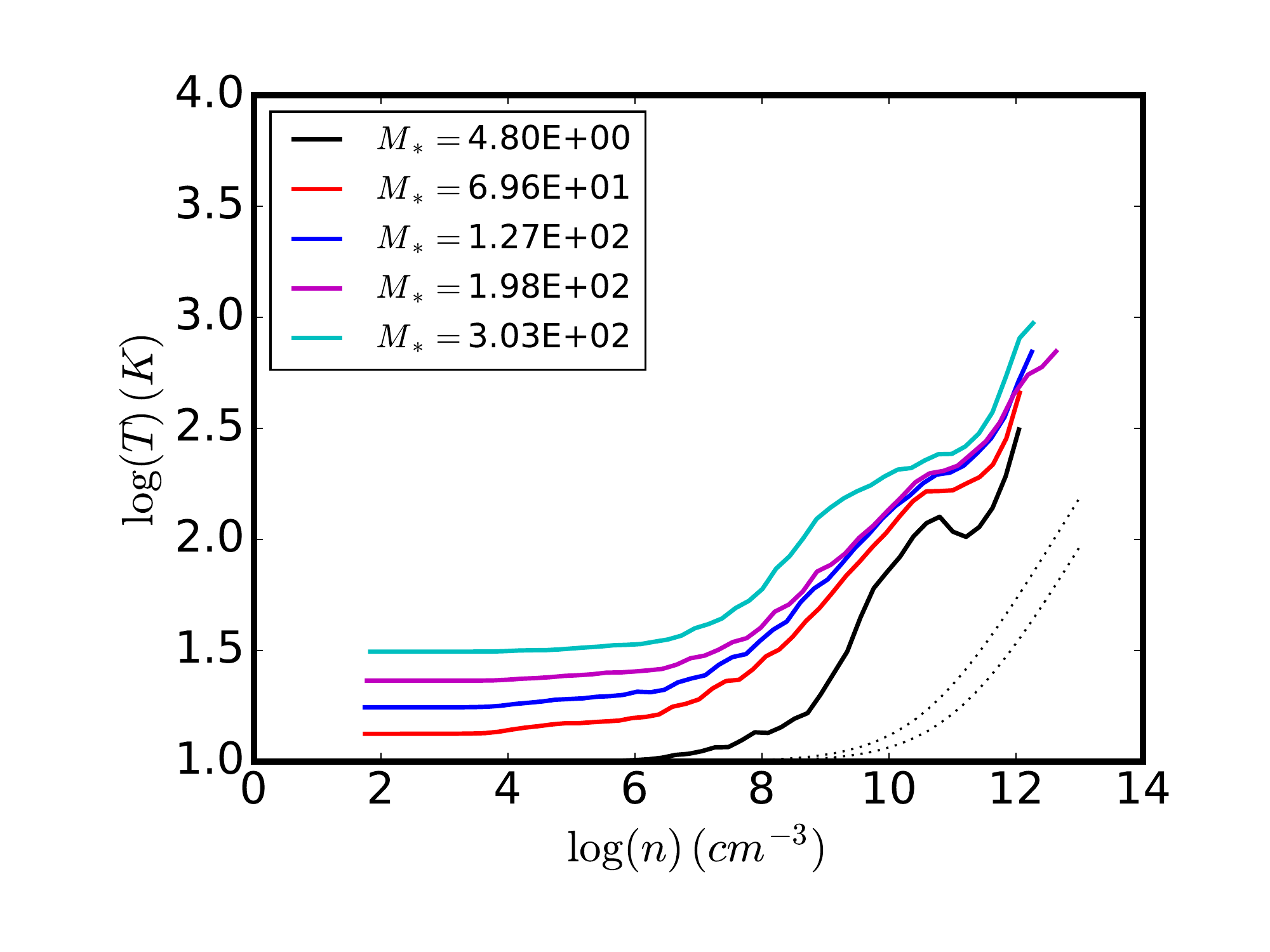}}  
\put(2,11.1){STAN-ACLUM}
\put(10,11.1){STAN-LOWACLUM}
\put(8,0){\includegraphics[width=8cm]{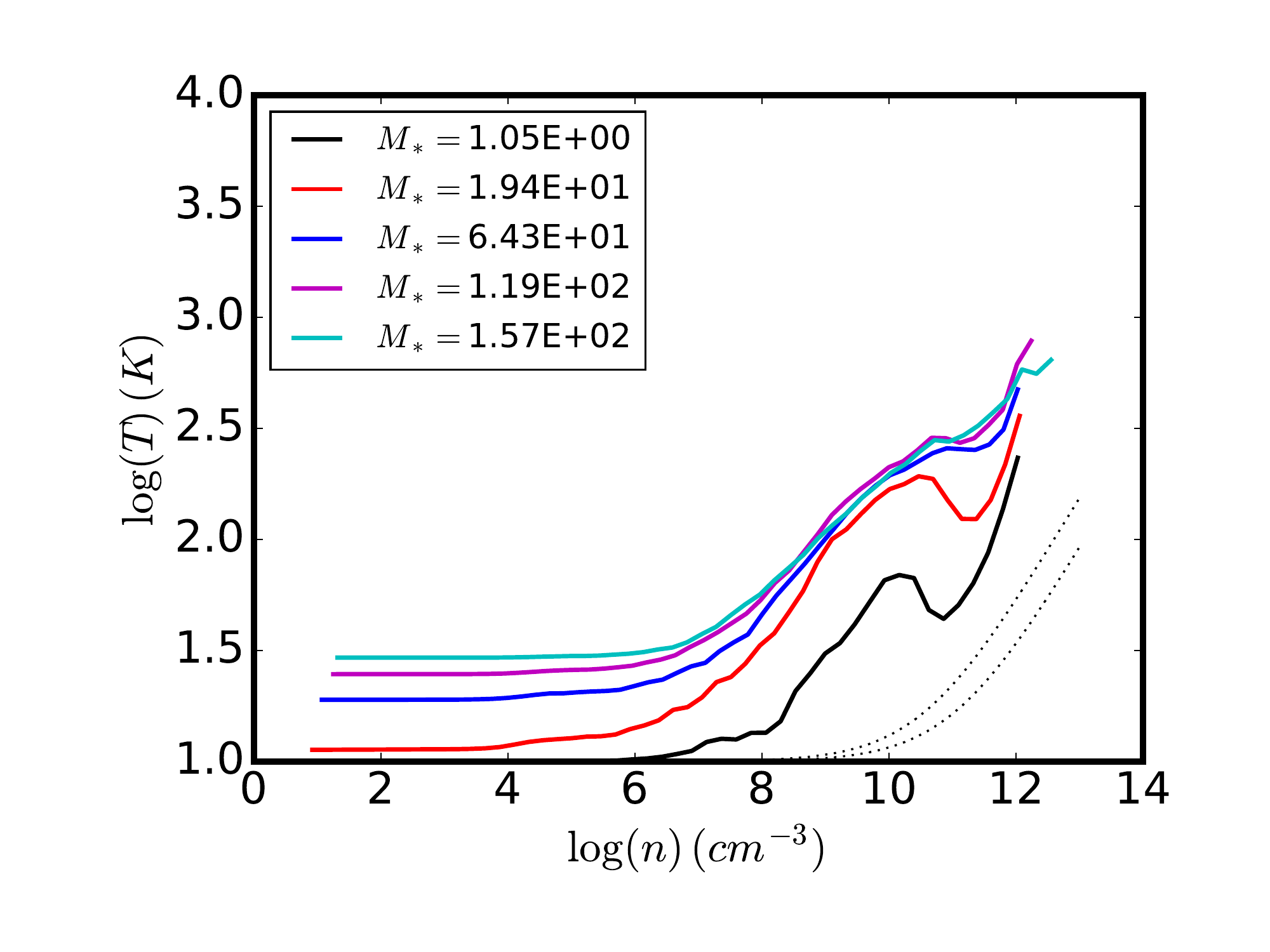}}  
\put(0,0){\includegraphics[width=8cm]{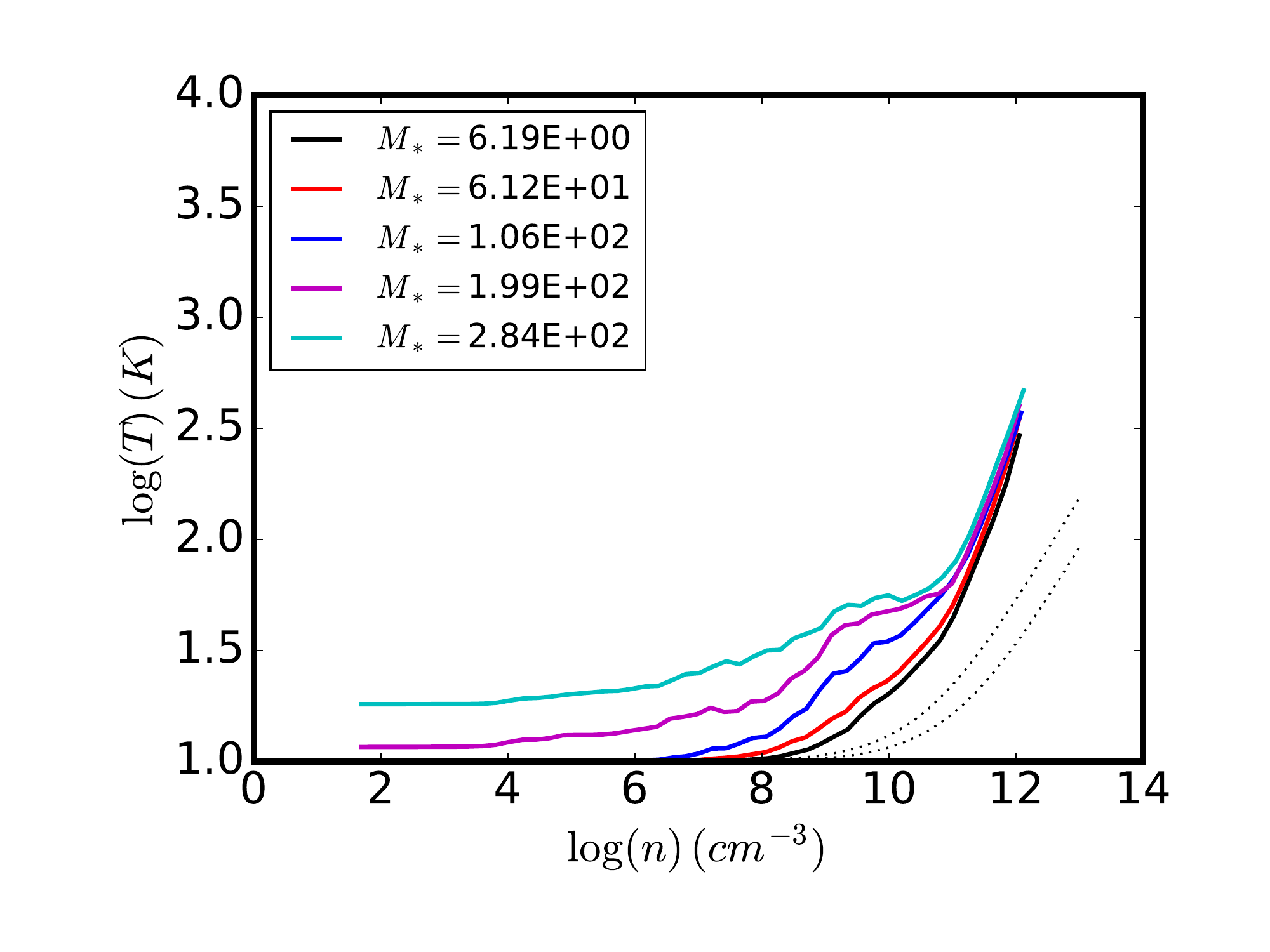}}  
\put(2,5.6){STAN-NOFEED}
\put(10,5.6){STANHMach-ACLUM}
\end{picture}
\caption{Same as Fig.~\ref{T_rho_comp} for the STAN runs. Clearly when accretion luminosity is taken into account
 temperature are significantly lower than for the COMP runs. This is due to lower accretion rate but also larger distance, on average 
from the source. The run STANHMach has a mach number equal to 10 initially instead of 5 and is globally accreting significantly less 
than the runs for which it is equal to 5 initially.  
}
\label{T_rho_diff}
\end{figure*}

\section{Numerical simulations}

\subsection{Numerical methods and setup}
All  simulations were run with the adaptive mesh refinement (AMR) magnetohydrodynamics (MHD) code RAMSES \citep{Teyssier02,Fromang06}
 though in this work magnetic field is not considered.
Two types of simulations have been carried out. 

The first type of  simulations use  radiative transfer using  
 the flux diffusion method and the gray approximation as described in \citet{commercon2011a,commercon2014}. 
At high density, the equation of state is the  one given by \citep{saumon1992} and \citep{saumon1995} which models the thermal properties of a
 gas containing the species H2, H, H+, He, He+, and He2+ (the He mass concentration is 0.27).
The opacities are as described in \citet{vaytet2013}. For the  range of temperature and densities covered 
in this work, the opacities are essentially the ones calculated in \citet{semenov2003}. 

The second type of simulations employs an effective equation of state and no radiative transfer. 
The prescription is the same as the one used in \citet{hetal2019} 
\begin{eqnarray}
T = T_0 \left\{1 \!+\! { (n / n _{\ad})^{(\gamma_1-1)} \over 
1+( n / n _{\ad,2})^{(\gamma_1-\gamma_2)}  }  \right\}, 
\label{eq_full_eos}
\end{eqnarray}
where $T_0=10$ K,  $n _{\ad,2}= 30 n_{\ad}$, 
$\gamma_1=5/3$ and $\gamma _2 = 7/5$. This equation of state (eos) mimics the thermal behaviour of the gas when it becomes
 non isothermal. Two values of $n _{\ad}$ have been explored, namely 
$(n _{\ad})_1=4 \times 10 ^{10}$ cm$^{-3}$ and $(n _{\ad})_2=1.2 \times 10 ^{11}$ cm$^{-3}$. The latter one has been chosen 
because it appears to be closer to the actual temperature of the radiative transfer calculations. 

 The boundary conditions used in this work are periodic and  we simulate a spherical cloud whose radius is 
four times lower than the computational domain size.
All simulations were run on a base grid of $2^8$ and typically 7  to  8 (up to 10)   AMR levels have been added leading to 
a total number of 15 or 16 AMR levels. The Jeans length is resolved with at least 10 points. In the appendix 
 less and more resolved runs are presented to investigate the issue of numerical convergence and test the influence of sink 
particle numerical parameters.

\subsection{Sink particles} 
We used the sink particle algorithm of \citet[]{Bleuler14}.
Sink particles are formed at the highest refinement level at the peak of clumps whose density 
is larger than $n \ge n_{\rm acc} / 10$ while the 
% $n \ge 10^{12}$  cm$^{-3}$. The 
sinks are introduced at a density $n = n_{\rm acc}$. Only clumps that satisfy a series 
of criteria indicating sufficient gravitational boundness \citep[see][]{Bleuler14} may lead to sink formation.
The value of  $n = n_{\rm acc}$ is equal to either 
  $n_{\rm acc} = 10^{12}$ cm$^{-3}$ or $n_{\rm acc} = 10^{13}$ cm$^{-3}$ and is discussed below. 
Typically, the value of  $n_{\rm acc}$ is chosen such that a computational cell having a density 
 equal to $n_{\rm acc}$ contains a mass that is about 1-2$\%$ of the mass of the first hydrostatic core, i.e. 
about $M_L=$0.03 M$_\odot$. 
At each time step, 10$\%$ of the  gas mass that is located within the sinks and 
has a density larger than $n_{\rm acc}$ 
is removed from the grid and transferred to the sink.  \citet{leeh2018b} and 
\citet{hetal2020} have modified  this value and conclude that it does not affect significantly the 
accretion rate onto the stars (essentially because the accretion is controlled by the larger scales) while 
on the other hand, it may affect the disk which forms around the sink.
The sinks are not allowed to merge. 
 We stress that according to us, it is necessary to describe sufficiently the FHSC in order to get the peak 
of the IMF. Introducing sink particles when the Jeans criterion is not satisfied for instance, is a viable approach 
in the isothermal phase only and is not suited to the physics of the FHSC that is essentially adiabatic.

\subsection{Stellar feedback and accretion luminosity}
An important aspect is the radiative feedback emitted by the sink particles. Two 
types of contributions have to be taken into account. 
First, the accretion luminosity, which is given by 
\begin{eqnarray}
L_{\rm acc} = {f_{\rm acc} G M_* \dot{M} \over R_*} . 
\label{accret_lum}
\end{eqnarray}
When assuming that all the accretion gravitational energy is radiated away, we have $ f_{\rm acc} \simeq 1$. 
This has been shown to be the dominant source of heating at early time and to have 
important consequences on the cloud \citep[e.g.][]{krumholz2007,offner2009}.
Second, the intrinsic luminosity of the protostars (mimicked by the sinks), $L_*$. 
The difficulty for star forming calculations is that 
the protostars are embedded and still heavily accreting. In our calculations, we use 
the radius and stellar luminosity given by \citet{kuiper2013} 
\citep[see also][]{hosokawa2009} which have developed models that take into account the accretion. 
As will be seen later, 
both effects can have significant influence on the outcome of the calculations. There are however 
serious uncertainties here. These radiations are emitted  at the stellar surface, i.e. at a 
few solar  radii. To what extent is it accurate to introduce them isotropically at a scale of a few AU, is
highly uncertain. Indeed \citet{krumholz2012} concluded that in the presence of a wind cavity, much of the radiation 
may escape and this would limit the impact of the radiation heating. Performing collapse
up to stellar densities, \citet{bate2010} found that the accretion luminosity may halt the accretion 
onto the young protostar and drive a jet. Therefore investigating the exact consequences of 
accretion luminosity requires detailed small scale calculations that are not reachable in calculations aiming 
at getting the mass distribution of stars. 
More generally, it is now firmly established that the  luminosity  of protostars
is significantly below the expected values \citep{kenyon1995,evans2009,stamatellos2011,offner2011} and present a considerable 
scatter, which has been interpreted as a signature of episodic accretion \citep{baraffe2009,baraffe2012,baraffe2017}.  
Indeed bursts of accretion in young protostars have been reported and several attempts to quantify their frequencies
have been made \citep[e.g.][]{frimann2016,hsieh2018,fischer2019}.

  A fundamental question is: what is the value of the parameter $f_{acc}$ in eq.~(\ref{accret_lum}), that is to say what 
is the value of the {\it effective} accretion luminosity? 
Based on accretion burst frequency estimate, \citet{offner2011} following  \citet{dunham2010} estimated $f_{acc} \simeq 0.23$.
We note that if some of the accretion energy is radiated away through mechanical processes, such as jets, or if 
the emission of the accretion luminosity is isotropic and concentrated in the direction of the jets for instance, 
$f_{acc}$ could be further reduced. 

Since it is likely that a substantial fraction of the accretion energy is radiated in short burst,  
a key aspect lies in the comparison between  the cooling and the dynamical time. The former can be estimated as 
\begin{eqnarray}
\tau _{cool} \simeq { E_{therm} \over \partial_ R F_R } \simeq { k_B T n r ^2 \kappa \rho \over c  a T^4 }
\end{eqnarray}
and the latter is simply the freefall time $\tau_{ff} = \sqrt{3 \pi / (32 G \rho) }$, where all expressions have their 
usual meaning,  $\kappa$ is the opacity, $c$ the speed of light, $a$ the radiative constant, $r$ the radius, 
$k_B$ the Bolzman constant, $F_R$ the radiative flux and $T$ the temperature.
Computing the ratio of these two times for physical  conditions corresponding to 
left panel of Fig.~\ref{masstemp_msink} we found that $\tau_{cool} / \tau_{ff}$ is typically between $10^{-2}$ and $10^{-6}$, therefore the gas 
temperature adjusts instantaneously to the source luminosity and thus if it is experiencing short bursts, 
as inferred from observations, the gas dynamics 
is not affected.
For this reason we have performed runs in which the {\it effective} accretion luminosity has been multiplied by a factor, $f_{\rm acc}=0.5$, 
 0.1 and even 0 (that is to say $L_{\rm acc}$ is ignored). 

 Another parameter that needs to be determined is when, i.e. for which mass of the sink, to start injecting the accretion luminosity
onto the sink particle and we choose to do so when the sink has a mass 
of about 2 $M_L$, i.e. 0.07 M$_\odot$. The reason is that due to the limited spatial resolution, when the sink 
is introduced the protostar is not formed yet. Since the size of the sink particles is not very different 
from the radius of the FHSC, it seems reasonable to assume that the protostar is formed only when the 
sink reaches a mass equal at least to $M_L$. Since in the time delay, more gas falls into the sink and since 
it is not very clear what is the minimum mass of the protostar for which the accretion luminosity 
can be described by eq.~(\ref{accret_lum}), in particular because the accretion rate is measured at a scale of 
the sink particle, we choose to start the accretion luminosity at 2 $M_L$.

\subsection{Initial conditions and runs performed}
We consider  spherical clouds in which turbulence has been added and is freely decaying. 
 The velocity perturbations present a powerspectrum that is equal to 11/3 and aim at reproducing 
a standard turbulent flow while the phases are random. Note that we do not start by running the code 
without gravity as \citet{leeh2018a} found that it makes little difference at least for the 
cases they explored.
To get relevant initial conditions, we looked at distributions of observed star forming clumps
such as the ones of the ATLASGAL \citep{Urquhart14} and Hi-GAL surveys \citep{elia2017}.
In both surveys the clump mass spans a range that typically goes from 100 to 10$^4$ M$_\odot$, with a few 
clumps that have lower or larger values. The radius has been found to depend on the mass, typically one has 
$M \propto R^2$ \citep[see][for an explanation of this relation]{lee2016a,lee2016b} but for a given mass, there is a spread 
in radius. Typically for 1000 solar mass clumps, the observed radius goes from 0.1 to 1 pc. We stress that 
the final galactic IMF should definitely be obtained by summing the stellar mass distribution of a clump distribution 
which reflects these observations \citep[see][]{lhc2017}. 

In this work, we consider clumps having a uniform density initially
  of mass $10^3 M_\odot$. They have initially either a radius of about 0.1 pc corresponding to an 
initial density of  5$\times 10^6$ cm$^{-3}$  or to 0.4 pc corresponding to a density of about 8$\times 10^4$ cm$^{-3}$.
Observationally, this seems to correspond to a very compact star forming clump and to a standard one. 
Below we refer to the first type of initial condition as COMP (for compact) and to the second as STAN (for standard).
 Indeed observations of massive star forming clumps found that a radius of 0.4 pc  is typical for 
a clump of $10^3 M_\odot$ while a radius of 0.1 pc corresponds to more extreme clouds \citep[e.g.][]{Urquhart14,elia2017}. 
The COMP runs have a freefall time of about 14 kyr while for the STAN ones it is about 110 kyr.
 The initial temperature is equal to 10 K and the ratio of thermal over gravitational 
energy is about 0.002 for the COMP cases and 0.008 for the STAN ones.
The initial value of the Mach number is 10 for the  COMP runs and 5 for the STAN runs which leads to the same 
turbulent over gravitational energy ratio. 
To investigate the influence of the initial Mach number, a STAN runs with ${\mathcal M}=10$ is also performed (STANHMACH). 
 This makes that the turbulent over gravitational energy ratio is about 0.2 for all clumps 
except for STANHMACH for which it is 0.8. Let us remind that \citet{leeh2018a} exploring the influence 
of the initial velocity dispersion onto the stellar mass spectrum concluded that its initial amplitude 
has a modest influence as long as the cloud is bound and that it is not too small \citep[][infer that it should be larger than $\simeq$0.1]{leeh2018a}.

To understand the impact of the radiative feedback processes, we perform various runs which  
include either none of them (NOFEED),  only the stellar luminosity  (NOACLUM) or both the stellar and the 
accretion luminosity  (ACLUM). Moreover as already mentioned, since the actual value of the effective accretion luminosity
that must be used is unclear, we perform runs for which the accretion luminosity is 
$G M_* d M/dt / R_*$ divided by 2 and by 10 (LOWACLUM). 
 
For comparisons with the  radiative transfer calculations, two runs with a barotropic equation of state are also done with 
two different values of the parameter $n_{\rm ad}$. 

Note that since the STAN type clouds are four times more spatially extended than the COMP type ones, it 
has not been possible to run the STAN simulations with the same spatial resolution as the COMP ones except for 
two runs (STAN-ACLUMhrhs and STAN-ACLUMvhrhs which are employed to investigate the issue of numerical convergence).
The COMP simulations have a nominal spatial resolution of about 2 AU while the STAN ones have 4 AU. 
Because of this difference the value of $n_{\rm acc}$ in runs STAN is chosen to be $10^{12}$ cm$^{-3}$ while it is 
equal to  $10^{13}$ cm$^{-3}$ for the runs COMP.

Finally,  to test the influence of numerical parameters, 
we have also performed  runs which have both lower and higher  spatial resolutions as well as runs which investigate  
the influence of  $n_{\rm acc}$.
 These runs are discussed in 
the appendix \S~\ref{num_res}.

Table~\ref{table_param_num} summarizes the various runs performed.

      \begin{table}
         \begin{center}
            \begin{tabular}{lcccccccc}
               \hline\hline
               Name & $R_c$ (pc) & $\mathcal{M}$ & lmax & $dx$ (AU)   & $n_{\rm ad}$ (cm$^{-3}$)  & $n_{\rm acc}$ (cm$^{-3}$)    & stellar lum & $f_{\rm acc}$  \\
               \hline
               COMP-ACLUM  & 0.1 & 10 & 15  & 2.3  &  NA & $10^{13}$  & yes & 0.5 \\ %CLUSTER_IMF_feedinter_eos_hres_hsink2_nobug
               COMP-LOWACLUM & 0.1 & 10   & 15   & 2.3  &  NA & $10^{13}$    & yes & 0.1 \\ %CLUSTER_IMF_feedinterlow_eos_hres_modcrit_hsink2
               COMP-NOACLUM & 0.1 & 10 & 15  & 2.3  &  NA   &  $10^{13}$  & yes & 0 \\ %CLUSTER_IMF_feedback_eos_hres_modcrit_hsink2
%               $R1hf$ & 15  & 2.3  &  NA   & 1 & 0 \\ %CLUSTER_IMF_feedback2_eos_res_modcrit_hsink2
               COMP-NOFEED  & 0.1 & 10 & 15  & 2.3  &  NA  & $10^{13}$   & no & 0 \\ %CLUSTER_IMF_nofeed2_eos_res_modcrit_hsink2
%               $R1acclum$ & 15  & 2.3  &  NA   & yes & yes \\ %CLUSTER_IMF_feednobug_eos_hres_modcrit_hsink2
               COMP-bar1  & 0.1 & 10 & 15  & 2.3 &  1.2 10$^{11}$   &  $10^{13}$  & NA & NA \\ %CLUSTER_IMF_baro_eos_hres_modcrit_hsink2
               COMP-bar2 & 0.1 & 10 & 15    & 2.3  &  4 10$^{10}$    &  $10^{13}$  & NA & NA \\ %CLUSTER_IMF_baro2_eos_hres_modcrit_hsink2
               \hline
               STAN-ACLUM  & 0.4 & 5 & 16   & 4.6  &  NA   &  $10^{12}$  & yes & 0.5 \\ %CLUSTER_IMF_feedinter_lowcd_eos_hres_hsink_nobug
               STAN-LOWACLUM & 0.4 & 5 & 16   & 4.6  &  NA   &  $10^{12}$  & yes & 0.1 \\ %CLUSTER_IMF_feedinter_lowcd_eos_hres_hsink_nobug
               STAN-NOFEED & 0.4 & 5 & 16  & 4.6  &  NA  &  $10^{12}$  & no & 0 \\ %CLUSTER_IMF_nofeed_lowcd_eos_hres_hsink_nobug
               STANHMACH-ACLUM  & 0.4 & 10 & 16   & 4.6  &  NA   &  $10^{12}$  & yes & 0.5 \\ %CLUSTER_IMF_feedinter_lowcd_eos_hres_highmach_hsink
               \hline
               \hline
               COMP-NOACLUMlrls  & 0.1 & 10 & 14   & 4.6  &  NA   &  $10^{12}$  & yes & 0 \\ %CLUSTER_IMF_feedback_eos_res_modcrit_hsink
               COMP-NOACLUMls  & 0.1 & 10 & 15   & 2.3  &  NA   &  $10^{12}$  & yes & 0 \\ %CLUSTER_IMF_feedback_eos_hres_modcrit_hsink
               COMP-NOACLUMhr  & 0.1 & 10 & 16   & 1.15  &  NA   &  $10^{13}$  & yes & 0 \\ %CLUSTER_IMF_feedback_eos_vhres_modcrit_hsink2
               \hline
               STAN-ACLUMhrhs  & 0.4 & 5 & 17   & 2.3  &  NA   &  $10^{13}$  & yes & 0.5 \\ %CLUSTER_IMF_feedinter_lowcd_eos_vhres_hsink_nobug
               STAN-ACLUMvhrhs  & 0.4 & 5 & 18   & 1.15  &  NA   &  $10^{13}$  & yes & 0.5 \\ %CLUSTER_IMF_feedinter_lowcd_eos_xhres_hsink2_nobug
            \end{tabular}
         \end{center}
         \caption{Summary of the runs performed.  lmax is the maximum level of grid used. For the COMP runs  
$lmax=15$ corresponds to about 2.3 AU of resolution while for the STAN runs $lmax=16$ corresponds to a resolution of about 4.6 AU. 
$n_{\rm ad}$ is the density at which the 
gas becomes adiabatic. Stellar  luminosity indicates whether it is taken into account and $f_{\rm acc}$ gives the fraction of the accretion 
luminosity which is taken into account in the calculation. 
NA stands for non-applicable.}
\label{table_param_num}
      \end{table}

\section{General description}

\subsection{Illustrations for a specific case}
To illustrate the simulation results,  large scale and small scale images are displayed.
Figure~\ref{coldens_diff-aclumhrhs} portrays two snapshots of runs STAN-ACLUMhrhs at early time after only a few sink
particles (black dots) have formed and later when a significant fraction of the gas has turned into stars. 
The stars tend to form in dense filaments and are strongly clustered. 
A small scale view is given in Fig.~\ref{coldens_diff-aclumhrhs_sink}, where the column density around 
the second sink particle that has formed in the simulation, is displayed at three snapshots.
On the second and third snapshots, the clustering is also clear. The object distribution is 
clearly hierarchical. A disk like structure is seen in the second snapshot and two objects appear to have formed 
as a consequence of disk fragmentation. Two objects have formed slightly further away and at least 
one of them is surrounded by a disk. Two more objects can be seen at time 0.104 Myr which are already 
decoupled from their gas reservoir. 
As expected, more objects weakly correlated to the gas, appear in the last snapshot.
It is worth remaining here that since magnetic field is not included here, the disks
are likely too large and fragment too easily \citep[see e.g.][]{liwurster2019,hetal2020,zhao2020}.

Temperature cuts through the yz plane are also given (right column).
As can be seen the central region around the sink presents temperatures that peak around 1500 K, which is 
typical of the temperatures reached at scales smaller than a few AU in a collapsing protostar. 
At a distance of a few hundreds of AU, the temperature drops below or becomes comparable to 100 K. 
Overall the temperature 
distribution remains less structured than the column density for instance. 
As time goes on the mean temperature increases which is due to the larger number of stars that 
have formed as quantified below.

\subsection{Accretion as a function of time}
As accreted mass onto sink/star is of primordial importance in these simulations, 
Fig.~\ref{time_mass} displays the total sink mass, $M_{*,tot}$, as a function of time for the 
COMP and STAN-type runs. As the initial densities in the two series of runs differ 
by almost two orders of magnitude (a factor 64), the freefall time and therefore 
the accretion times differ by a factor of about 8-10.   
As can be seen the accretion luminosity has only a modest influence 
on the global accretion except initially for COMP-type runs where we see that it 
almost stops accretion for a brief period of time before the first solar mass 
of gas has been accreted. As time goes on and after a few solar mass of gas  
is accreted, the mass ratio is roughly a factor of 2 and this ratio keeps decreasing 
with time. The difference between NOFEED and ACLUM runs for the more diffuse clumps 
(STAN-type) is weaker with mass differences of only a few tens of percents. 
These curves constitute a first indication that the accretion luminosity is 
playing some role during the collapse of a massive clump without changing 
drastically the final result. This is  extensively discussed below.

Also plotted is STANHM-ACLUM that we remind has a Mach number of 10 initially instead 
of 5 for STAN-ACLUM.  Clearly, the higher turbulence modifies the accretion history. 
Star formation starts a bit earlier and this is because some velocity fluctuations 
help compressing the gas locally. However globally the accretion rate is a bit lower 
(by 20-30$\%$) and this is because turbulence exerts some support on the cloud 
at large scale.

\section{Temperature distributions}
The gas temperature is strongly influenced by the radiative feedback and here we investigate 
its distribution within the clumps. 

\subsection{Temperature-density histogram}
To get insight on the physical conditions that prevail within the simulated clouds, 
we now present in Figs.~\ref{T_rho_comp} and~\ref{T_rho_diff} the mean temperature as a function of gas density
for several timesteps (which are more easily referenced by their accreted mass $M_{*,tot}$). 
This is obtained by simply computing the mass weighted temperature in all density intervals. 
While the information carried by the mean temperature is  incomplete, it is relatively simple, which 
facilitates the comparisons between runs. Bidimensional histograms, that contains much more detailed 
information are given in \S~\ref{bidim}. The dotted lines visible in Figs.~\ref{T_rho_comp} and~\ref{T_rho_diff}
represent the analytical expression stated by eq.~(\ref{eq_full_eos}).

In all runs, we observe a change between an isothermal and a non-isothermal, adiabatic-like, regime around 
10$^{8-9}$ cm$^{-3}$. Note that the adiabatic regime appears to be only poorly described by the analytic functions.
This clearly is a consequence of the heating that results from the emitted radiation. However the 
discrepancy is amplified 
 partly by the averaging procedure and partly due to insufficient resolution (see \S~\ref{bidim}).
The non-monotonic behaviours, in particular the bump located around 10$^{10}$ cm$^{-3}$,  are due to 
the averaging procedure and to the presence of high temperature 
gas as revealed in Fig.~\ref{Tcoldens}. 
 
As expected, the temperatures in run COMP-LOWACLUM are lower than the ones of run COMP-ACLUM, typically 
by a factor on the order of 1.5-2. Anticipating the analytical development made below, this is expected since 
the temperature typically varies like $L_{\rm acc}^{1/3-1/4}$ and the luminosities of the two simulations, differ by a factor 5. 

The comparison with run COMP-NOACLUMN (that we remind takes into account the stellar luminosity but not the accretion one), 
shows that in a first phase ($M_{*,tot} < 100$ M$_\odot$) the temperature in run COMP-NOACLUMN remains typically 3-4 times
below the temperature of
the runs which take the accretion luminosity into account. However, at later times, when more gas has been turned into stars, 
several stars more massive than a few solar mass formed and the stellar luminosity leads to temperatures that are roughly 
only a factor 2 below the ones of runs COMP-LOWACLUM.

Finally, the bulk temperatures of run COMP-NOFEED  remains low, typically around 20-30 K, even when $M_{*,tot}$ is larger than 200-300 M$_\odot$.

Overall the temperatures of the series of STAN-type runs are up to  three times lower. For instance 
in run STAN-ACLUM the temperature at low density in on the order of 50 K for $M_{*,tot} = 330$ M$_\odot$, while for 
run STAN-LOWACLUM, it is roughly 30 K. This clearly is because $i)$ the cloud is more extended so the distances from the sources
are more important in STAN-type runs than in the COMP-type ones and $ii)$ the accretion rate is lower for the former than for the 
latter. This is quantified in the next section. 

As for the COMP-NOFEED run,  the temperature of the STAN-NOFEED run is significantly lower than when the accretion luminosity 
is taken into account, the largest temperature obtained at low density is only about 20 K.

To explore further how initial conditions influence the temperature distribution, right-bottom panel of Fig.~\ref{T_rho_diff}
presents run STANHmach-ACLUM, which initially has a Mach number of 10 instead of 5 for run STAN-ACLUM. The accreted mass remains 
below 200 $M_\odot$ because the clump is marginally bound. Comparing the temperatures of run STAN-ACLUM and STANHMach-ACLUM when the same
amount of mass has been accreted, we see that the temperatures are slightly lower for run STANHmach-ACLUM. This is because the 
accretion rate is lower in this run than in run STAN-ACLUM.

\subsection{Analytical developments: predicting the temperature distribution}
As temperature distribution  plays an important role in the clump evolution
both regarding its fragmentation and its chemical composition, we provide here analytical 
estimates. More specifically, we will estimate here  
the clump mass per units of accreted mass, which   lays above a certain temperature threshold
chosen to be 100 K. 
The calculation entails several steps. First, we estimate the temperature profile
of an envelope (with a density profile assumed to be $\propto r^{-2}$) around a source that 
is emitting a flux $f_{\rm acc} G M_* dM/dt / R_*$. Second we obtain the 
accretion rate for a source of mass $M_*$ and third we choose (based on 
former studies) the mass spectrum of the stars. Finally, we perform an integration over
the mass spectrum to get the heated mass of gas per units of accreted mass.

\subsubsection{Temperature distribution around a single source}
We consider a spherically symmetric clump with a central source of mass $M_*$ 
accreting at a rate $\dot{M}_*$. The gas density is further assumed to be 
\begin{eqnarray}
\label{rho_sis}
\rho (r) = {\delta _\rho  C_{s,0}^2 \over 2 \pi G r^2},
\end{eqnarray}
where $\delta _\rho$ is a dimensionless factor which typically is equal to 10-200 as discussed in \S~\ref{delta_sink}. 
We further assume that gas, dust and radiation, have the same temperature and are all 
stationary. A single radiation frequency is considered and since the medium is optically thick, we 
have 
\begin{eqnarray}
\label{transfert}
- 4 \pi r^2  {c \over 3 \kappa (T) \rho (r) } \partial _r (a T^4) = f_{\rm acc} {G M_* \dot{M}_* \over R_*}, 
\end{eqnarray}
Following \citet{semenov2003}, we can distinguish two regimes of temperature, 
\begin{eqnarray}
\label{kappa}
\kappa (T) \simeq 5 \, \rm{cm}^2 \rm{g}^{-1} \; \rm{for} \; T > T_{\rm crit} \simeq 100 \,  \rm{K} , \\
\kappa (T) \simeq 5 \, \rm{cm}^2 \rm{g}^{-1} \left( { T \over T_{\rm crit}} \right)^{\alpha} \; \rm{for} \; T < T_{\rm crit}.
\nonumber
\end{eqnarray}
where $\alpha$ is typically between 1 and 2. In this work we adopted $\alpha=1.5$.
Combining eqs.~(\ref{rho_sis}),~(\ref{transfert}) and~(\ref{kappa}), we get
\begin{eqnarray}
\label{temp}
 T(r) = \left( T_{\rm crit}^4 + K \left( {1 \over r^3} - {1 \over r_{\rm crit}^3}  \right)  \right)^{1/4} \; \rm{for} \; T > T_{\rm crit}, \\
 T(r) = \left( T_{\infty}^{4-\alpha} + K T_{\rm crit}^{-\alpha} {4 - \alpha \over 4}  {1 \over r^3}    \right)^{1/(4-\alpha)} \; \rm{for} \; T < T_{\rm crit},
\nonumber
\end{eqnarray}
where $T_{\infty}$ is the temperature at infinity and  $T_{\infty}= T_0 = 10$ K initially, 
\begin{eqnarray}
K = B_{\rm rad} f_{\rm acc} \delta _\rho C_{s,0}^2  M_* \dot{M}_*, \\
B_{\rm rad} =   {3 \kappa  \over 24 \pi ^2 R_* a c  },
\label{Brad}
\end{eqnarray}
and $r_{\rm crit}$ is the radius at which $T=T_{\rm crit}$ and is given by
\begin{eqnarray}
\label{rcrit}
r_{\rm crit} ^3 =   K   {4 - \alpha \over 4}   { T_{\rm crit}^{-\alpha} \over   T_{\rm crit}^{4-\alpha}  -  T_{\infty}^{4-\alpha}  } 
\simeq  K   {4 - \alpha \over 4}    T_{\rm crit}^{-4}
\end{eqnarray}

Left panel of Fig.~\ref{masstemp_msink} displays $T(r)$ for three cases corresponding 
roughly to a low mass protostar ($M_*=$0.1 M$_\odot$, $dM/dt$=$10^{-5}$ M$_\odot$ yr$^{-1}$ and $\delta_\rho=10$, red line), 
a protostar with intermediate mass 
(1 M$_\odot$, $10^{-4}$ M$_\odot$ yr$^{-1}$, 30, dark line) and a more massive protostar (5 M$_\odot$, $10^{-3}$ M$_\odot$ yr$^{-1}$, 100, blue line).

\subsubsection{Accretion rate}
To estimate the accretion rate on each star, we proceed like in \citet{leeh2018a} who have estimated 
it to be $\dot{M} \simeq M / \tau_{\rm ff}$, where $M$ is the mass of the reservoir from which 
the star is building its mass and $\tau_ {ff}$ the associated freefall time, $\sqrt{3 \pi / (32 G \rho)}$. To get the 
accretion reservoir, for simplicity we assume that its mass is nearly the one of the star (for instance jets that 
may change this efficiency are not considered in the present work), meaning that all 
mass losts are neglected,  while the reservoir radius is determined 
by the virial theorem which leads to 
\begin{eqnarray}
M  = {\pi^{5/2} \over 6} { \Bigl[ (C_s)^2 + (\sigma_c^2 / 3) (R /  R_c )^{2 \eta} \Bigr]^{3 \over 2}   \over \sqrt{G^3 \rho  }  },
\label{mass_jeans}
\end{eqnarray}
where $\sigma_c$ is the velocity dispersion at the cloud scale, $R_c$ is the clump radius and $\eta \simeq 0.5$ is 
the exponent through which the velocity dispersion varies with spatial scale,
 $\sigma \propto R ^\eta$ as expected for a turbulent fluid. 
This leads to 
\begin{eqnarray}
\tau _{\rm ff}  = \sqrt{ 3 \over 2} \pi^{-1/4} { R  \over \left( (C_s)^2 + (\sigma_c^2 / 3) (R /  R_c )^{2 \eta}   \right)^{1/2} }.
\label{freefall}
\end{eqnarray}
Combining eq.~(\ref{mass_jeans}) and eq.~(\ref{freefall}), we have the freefall time and therefore the accretion
rate as a function of the mass. 

To get further physical hint, it is worth simplifying this expression, which can be achieved by neglecting the 
sound speed with respect to the turbulent dispersion in eqs.~(\ref{mass_jeans}) and (\ref{freefall}). 
This leads to 
\begin{eqnarray}
\nonumber
\tau _{\rm ff}  &=&  {3  \over \sqrt{2}} \pi^{-1/4}     \sigma_c^{-1}  R \left( {R \over  R_c } \right)^{ -1/2} \\
 &=&  K_{\rm ff}  G^{1/4} \sigma_c ^{-3/2} R_c ^{3/4} M^{1/4}, 
\label{freefall_approx}
\end{eqnarray}
where we have assumed $\eta=0.5$ and
where $K_{\rm ff} = 3  \pi^{-1/4}  \left( 6  \pi  \right) ^{1/3} / \sqrt{2}$.
This expression, which will be used later in the final expression of the heated gas mass per mass of stars, $f_M$,  
implies that only the contribution of the most massive stars will be accurately represented. However, by 
comparing the two estimates inferred from eq.~(\ref{freefall}) and eq.~(\ref{freefall_approx}), 
we found that this is a valid approximation.

\subsubsection{Source distribution}
To get the temperature distribution inside the clouds, we need to know the source 
distribution. We assume that the mass spectrum is given by
\begin{eqnarray}
{\cal N} = { d N  \over d \log M} = A M^{-\beta},
\label{mass_spectrum}
\end{eqnarray}
This mass spectrum 
applies between a minimum and maximum mass, respectively $M_{\rm min}$ and $M_{\rm max}$.
While the former is typically equal to 0.3 M$_\odot$, which corresponds to the 
peak of the IMF, the latter increases with the total accreted mass, $M_{tot,*}$ and 
is equal to a few solar mass. Obviously this is a simplification since one should 
sum over the full mass spectrum. However as seen below the dependence on
$M_{\rm min}$ is quite shallow and its exact value is not really consequential. 
We have 
\begin{eqnarray}
\int _{M_{\rm min}} ^{M_{\rm max}} M { d N  \over d \log M} = {A \over {-\beta+1} }  \left( M_{\rm max}^{-\beta+1} - M_{\rm min}^{-\beta+1} \right) = M_{tot,*},
\end{eqnarray}
which leads to 
\begin{eqnarray}
\label{Anorm}
 A  = { (-\beta+1) \over  \left( M_{\rm max}^{-\beta+1} - M_{\rm min}^{-\beta+1} \right) } M_{tot,*}.
\end{eqnarray}

\subsubsection{Heated mass fraction}
The mass enclosed in the sphere of radius $r_{\rm crit}$ is given by 
\begin{eqnarray}
m_{\rm crit} =  \int _0 ^{r_{\rm crit}} 4 \pi r^2  \delta _\rho {C_{s, 0}^2 \over 2 \pi G r^2 } dr   \simeq
{2 \delta _\rho C_{s,0}^2 \over G } r_{\rm crit}.
\end{eqnarray}

Thus the total mass heated above $T_{\rm crit}$ can be estimated as 
\begin{eqnarray}
\label{total}
\nonumber
M_{\rm crit} &=& \int _{m_{\rm min}} ^{m_{\rm max}} m_{\rm crit} {\cal N} d \log M \\
&=&  {2 f_{\rm acc} ^{1/3} 
\delta _\rho^{4/3} C_{s,0}^{8/3} \over G T_{\rm crit}^{4/3} }  A B _{\rm rad} ^{1/3}   \left( {4 - \alpha \over 4} \right)^{1/3}    
\int _{m_{\rm min}} ^{m_{\rm max}} \left( {M^2 \over \tau_{\rm ff}(M) } \right)^{1/3}  M^{-\beta}  d \log M, 
\end{eqnarray}
which is the expression that we will use below to confront with the simulation results. 
Note at this stage that a difficulty arises regarding the choice of $\beta$, the exponent 
of the stellar mass spectrum (as stated by eq.~\ref{mass_spectrum}). Most observations 
found that above a mass of few solar mass, the exponent is close to 1.3, which is the value 
originally inferred by Salpeter. However, as explained below and in \citet{leeh2018a}, the mass spectra 
obtained in numerical simulations of massive collapsing clumps, tend to be a bit flatter and are more accurately described 
by an exponent of 3/4-1 (followed by an exponential cut-off at highest masses). 
Therefore for the purpose of comparing with the numerical results, we will use from 
this point the value of 3/4 although observationally it would be more logical to use the value of 1.3.
Fortunately, it makes little difference as the results with the two values of $\beta$ vary by 
a few tens of percents, which is far below the expected accuracy of our analytical approach.

As it is useful to obtain a simpler expression, we 
use the simplified expression for the freefall time as 
stated by eq.~(\ref{freefall_approx}). Assuming further than 
$\eta=0.5$ and $\beta=3/4$, we then get

\begin{eqnarray}
\label{total_approx}
f_{M,crit} = {M_{\rm crit} \over M_{tot,*} }  = 
  3^{2/3}   K_{\rm ff}^{-1/3}  \left( 4 - \alpha \right)^{1/3}  f_{\rm acc} ^{1/3}
 {   B _{\rm rad} ^{1/3}  \over G^{13/12}  }  
{  M_{\rm min}^{-1/6} - M_{\rm max}^{-1/6}  \over  M_{\rm max}^{1/4} - M_{\rm min}^{1/4}  }  
{\delta _\rho ^{4/3}  C_{s,0}^{8/3} \over T_{\rm crit}^{4/3} } \sigma_c ^{1 / 2 } R_c^{-1/4},
\end{eqnarray}
where $f_{M,crit}$ is the mass of gas having a temperature larger than $T_{\rm crit}$ per units of 
accreted mass. As we see, it weakly depends on the minimum and maximum stellar mass present in the sample. 
It depends on the clump physical conditions through the radius $R_c$, the velocity dispersion, $\sigma_c$ and 
the over-density, $\delta_\rho$. The  values of this latter vary with the clump parameters and is typically 
 on the order of 10-100. Through $B_{\rm rad}$ and eq.~(\ref{Brad}), we see that 
 $f _{M,crit} \propto f_{\rm acc} ^{1/3}$. Thus variations of $f_{\rm acc}$ induce limited changes of $f _{M,crit}$.

If we further assume that $\sigma_c = \sqrt{G M_c / R_c}$, which is close to the initial 
value that has been chosen that simply reflects energy equipartition,   we obtain 
\begin{eqnarray}
\label{total_approx_num}
f _{M,crit} = {M_{\rm crit} \over M_{tot,*} }  \simeq 10^{-2} \times \delta _\rho ^{4/3}   f_{\rm acc} ^{1/3}
\left( {M_c \over 1000 \, {\rm M}_\odot} \right) ^{1 / 4 } \left( {R_c \over 1 \, {\rm pc} } \right)^{-1/2},
\end{eqnarray}

\subsubsection{Feedback efficiency: comparisons between simulations and theory}

\setlength{\unitlength}{1cm}
\begin{figure*}%[h!]
%\centering
\begin{picture} (0,8)
\put(0,0){\includegraphics[width=8cm]{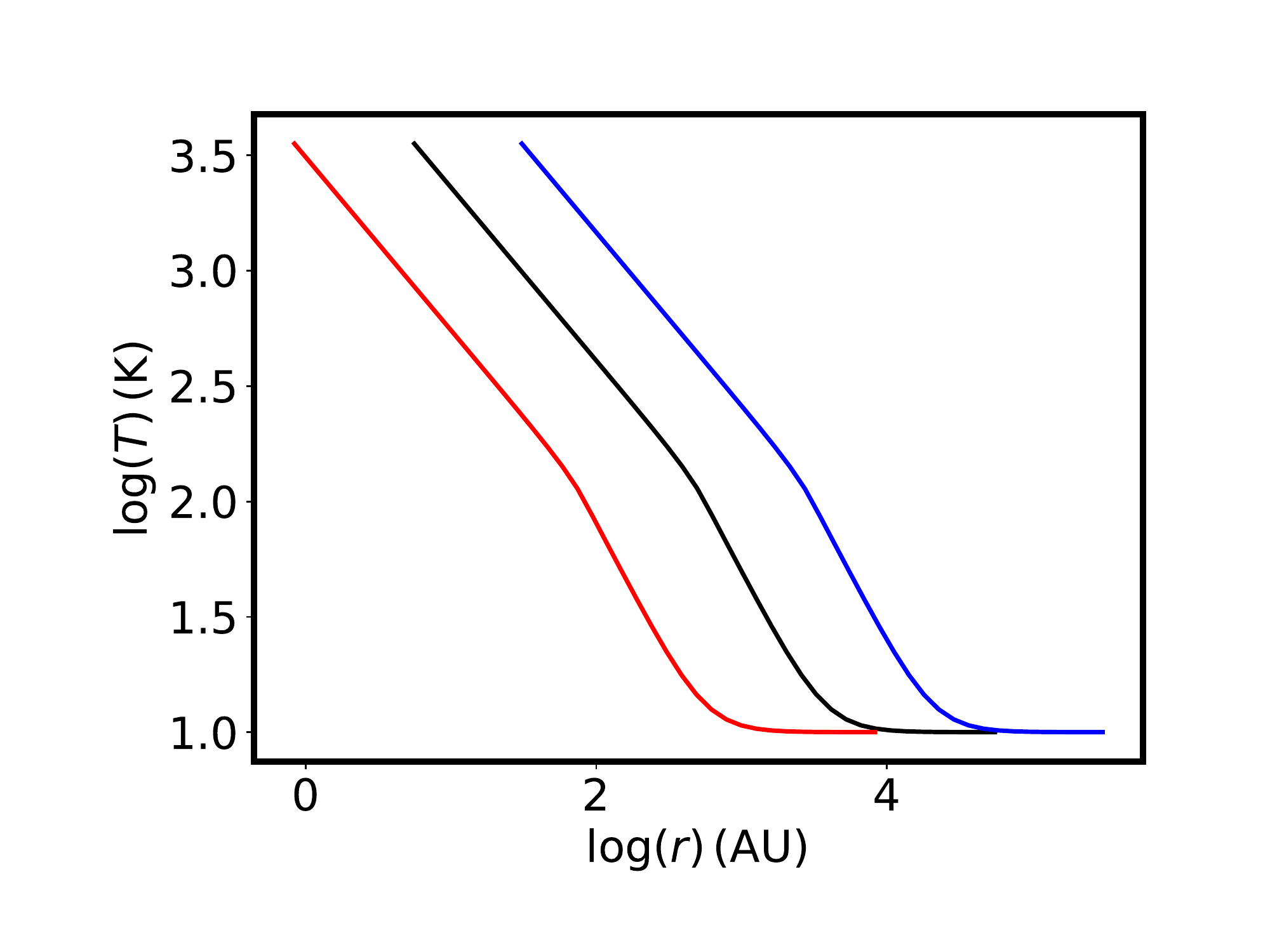}}  
\put(8,0){\includegraphics[width=8cm]{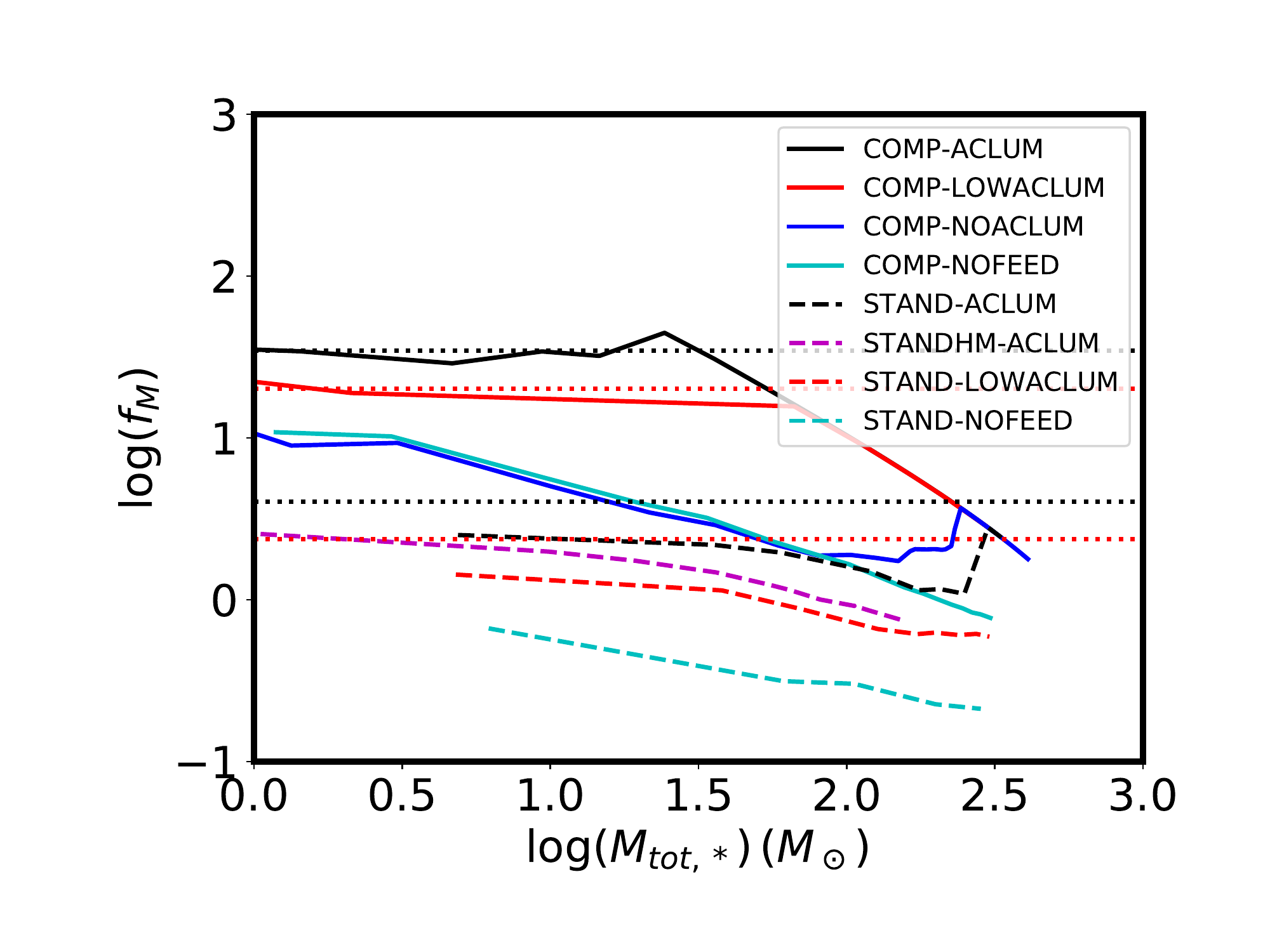}}  
\end{picture}
\caption{Left panel shows the temperature stated by eqs.~(\ref{temp}) for three cases  corresponding to 
three values of $M_*$, $dM/dt$ and $\delta_\rho$ namely 0.1 M$_\odot$, $10^{-5}$ M$_\odot$ yr$^{-1}$, 10 (red line); 
1 M$_\odot$, $10^{-4}$ M$_\odot$ yr$^{-1}$, 30 (dark line) and 5 M$_\odot$, $10^{-3}$ M$_\odot$ yr$^{-1}$, 100 (blue line).
 Right panel displays the mass above 100 K per mass of gas turned into stars for the various runs. The dotted lines correspond to the 
analytical expression stated by eq.~(\ref{total_approx_num}) with values of $R_c$, $f_{\rm acc}$ and $\delta _\rho$ 
that correspond to each specific run. While the values of  value of $R_c$ and $f_{\rm acc}$ are specified, 
$\delta _\rho$ (stated by eq.~\ref{rho_sis}) is measured in the simulations.
As can be seen the agreement between the analytical expression and values inferred from the simulation is entirely reasonable.
}
\label{masstemp_msink}
\end{figure*}

Figure~\ref{masstemp_msink} portrays the values of $f _{M,crit}$, i.e. the mass of gas having 
a temperature higher than $T_{\rm crit}$ per units of accreted mass, for the various runs performed 
as a function of the accreted mass, $M_{tot,*}$. Solid lines represent the COMP-type runs 
while dashed lines display the STAN-type ones. The dotted lines represent the analytical models 
where the parameters entering in eq.~(\ref{total_approx_num}), namely, $R_c$, $f_{\rm acc}$ are taken 
from Table~\ref{table_param_num}. On the other hand the parameter $\delta_\rho$ is estimated from the 
simulations (see \S~\ref{delta_sink}). 
For COMP-type runs we have $\delta_\rho \simeq 30-300$ and $\delta_\rho \simeq 10-100$ for the STAN-type ones. 
To perform the calculations of the models in Fig.~\ref{masstemp_msink}, we have used the 
values $\delta_\rho =150$ and 50 respectively. Note that from Fig.~\ref{sink_delta}, it is 
seen that $\delta_\rho$ tends to increase with $M_*$, the dependence on $M_{\rm max}$ 
stated in eq.~(\ref{total_approx_num}) may actually be underestimated. 

As expected for the runs where $f_{\rm acc} \neq 0$ (i.e. ACLUM and LOWACLUM-type runs),
 $f_M$ is almost independent of $M_{tot,*}$.
The change of slope at 30-100 M$_\odot$ for the COMP-ACLUM runs is due to the fact that all the gas is warm in 
the computational box and so $M_{\rm crit}$ does not increase while 
 $M_{tot,*}$, the mass within the sinks, keeps increasing. The agreement between the model (dotted lines)
and the simulations (solid and dashed lines) is reasonable. In particular, the trends are well reproduced.  

Typically we have $f_M \simeq 30$ for COMP-ACLUM run while 
$f_M \simeq 3$ for STAN-ACLUM run. Altogether a star of  mass, $M_*$, is able to hit above $T_{\rm crit}$ 
almost ten times more gas in the compact cloud than in the diffuse one. This is a consequence of 
the density which is   3-4 times larger for the former compared to the latter. As expected 
$f_M$ is considerably smaller, about a factor of ten, in runs where there is no accretion luminosity ($f_{\rm acc}=0$),
than in run COMP-ACLUM. We also note that $f_M$ is a decreasing function of $M_{tot,*}$ for runs for which  
$f_{\rm acc}=0$.

We therefore conclude that eq.~(\ref{total_approx_num}), which gives the expression of $f_M$, is accurate within a factor of a few 
and reproduces the qualitative behaviour observed in the simulations.

\subsection{Qualitative comparison with observations}
As discussed above the temperature distribution reflects the gas mass distribution and the
evolution of the clumps, i.e. the fraction of gas that has been converted into stars
assuming, as mentioned previously,
that all the mass of the star is about the mass of the reservoir.
Therefore comparing with observations is not an easy task as these quantities are generally poorly 
known. 
There are also various techniques such as sed fitting and molecular spectroscopy which 
provide  different results depending on which regions of the clump is actually probed
\citep[see for instance Fig.~11 of][]{giannetti2017}.

Both the ATLASGAL \citep{Urquhart14} and Hi-GAL surveys \citep{elia2017} provide 
mean temperature distributions. Looking for instance at Fig.~5 of \citet{elia2017}, 
we see that the temperatures of protostellar sources is higher than the ones of 
the prestellar clumps indicating internal heating. However the peak of the distribution 
is about 13 K and only 
few protostellar sources present temperatures above 30 K. Since the sample contains both 
massive and compact clumps, comparable to the ones simulated here, this may 
place constrains on the effective $f_{\rm acc}$ although the sed fitting has been 
restricted to temperature below 40 K. Similar numbers are provided by 
\citet{Urquhart14} (Fig.~10) where the NH$_3$ molecules has been used, although sources with 
temperatures larger than 45 K have been discarded. 
Using other molecular tracers (such as CH$_3$OH for instance), \citet{giannetti2017} infer temperatures 
for massive star forming clumps 
selected for the TOP100 sample \citep{csengeri2016}. Temperatures as high as few hundreds of K are reported 
but this may correspond to the inner part of the clumps. Indeed the temperature spatial 
distribution is a clue to assess the importance of thermal feedback.

In this respect, an interesting set of observations has been undertaken by \citet{ginsburg2017} who mapped 
several massive star forming clumps and infer temperatures using rotational diagram of 
CH$_3$OH (see also Fig. 3 of \citet{motte2018} where temperature above 60-80 K are obtained 
for massive clumps at scales of several thousands of AU). 
The data of  \citet{ginsburg2017} reveal temperatures exceeding 100 K extending up to 5000 AU. 
For instance Fig.~6 of \citet{ginsburg2017} shows for the clump e2, temperatures of 100-200 K at distances 
larger than 10$^4$ AU from the center of the source. While the mass in the region around e2 is estimated 
to be on the  order of 10$^4$ M$_\odot$, it contains about 500 M$_\odot$ within the central 10$^4$ AU. 
Since it contains massive stars and a total stellar mass higher than 50 M$_\odot$ \citep[Ginsburg private communication,][]{goddi2018}, 
it is broadly comparable to our COMP-type clumps at an age where at least 50 M$_\odot$ have been accreted. 
This may be comparable (within a factor 2-3) with what has been inferred for run 
COMP-ACLUM but also for run COMP-LOWACLUM. 
At this stage because of the broad uncertainties, it does not seem possible to draw 
strong conclusions and this remains a challenge for future studies.

\setlength{\unitlength}{1cm}
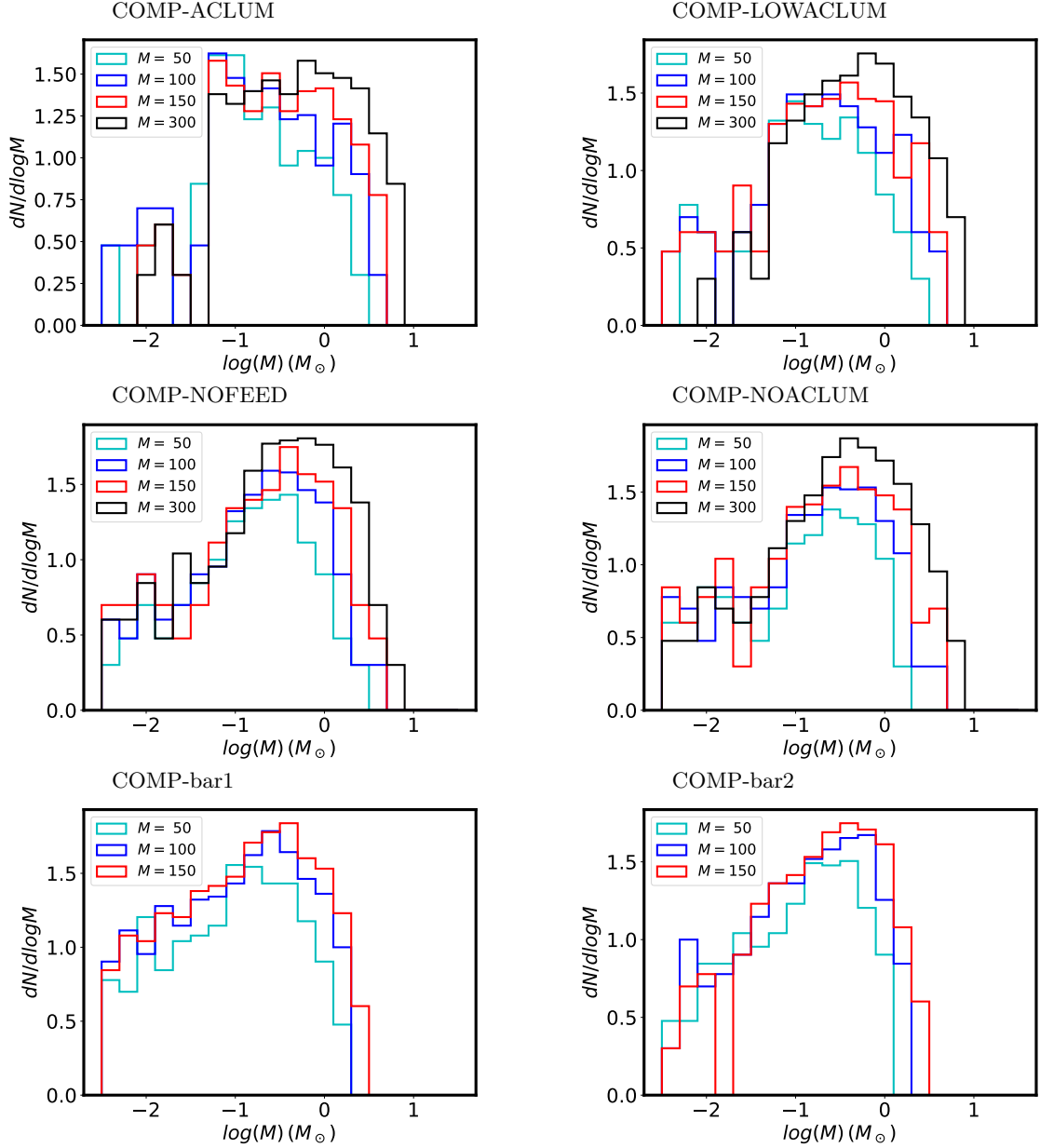
\begin{figure*}%[h!]
%\centering
\begin{picture} (0,17)
\put(0,11){\includegraphics[width=8cm]{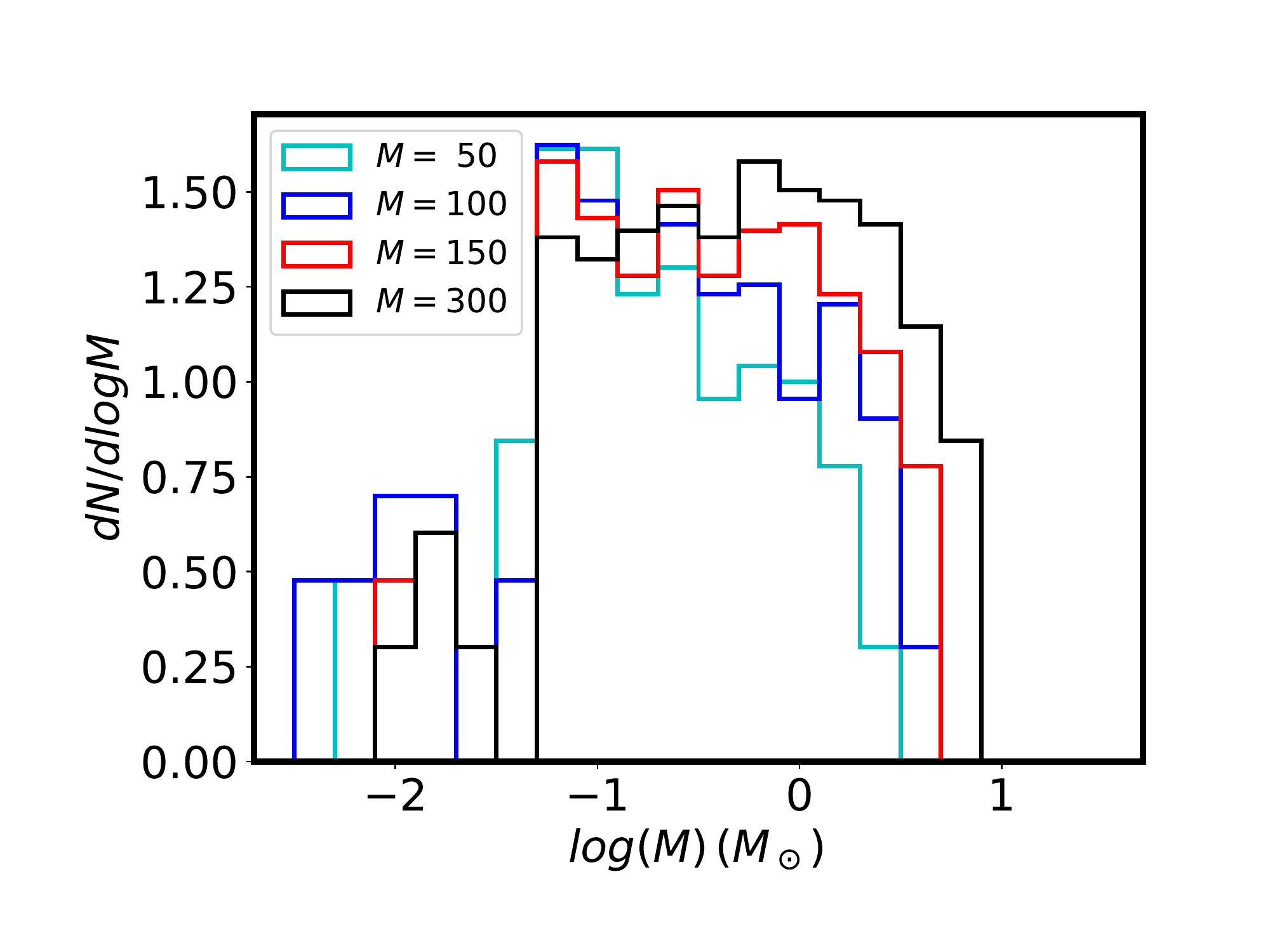}}  
\put(2,16.6){COMP-ACLUM}
%\put(8,11){\includegraphics[width=8cm]{FIG_PAPIER_RHDIMF/CLUSTER_IMF_feedback2_eos_hres_modcrit_hsink2/sink_hist_mass_serie.pdf}}  
\put(8,11){\includegraphics[width=8cm]{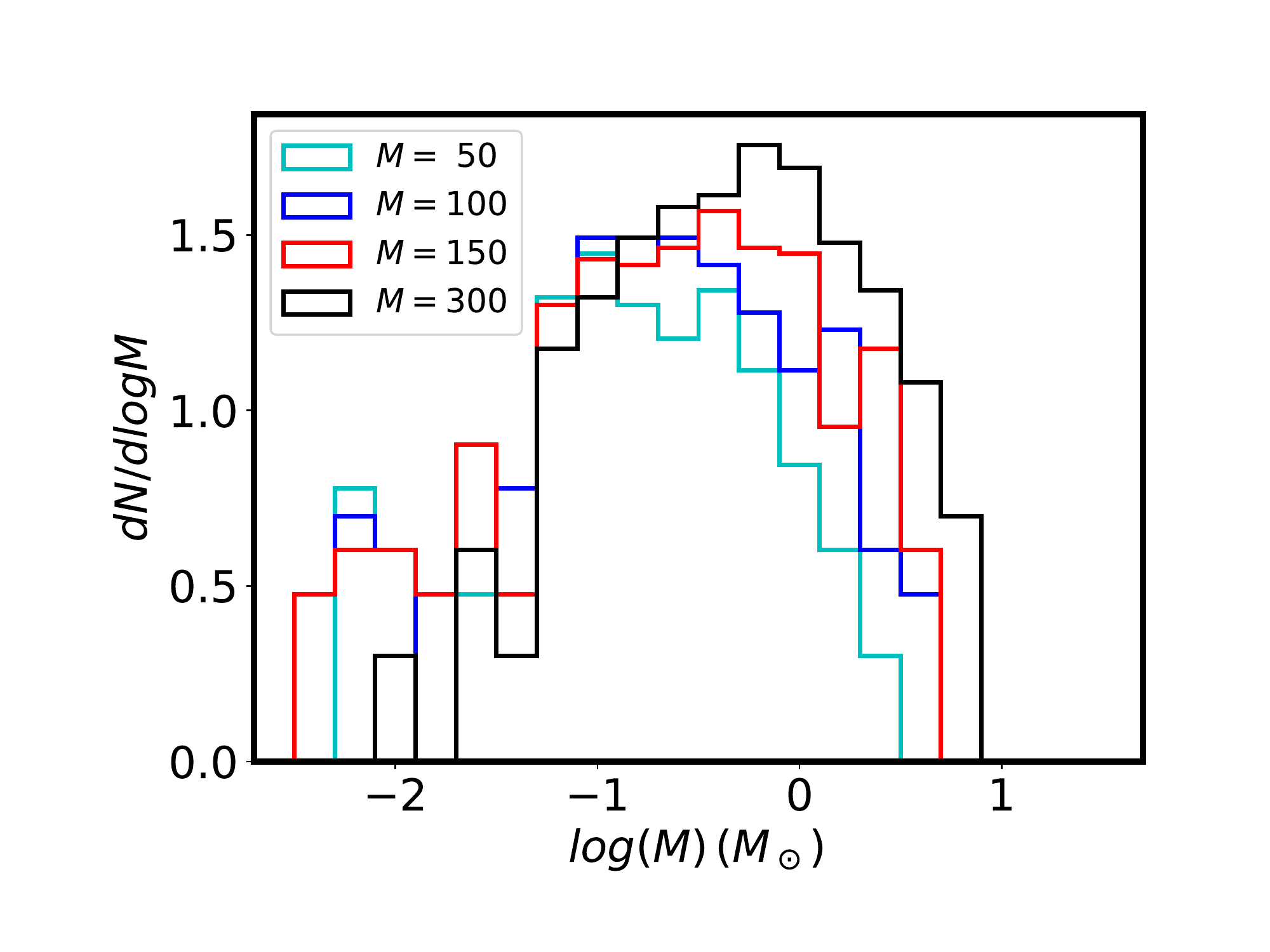}}
\put(10,16.6){COMP-LOWACLUM}
\put(0,5.5){\includegraphics[width=8cm]{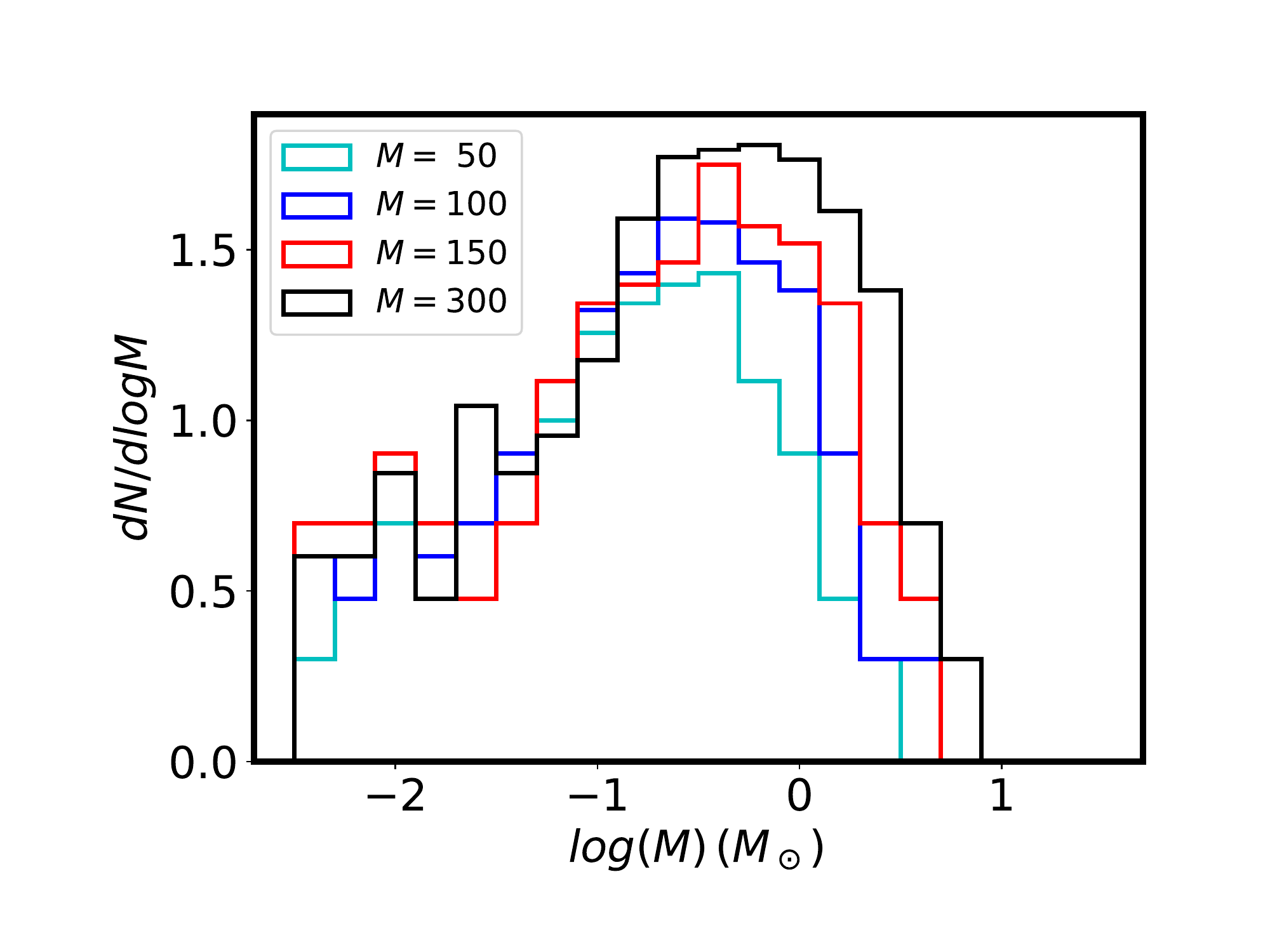}}  
\put(8,5.5){\includegraphics[width=8cm]{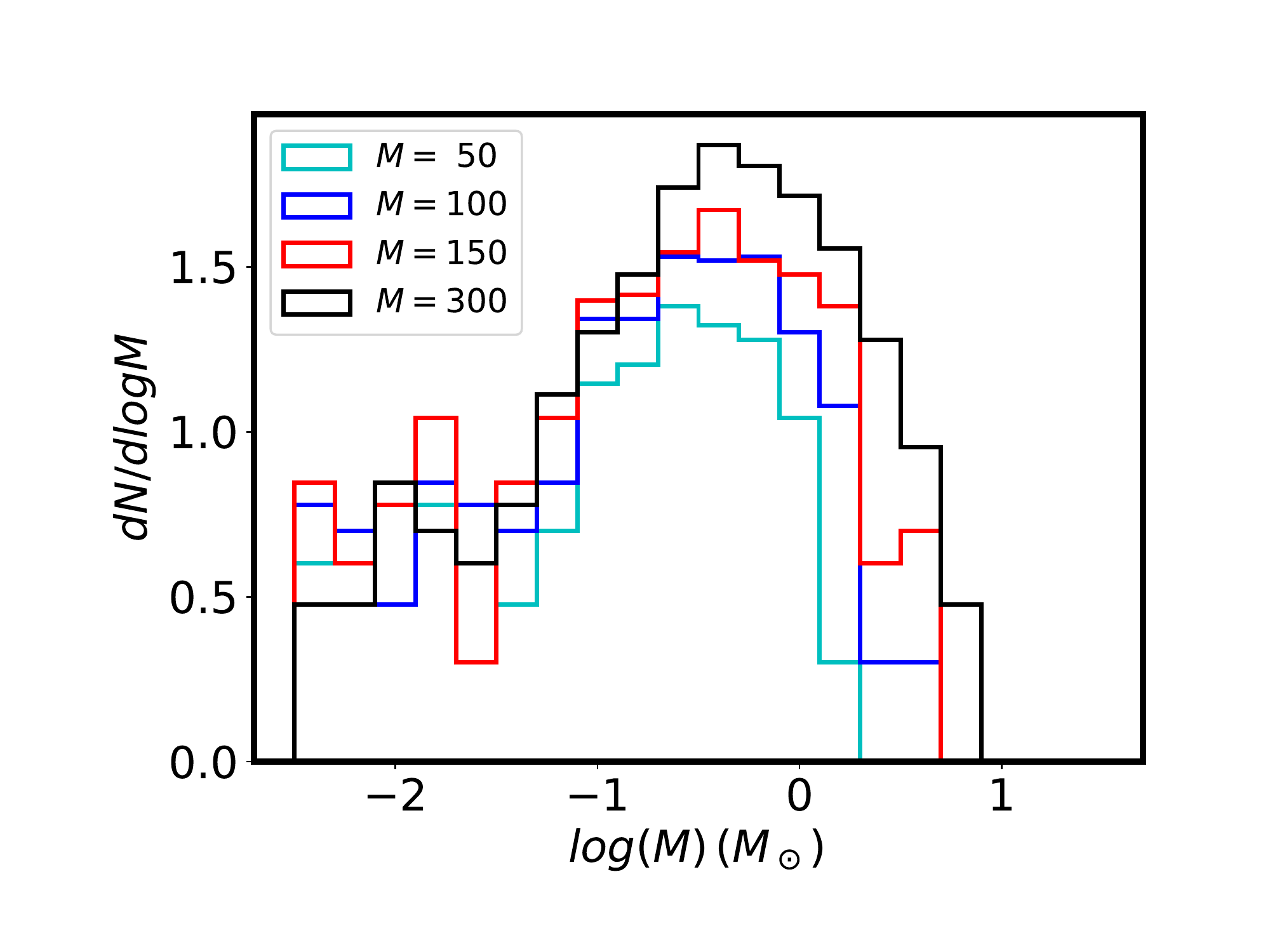}}  
\put(2,11.1){COMP-NOFEED}
\put(10,11.1){COMP-NOACLUM}
\put(8,0){\includegraphics[width=8cm]{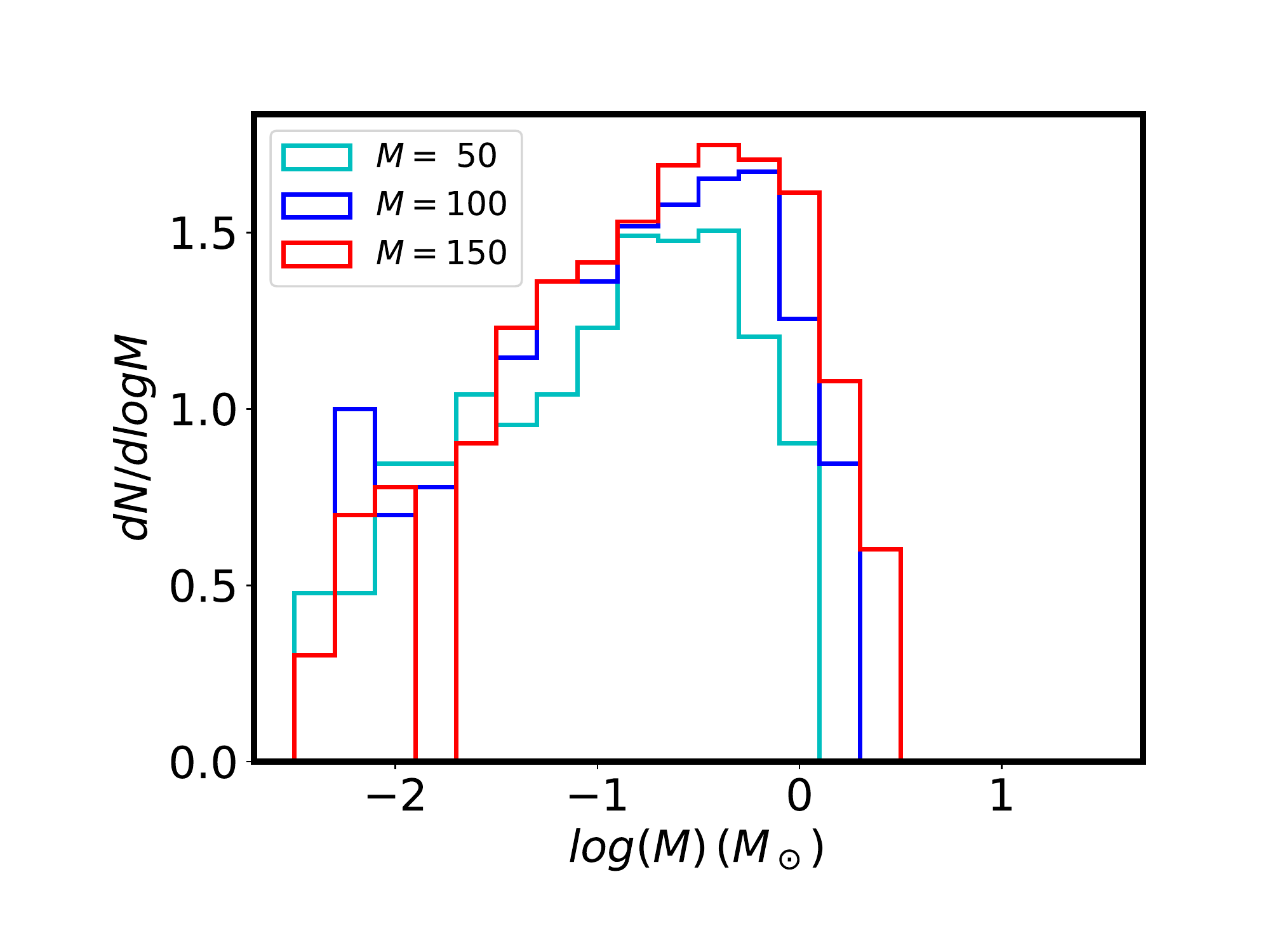}}  
\put(0,0){\includegraphics[width=8cm]{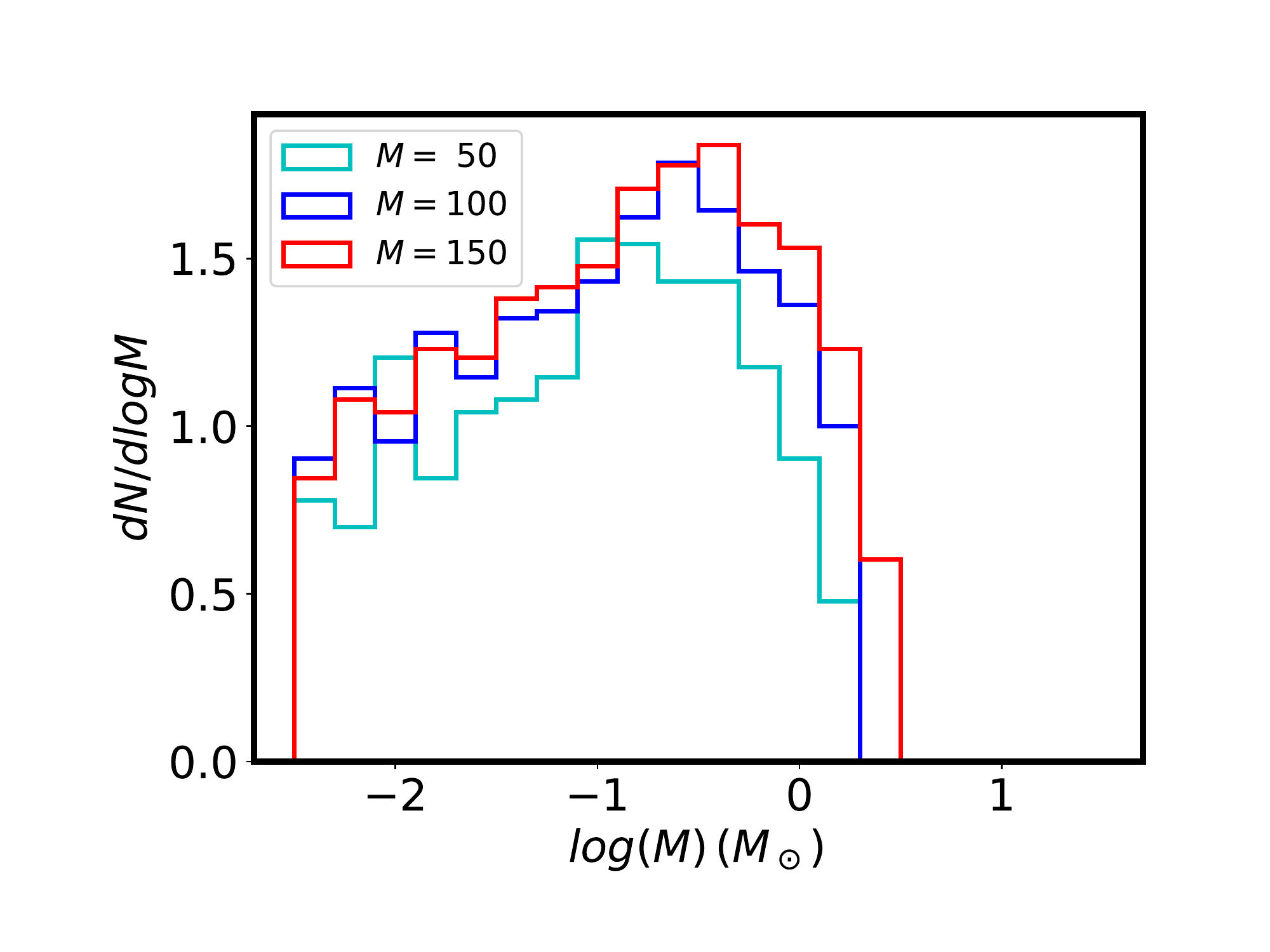}}  
\put(2,5.6){COMP-bar1}
\put(10,5.6){COMP-bar2}
\end{picture}
\caption{Mass spectra at various times for COMP-type runs. We remind that 
 COMP-bar1 and COMP-bar2 use a barotropic eos and no radiative transfer.
The total mass accreted by the sink particles (expressed in solar mass) is indicated in the legend.
 All distributions but the ones of COMP-ACLUM,
 peak at about 0.2-0.3 $M _\odot$ and have generally a very similar shape. This clearly 
demonstrates that up to the time investigated here, stellar feedback has a very limited impact on the 
mass spectrum.  What is more important is the accretion luminosity but it is subject to uncertainties (see text).
}
\label{run_IMF_comp}
\end{figure*}

\setlength{\unitlength}{1cm}
\begin{figure*}%[h!]
%\centering
\begin{picture} (0,12)
%\put(0,11){\includegraphics[width=8cm]{FIG_PAPIER_RHDIMF/CLUSTER_IMF_feedinter_eos_hres_hsink2_nobug/sink_hist_mass_serie.pdf}}  
%\put(2,16.6){COMP-ACLUM}
%\put(8,11){\includegraphics[width=8cm]{FIG_PAPIER_RHDIMF/CLUSTER_IMF_feedback2_eos_hres_modcrit_hsink2/sink_hist_mass_serie.pdf}}  
%\put(8,11){\includegraphics[width=8cm]{FIG_PAPIER_RHDIMF/CLUSTER_IMF_feedinterlow_eos_hres_modcrit_hsink2/sink_hist_mass_serie.pdf}}
%\put(10,16.6){COMP-LOWACLUM}
\put(0,5.5){\includegraphics[width=8cm]{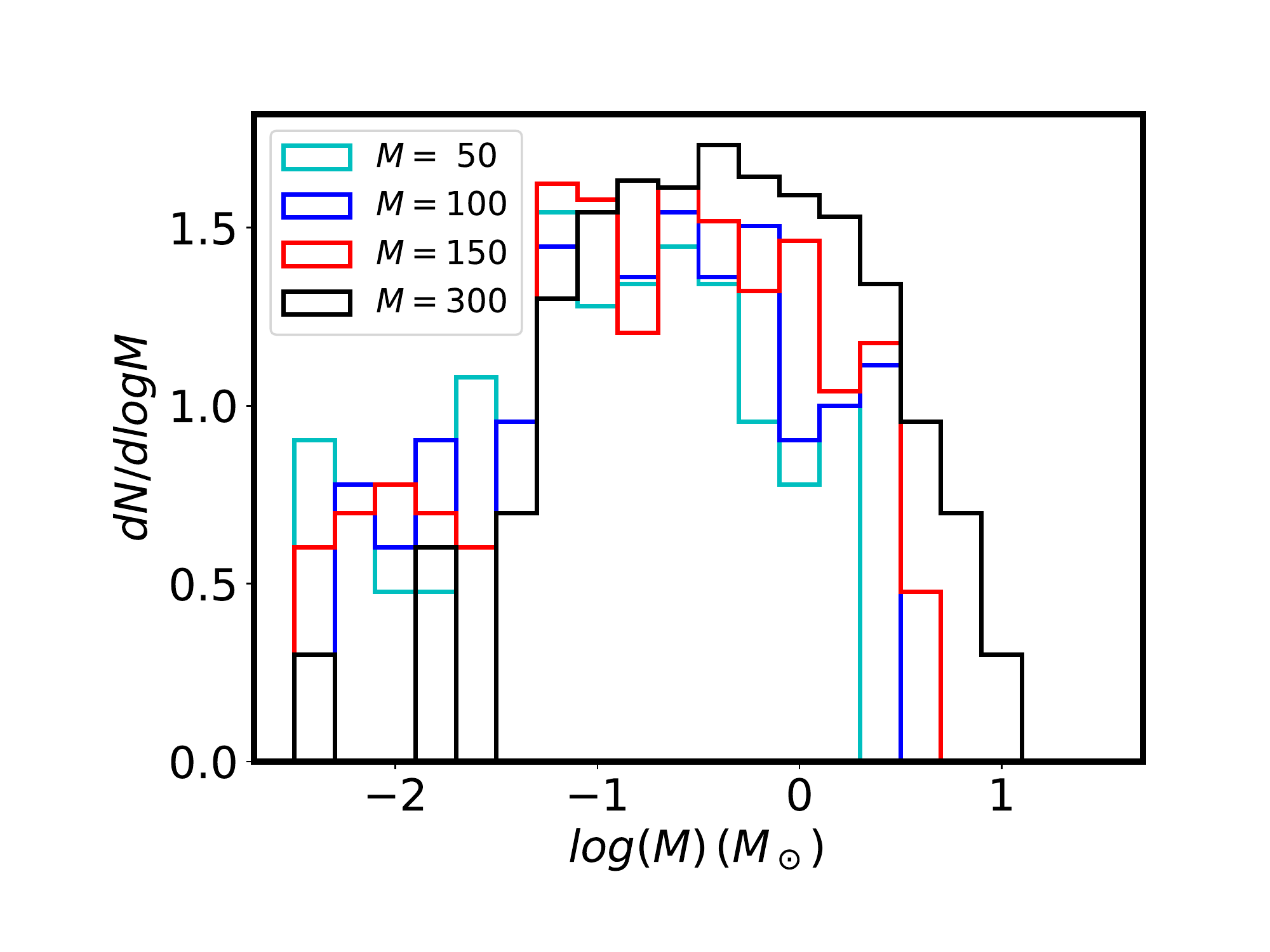}}  
\put(8,5.5){\includegraphics[width=8cm]{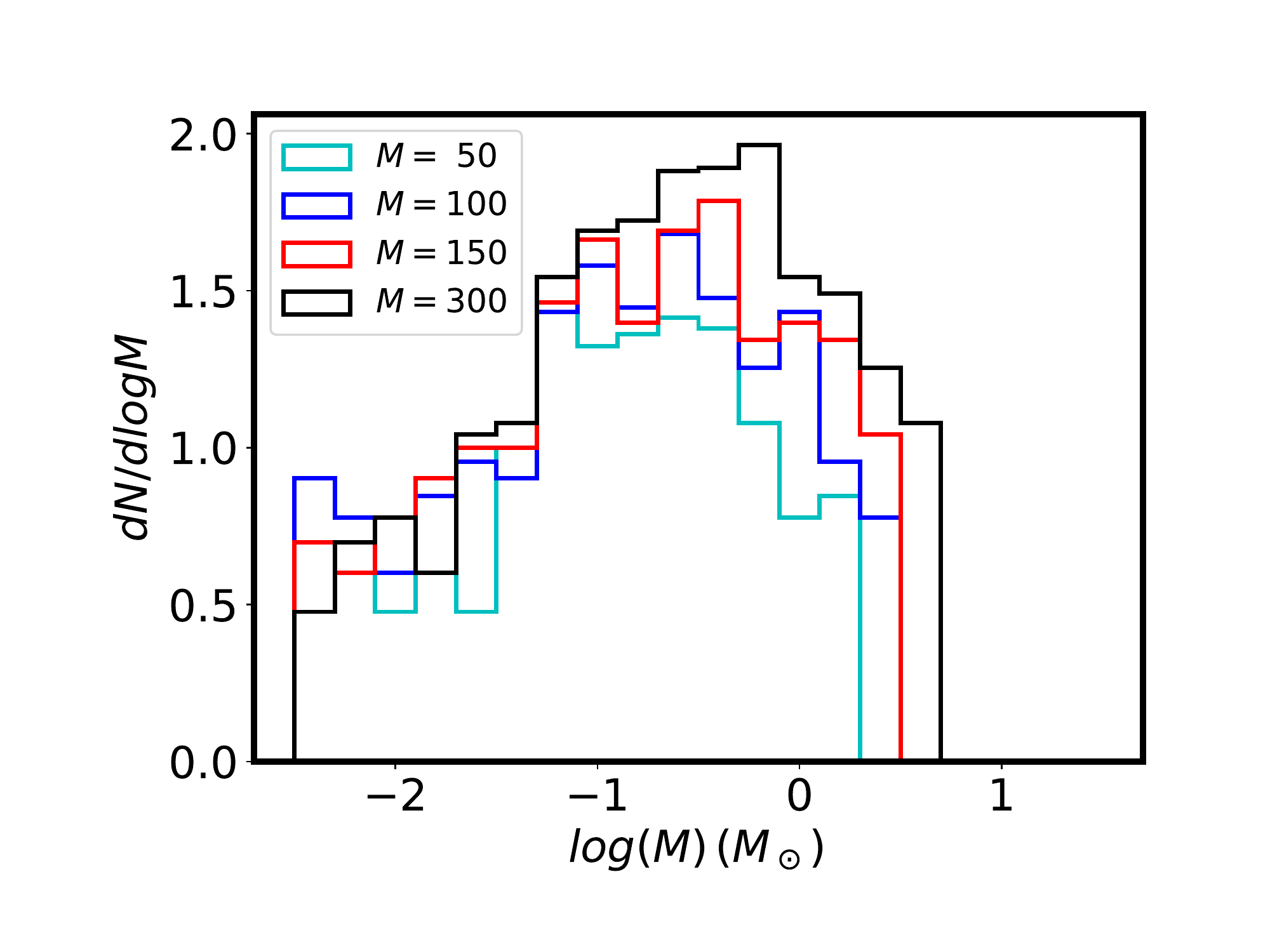}}  
\put(2,11.1){STAN-ACLUM}
\put(10,11.1){STAN-LOWACLUM}
\put(8,0){\includegraphics[width=8cm]{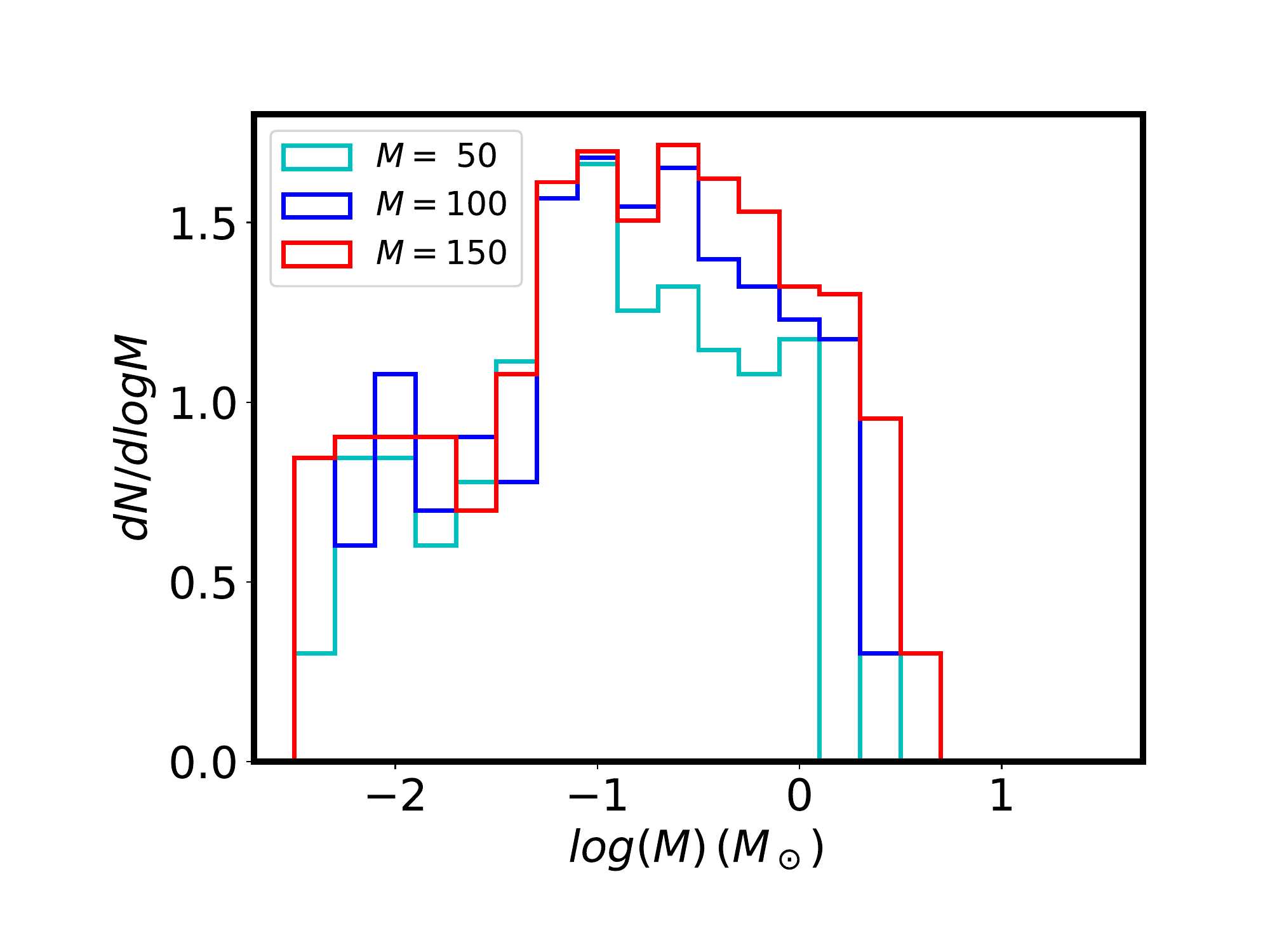}}  
\put(0,0){\includegraphics[width=8cm]{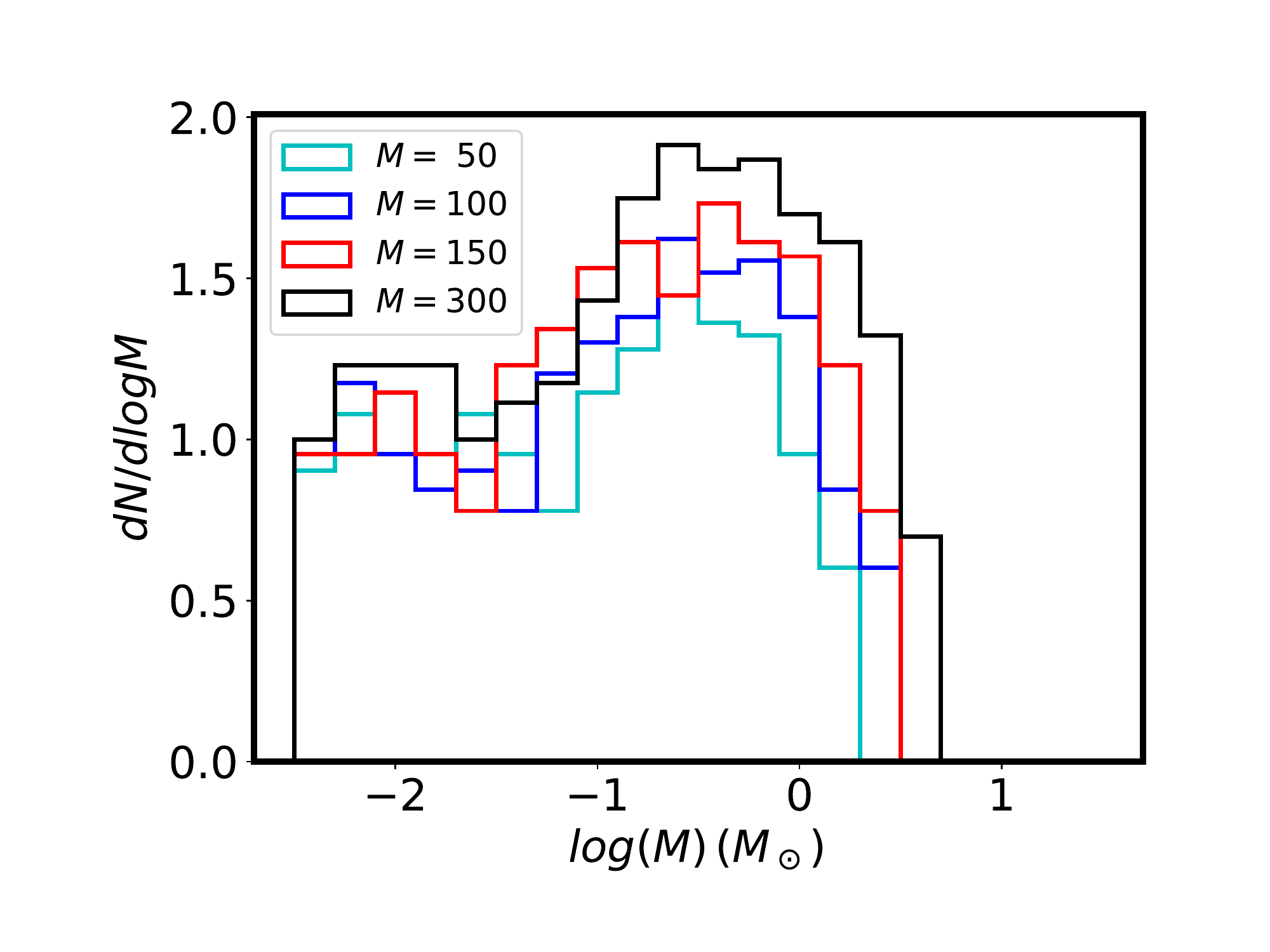}}  
\put(2,5.6){STAN-NOFEED}
\put(10,5.6){STANHMach-ACLUM}
\end{picture}
\caption{Mass spectra at various times for STAN-type runs.
 All distributions 
 peak at about 0.2-0.3 $M _\odot$ and have generally a very similar shape.
Only run STAN-ACLUM which considers the accretion luminosity is broader in shape.
This shows that for moderately dense clumps, radiative feedback, even when 
it includes high accretion luminosity, does not strongly modify the shape of 
the mass spectrum.
}
\label{run_IMF_diff}
\end{figure*}
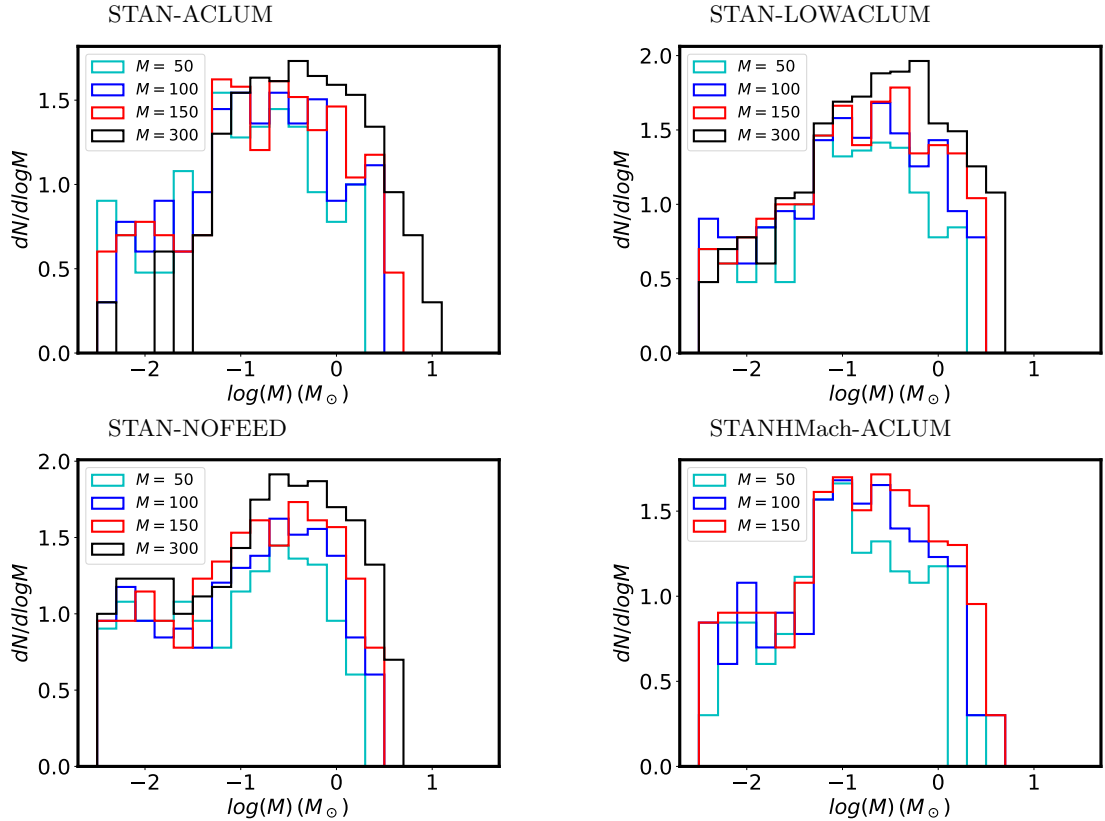

\setlength{\unitlength}{1cm}
\begin{figure}%[h!]
%\centering
\begin{picture} (0,12)
\put(0,0){\includegraphics[width=8cm]{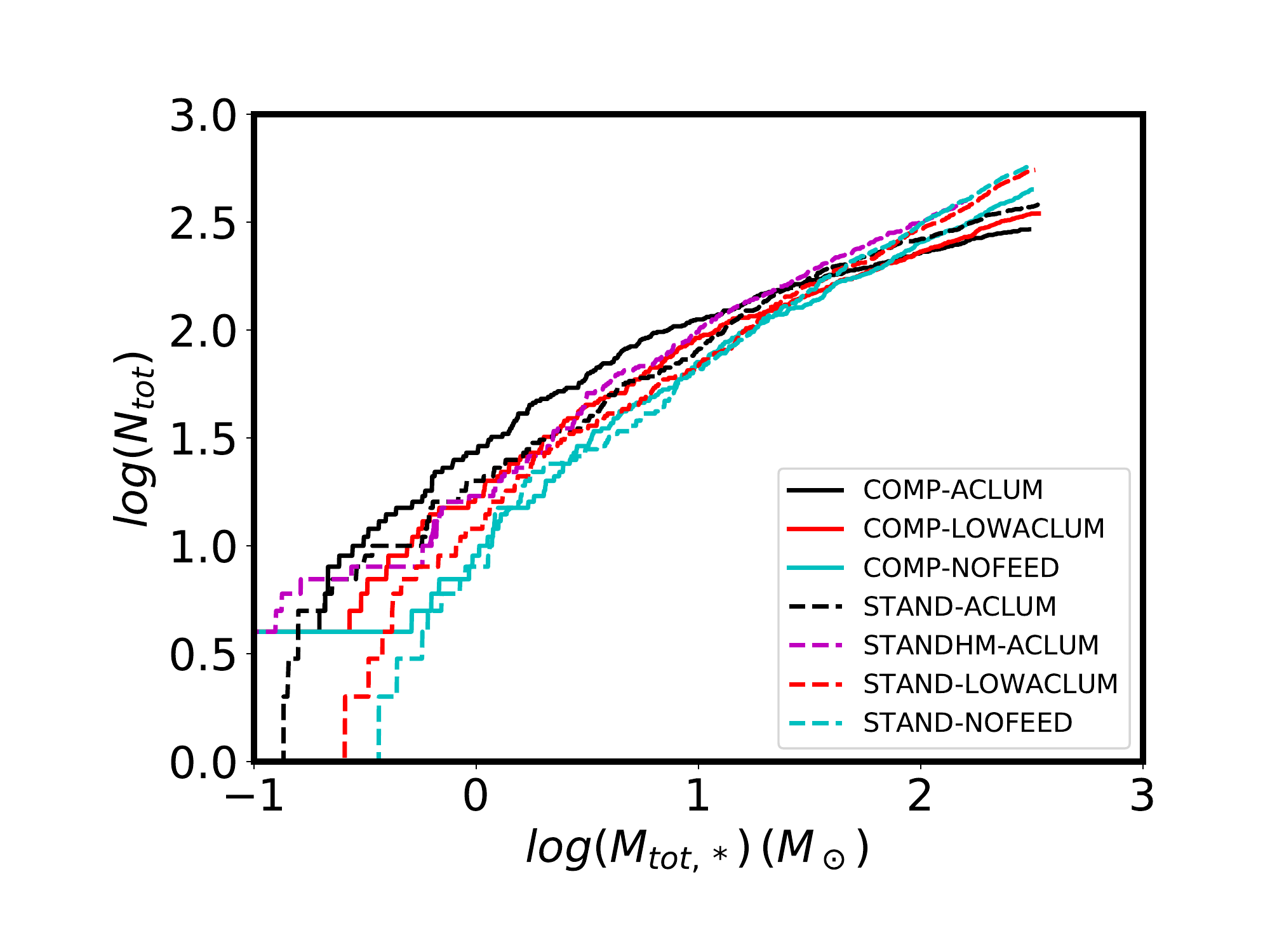}}  
\put(0,6){\includegraphics[width=8cm]{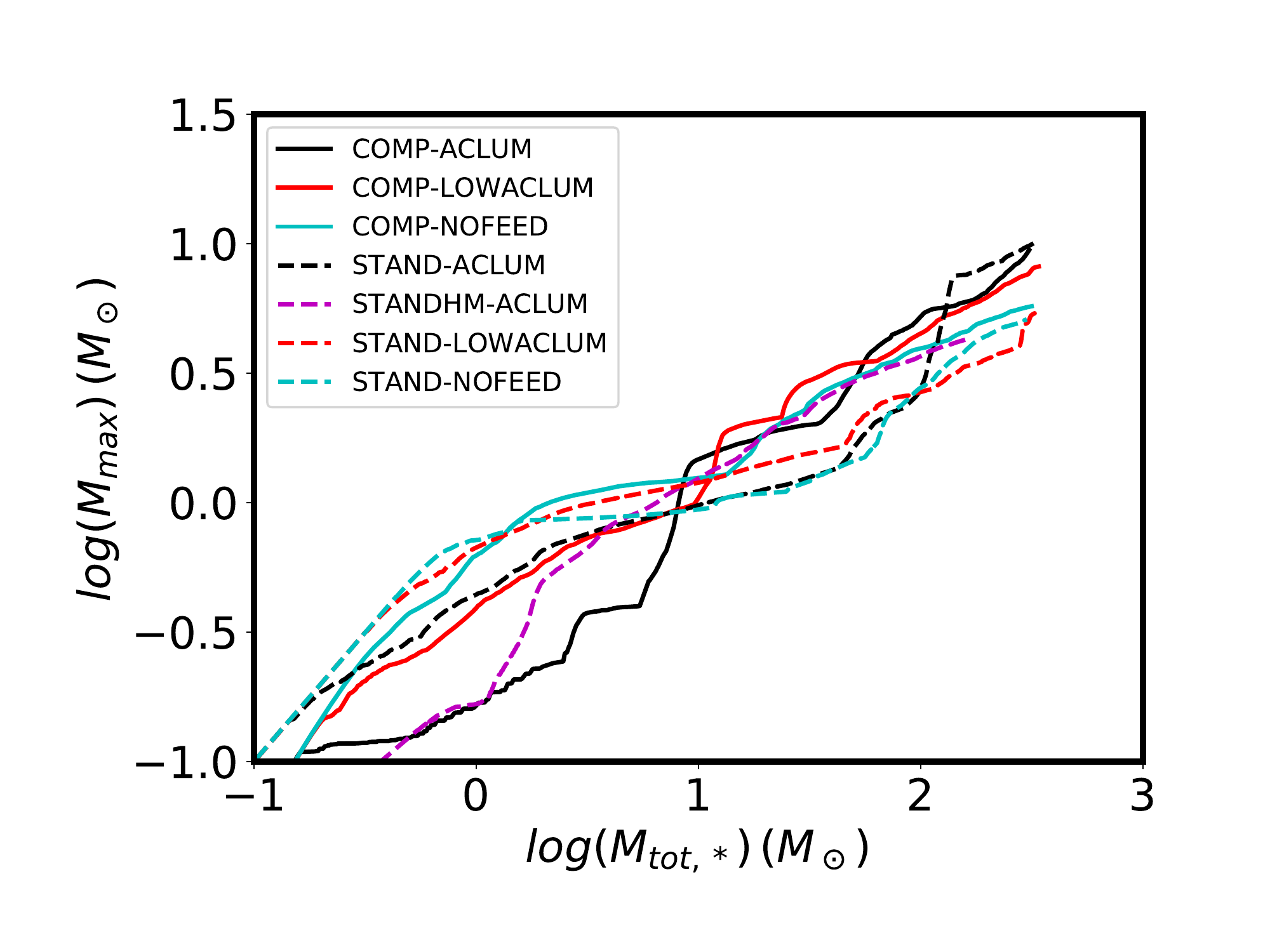}}  
\end{picture}
\caption{The largest sink mass (upper panel)  and 
the number of sinks, $N _{tot} $, 
as a function of the total accreted mass, $M_{tot}$, in several runs. 
}
\label{ntot_massmax}
\end{figure}

\section{Mass spectra}
We now turn to the  stellar mass spectra formed in these calculations.

\subsection{Results}
Figures~\ref{run_IMF_comp} and \ref{run_IMF_diff} portray the mass distribution of the sink particles respectively 
for the six COMP-type runs and four STAN-type runs listed
in Table~\ref{table_param_num} (10 first runs). 
To follow the evolution, the mass spectra are shown at various timesteps, which correspond to various amounts of accreted mass.
The mass spectra are complemented by Fig.~\ref{ntot_massmax} which portrays for six of the runs the number 
of objects formed and the mass of the most massive object as a function of $M_{tot,*}$, the total accreted mass.

Let us start with run COMP-NOFEED, which we remind has no stellar feedback and no accretion luminosity.
The mass spectra present a clear peak around 0.3 M$_\odot$ at early time i.e. when less than 100 M$_\odot$
have been accreted. At later times, the peak broadens and shift towards $\simeq$0.5 M$_\odot$. 
The high mass part presents a powerlaw-like 
shape with an exponent around $\simeq$1. This behaviour is very similar to several  mass 
spectra published in the literature  using either a barotropic equation of state \citep[e.g.][]{bate03,leeh2018a}
\footnote{Note that most published calculations, which use a barotropic equation of state, generally have mass spectra that 
peak at mass smaller than 0.3 M$_\odot$. This is simply a consequence of the chosen equation of state. 
Some of the simulations presented in
\citet{leeh2018b} and in 
this work, the 
 uses a barotropic eos but nevertheless present a peak of the mass spectrum near 0.3 M$_\odot$.}
or  radiative transfer calculations but no accretion luminosity \citep[e.g.][]{bate2009,bate2012}. 
The run COMP-NOACLUM (which considers stellar feedback) presents a very similar behaviour, although at late 
times the powerlaw behaviour for the high mass is better defined. Interestingly, we see that the run COMP-bar1, which 
has an equation of state that broadly reproduces the density-temperature relation of run COMP-NOFEED at the transition 
point between the isothermal and adiabatic regimes (that is to say for  $n \simeq 10^{10-11}$ cm$^{-3}$), presents
mass spectra that are very similar, with a peak located at 0.3 M$_\odot$ and a powerlaw-like 
mass spectrum at high mass. In run COMP-bar2, for which the transition from isothermal to adiabatic occurs at slightly 
lower density, the peak occurs at slightly larger mass.  Indeed, if the gas becomes adiabatic at lower density, then more gas 
piles up before a density of 10$^{13}$ cm$^{-3}$ is reached and the sink particle is being introduced. Physically, this would correspond 
to a more massive FHSC. Therefore the sink has more gas to accrete and the peak of the IMF is shifted toward larger masses.

The inclusion of the accretion luminosity (run COMP-ACLUM) leads to significant differences. 
We recall that in these runs $f_ {\rm acc}=0.5$
for the accretion luminosity.
First at early time when $M_{tot,*}=50$ M$_\odot$, the peak of the distribution is located at about 0.07 M$_\odot$.
The reason is that this is precisely the mass at which the accretion luminosity starts being applied. This 
choice, which simply corresponds to roughly two times the mass of the first hydrostatic core, is somehow 
arbitrary and therefore this feature remains rather uncertain. 
The accretion luminosity feedback is so strong that when it is taken into account, further accretion is abruptly stopped for a 
moment. This is also very clearly seen in Fig.~\ref{ntot_massmax}, which reveals (upper panel) that 
when $M_{tot,*} \simeq 3 M_\odot$, 
the mass of the most massive object remains up to a factor of five below its value in run COMP-NOFEED. 
Correspondingly, the number of objects for a given value of $M_{tot,*}$, is initially the largest 
in run COMP-ACLUM. 
As time goes on, the sinks eventually grow in mass and their numbers, as well as 
the largest sink mass, become similar to the other runs (COMP-NOFEED and COMP-LOWACLUM)  when 
$M_{tot,*} \simeq 30 M_\odot$.
The  peak at 0.07 $M_\odot$ 
is not visible any more when $M_{tot,*}=300$ M$_\odot$. Instead a weak peak located around 0.5 M$_\odot$
has developed. The mass spectrum tends however to be nearly flat between 0.07 and 3 M$_\odot$. 
Flat mass spectra (in $dN/dlog M$) are typical of clumps for which the 
thermal support is very high, so that at the scale of the mass reservoir of individual sink/star, it 
dominates over the turbulent dispersion \citep[see the discussion in][]{leeh2018a}.
Indeed Fig.~\ref{T_rho_comp} clearly shows that the thermal support is very large and up to 20 times 
its value in an isothermal run. 
Let us stress that a similar effect is also probably present in the calculations of 
\citet{krumholz2012} where a flat mass spectrum is also inferred. 

When accretion luminosity is present but weaker (COMP-ACLUMLOW which we remind has $f_{\rm acc}=0.1$), 
the peak at 0.07 M$_\odot$ is much less pronounced and it quickly disappears. A pronounced 
peak located at 0.3-0.5 M$_\odot$ appears and when $M_{tot,*}=300$ M$_\odot$, a relatively clear 
mass spectrum, whose slope is around -1-1.2, has developed. \\ \\

The STAN-NOFEED runs (Fig.~\ref{T_rho_comp}) present mass spectra that resemble the 
COMP-NOFEED ones, with a peak around 0.5 M$_\odot$. This is in good agreement with 
what has been concluded in \citet{leeh2018a}, beyond a certain threshold of 
initial density, the peak of the mass spectrum does not depend on the initial 
density, illustrating in particular that it is not related to the initial 
Jeans mass  \citep[see also][]{lee2019}.

The influence of the accretion luminosity on the mass spectrum is significantly 
less pronounced than for the COMP-type runs. The main difference between 
runs STAN-ACLUM and STAN-NOFEED is that the mass spectrum is a bit broader 
for run STAN-ACLUM. There are more objects of mass 0.07 M$_\odot$ for 
STAN-ACLUM than for STAN-NOFEED. This is certainly due to the sudden 
introduction of the accretion luminosity at 0.07 M$_\odot$. This does not 
lead however to a pronounced peak as for run COMP-ACLUM. This is likely 
because the accretion rate is much lower in the STAN-type runs than 
for the COMP-type ones. As expected the impact of the accretion luminosity 
is even smaller for runs STAN-LOWACLUM, which presents mass spectra 
that are very similar to the STAN-NOFEED ones. 
The moderate influence of the accretion luminosity is also visible in 
Fig.~\ref{ntot_massmax} where it is seen that both the most massive star and the number 
of stars are very similar between all STAN-type runs. 
The most significant difference is found at late times, where the largest mass is 
about 10 M$_\odot$, 
roughly 2 times larger for run STAN-ACLUM than for run STAN-NOFEED for instance. 
Clearly the increase of temperature favors the growth of existing 
protostars by reducing the amount of fragmentation. 

Finally, it is interesting to see that run STANHMACH-ACLUM presents  
mass spectra that peak at smaller mass than run STAN-ACLUM and 
remains slightly narrower. This is likely because the higher turbulence
makes the 
the accretion rate  lower and the clump radius larger. Both effects
reduce the influence of the accretion luminosity.

\subsection{Interpretation and discussion}
By performing numerical simulations of a massive collapsing clump
 with a barotropic eos and comparing with an analytical model,  \citet{leeh2018a} have identified two regimes 
resulting in two different mass spectra. If the thermal support is high, that is to say if the 
initial density is low or equivalently if the temperature is high,  the mass 
spectrum tends to be flat, that is to say $dN / \log M \propto M^0$ (see run $A$ of \citet{leeh2018a}
and runs presented in  bottom panels of Fig.~2 of \citet{jones2018}). When, on the other 
hand, turbulent support dominates, the mass spectrum presents a peak at about 
ten times the mass of the FHSC and 
a powerlaw at high mass which is found to be $dN / \log M \propto M^{-3/4}$
(see runs $B$, $C$ and $D$ of \citet{leeh2018a}). Given the similarity found between the mass 
spectra obtained with radiative transfer (except for run COMP-ACLUM) and with the barotropic eos, it seems likely that the physical 
interpretation developed in \citet{leeh2018a} and \citet{leeh2018b} for the peak of the IMF, namely the impact of tidal effects, remains valid when  
radiative transfer is considered. 

The different behaviour obtained for run COMP-ACLUM, for which the mass spectrum 
is nearly flat for mass between $\simeq 0.3$ and $\simeq 3$ M$_\odot$, is most likely 
a consequence of the high thermal support induced by the large temperatures due 
to the high accretion luminosities.
As mentioned earlier, however, the {\it effective} value of $f_{\rm acc}$ for real protostars, is not well established. 
 Indeed, the observed luminosities of protostars are much fainter than the values corresponding to such accretion
luminosities \citep[the luminosity problem, ][]{evans2009}, 
 A  plausible explanation to the luminosity problem is episodic accretion \citep[see e.g.][]{baraffe2009,baraffe2012},
evenso this issue is not entirely settled yet. If this indeed happens, it means that the 
heating of the gas through accretion luminosity may be intermittent as well and  so the {\it effective} value of $f_{\rm acc}$, i.e. 
the value that the clump is experiencing most of the time,  may be on average small. 
Interestingly, \citet{stamatellos2011} developed a model of intermittent accretion to study disk fragmentation. 
By comparing simulations with and without  accretion luminosity, they concluded that the simulations 
with episodic accretion ressembles the one without accretion luminosity. 
While episodic accretion received strong support from observations \citep{evans2009} and 
theory \citep{baraffe2012}, it is currently unclear whether the proposed mechanism 
of gravitational instability \citep{vorobyov2010} is 
sufficiently universal. 
Another possible effect that could lead to small effective $f_{\rm acc}$  has been proposed 
by \citet{krumholz2012} who claimed that most of the radiation can escape through the wind cavities. 
It is therefore unclear whether the impact of the accretion luminosity is as found in run 
COMP-ACLUM. 
It should also be stressed  that the physical conditions corresponding to COMP-type runs, are 
not typical of most star forming clumps. In this respect, STAN-type runs are more typical.
Since there the accretion luminosity appears to have an impact that is more limited, in particular regarding the peak, 
it seems that, in most star forming clumps of the Milky-Way, the accretion luminosity is not drastically influencing
the stellar spectrum.

\section{Conclusions}
We have performed a series of  numerical simulations with a spatial resolution of a 
few AU (and up to 1 AU for convergence runs), to investigate the influence of radiative feedback on 
the mass distribution of stars which form during the collapse of a 1000 $M_\odot$ clump. 
We also performed two runs in which a barotropic eos  is employed. 
Two types of initial conditions have been explored, one corresponds to a very compact 
clump initially (with a radius of 0.1 pc) while the other is more typical of 
Milky-way star forming clumps (radius 0.4 pc initially). 

We found that as long as accretion luminosity is not considered, the stellar mass spectra
that form in the various runs, both with  radiative transfer and with an effective equation of state,
 present strong similarities and resemble the observed 
IMF.  This suggests that in this case, radiative transfer is not fundamental in setting the IMF, evenso 
the gas temperature of the dense star forming gas is 3-10 times higher than when an eos is employed. 

 When accretion luminosity  is included, its impact depends on the initial conditions.
For the case of the very compact clump, the  mass spectrum is initially strongly peaked
at small mass. As time goes on, it progressively becomes flat between  0.1 and 3 M$_\odot$. 
For the case of the less compact clump (0.4 pc of radii), the effect remains more limited particularly
at late time. The peak is located at roughly the same value than in the absence of accretion 
luminosity. An interesting effect though is that there are few more small and massive objects. 
While the latter are a consequence of the strong heating induced by the accretion luminosity which 
prevent fragmentation, 
the former are also a consequence on this heating, which when a low mass object has just been created 
prevent further accretion.  We stress that the exact history of accretion luminosity, 
in particular the mass at which it actually starts is largely unknown and therefore the increase 
of low mass objects we found, must be regarded with care.

We conclude that the accretion luminosity, {\it if} its effective value is equal to or at least comparable with
with the gravitational energy  released at the surface of the star,  is expected in most galactic 
star formation clumps to have an impact that mainly consists in producing both smaller and 
more massive objects that what would have been formed otherwise. In particular, it is 
likely that in most circumstances, the peak of the distribution is a consequence of 
the change of the effective equation of state that is responsible of 
the first hydrostatic core rather than due to the feedback heating of the 
collapsing envelope. It seems however that feedback heating, leads to the formation 
of more massive stars \citep{krumholz2007,urban2010}, which otherwise appear to be seldom.

%__________________________________________________________________

\begin{acknowledgements}
PH warmly thanks Timea Csengeri, Neal Evans, Adam Ginsburg, Fabien Louvet, Ana\"elle Maury, Sergio Molinari,
Fr\'ed\'erique Motte and Alessio Traficante for enlighting discussions on the 
temperature interpretations of massive star forming cores and the issue of accretion luminosity. 
We thank the anonymous referee for their constructive comments that have improved the paper.
PH acknowledges financial support from the European Research Council (ERC) via the ERC Synergy Grant {\em ECOGAL} (grant 855130).
%This research has received funding from the European Research Council under the European
% Community's Seventh Framework Programme (FP7/2007-2013 Grant Agreement no. 306483).
\end{acknowledgements}

% for the bibliography
\bibliography{lars}{}

\begin{thebibliography}{}
\expandafter\ifx\csname natexlab\endcsname\relax\def\natexlab#1{#1}\fi
\providecommand{\url}[1]{\href{#1}{#1}}
\providecommand{\dodoi}[1]{doi:~\href{http://doi.org/#1}{\nolinkurl{#1}}}
\providecommand{\doeprint}[1]{\href{http://ascl.net/#1}{\nolinkurl{http://ascl.net/#1}}}
\providecommand{\doarXiv}[1]{\href{https://arxiv.org/abs/#1}{\nolinkurl{https://arxiv.org/abs/#1}}}

\bibitem[{{Ballesteros-Paredes} {et~al.}(2015){Ballesteros-Paredes},
  {Hartmann}, {P{\'e}rez-Goytia}, \& {Kuznetsova}}]{BallesterosParedes15}
{Ballesteros-Paredes}, J., {Hartmann}, L.~W., {P{\'e}rez-Goytia}, N., \&
  {Kuznetsova}, A. 2015, \mnras, 452, 566, \dodoi{10.1093/mnras/stv1285}

\bibitem[{{Baraffe} {et~al.}(2009){Baraffe}, {Chabrier}, \&
  {Gallardo}}]{baraffe2009}
{Baraffe}, I., {Chabrier}, G., \& {Gallardo}, J. 2009, \apjl, 702, L27,
  \dodoi{10.1088/0004-637X/702/1/L27}

\bibitem[{{Baraffe} {et~al.}(2017){Baraffe}, {Elbakyan}, {Vorobyov}, \&
  {Chabrier}}]{baraffe2017}
{Baraffe}, I., {Elbakyan}, V.~G., {Vorobyov}, E.~I., \& {Chabrier}, G. 2017,
  \aap, 597, A19, \dodoi{10.1051/0004-6361/201629303}

\bibitem[{{Baraffe} {et~al.}(2012){Baraffe}, {Vorobyov}, \&
  {Chabrier}}]{baraffe2012}
{Baraffe}, I., {Vorobyov}, E., \& {Chabrier}, G. 2012, \apj, 756, 118,
  \dodoi{10.1088/0004-637X/756/2/118}

\bibitem[{{Bastian} {et~al.}(2010){Bastian}, {Covey}, \& {Meyer}}]{bastian2010}
{Bastian}, N., {Covey}, K.~R., \& {Meyer}, M.~R. 2010, \araa, 48, 339,
  \dodoi{10.1146/annurev-astro-082708-101642}

\bibitem[{{Basu} {et~al.}(2015){Basu}, {Gil}, \& {Auddy}}]{basu2015}
{Basu}, S., {Gil}, M., \& {Auddy}, S. 2015, \mnras, 449, 2413,
  \dodoi{10.1093/mnras/stv445}

\bibitem[{{Basu} \& {Jones}(2004)}]{basu2004}
{Basu}, S., \& {Jones}, C.~E. 2004, \mnras, 347, L47,
  \dodoi{10.1111/j.1365-2966.2004.07405.x}

\bibitem[{{Bate}(2009)}]{bate2009}
{Bate}, M.~R. 2009, \mnras, 392, 1363, \dodoi{10.1111/j.1365-2966.2008.14165.x}

\bibitem[{{Bate}(2010)}]{bate2010}
---. 2010, \mnras, 404, L79, \dodoi{10.1111/j.1745-3933.2010.00839.x}

\bibitem[{{Bate}(2012)}]{bate2012}
---. 2012, \mnras, 419, 3115, \dodoi{10.1111/j.1365-2966.2011.19955.x}

\bibitem[{{Bate} {et~al.}(2003){Bate}, {Bonnell}, \& {Bromm}}]{bate03}
{Bate}, M.~R., {Bonnell}, I.~A., \& {Bromm}, V. 2003, \mnras, 339, 577,
  \dodoi{10.1046/j.1365-8711.2003.06210.x}

\bibitem[{{Bhandare} {et~al.}(2020){Bhandare}, {Kuiper}, {Henning}, {Fendt},
  {Flock}, \& {Marleau}}]{bhandare2020}
{Bhandare}, A., {Kuiper}, R., {Henning}, T., {et~al.} 2020, \aap, 638, A86,
  \dodoi{10.1051/0004-6361/201937029}

\bibitem[{{Bhandare} {et~al.}(2018){Bhandare}, {Kuiper}, {Henning}, {Fendt},
  {Marleau}, \& {K{\"o}lligan}}]{bhandare2018}
---. 2018, \aap, 618, A95, \dodoi{10.1051/0004-6361/201832635}

\bibitem[{{Bleuler} \& {Teyssier}(2014)}]{Bleuler14}
{Bleuler}, A., \& {Teyssier}, R. 2014, \mnras, 445, 4015,
  \dodoi{10.1093/mnras/stu2005}

\bibitem[{{Bonnell} {et~al.}(2011){Bonnell}, {Smith}, {Clark}, \&
  {Bate}}]{Bonnell11}
{Bonnell}, I.~A., {Smith}, R.~J., {Clark}, P.~C., \& {Bate}, M.~R. 2011,
  \mnras, 410, 2339, \dodoi{10.1111/j.1365-2966.2010.17603.x}

\bibitem[{{Cappellari} {et~al.}(2012){Cappellari}, {McDermid}, {Alatalo},
  {Blitz}, {Bois}, {Bournaud}, {Bureau}, {Crocker}, {Davies}, {Davis}, {de
  Zeeuw}, {Duc}, {Emsellem}, {Khochfar}, {Krajnovi{\'c}}, {Kuntschner},
  {Lablanche}, {Morganti}, {Naab}, {Oosterloo}, {Sarzi}, {Scott}, {Serra},
  {Weijmans}, \& {Young}}]{cappellari2012}
{Cappellari}, M., {McDermid}, R.~M., {Alatalo}, K., {et~al.} 2012, \nat, 484,
  485, \dodoi{10.1038/nature10972}

\bibitem[{{Chabrier}(2003)}]{chabrier2003}
{Chabrier}, G. 2003, \pasp, 115, 763, \dodoi{10.1086/376392}

\bibitem[{{Chabrier} {et~al.}(2014){Chabrier}, {Hennebelle}, \&
  {Charlot}}]{chabrier2014}
{Chabrier}, G., {Hennebelle}, P., \& {Charlot}, S. 2014, \apj, 796, 75,
  \dodoi{10.1088/0004-637X/796/2/75}

\bibitem[{{Colman} \& {Teyssier}(2020)}]{colman2019}
{Colman}, T., \& {Teyssier}, R. 2020, \mnras, 492, 4727,
  \dodoi{10.1093/mnras/staa075}

\bibitem[{{Commer{\c{c}}on} {et~al.}(2014){Commer{\c{c}}on}, {Debout}, \&
  {Teyssier}}]{commercon2014}
{Commer{\c{c}}on}, B., {Debout}, V., \& {Teyssier}, R. 2014, \aap, 563, A11,
  \dodoi{10.1051/0004-6361/201322858}

\bibitem[{{Commer{\c{c}}on} {et~al.}(2011){Commer{\c{c}}on}, {Teyssier},
  {Audit}, {Hennebelle}, \& {Chabrier}}]{commercon2011a}
{Commer{\c{c}}on}, B., {Teyssier}, R., {Audit}, E., {Hennebelle}, P., \&
  {Chabrier}, G. 2011, \aap, 529, A35, \dodoi{10.1051/0004-6361/201015880}

\bibitem[{{Csengeri} {et~al.}(2016){Csengeri}, {Leurini}, {Wyrowski},
  {Urquhart}, {Menten}, {Walmsley}, {Bontemps}, {Wienen}, {Beuther}, {Motte},
  {Nguyen-Luong}, {Schilke}, {Schuller}, {Zavagno}, \& {Sanna}}]{csengeri2016}
{Csengeri}, T., {Leurini}, S., {Wyrowski}, F., {et~al.} 2016, \aap, 586, A149,
  \dodoi{10.1051/0004-6361/201425404}

\bibitem[{{Dunham} {et~al.}(2010){Dunham}, {Evans}, {Terebey}, {Dullemond}, \&
  {Young}}]{dunham2010}
{Dunham}, M.~M., {Evans}, Neal~J., I., {Terebey}, S., {Dullemond}, C.~P., \&
  {Young}, C.~H. 2010, \apj, 710, 470, \dodoi{10.1088/0004-637X/710/1/470}

\bibitem[{{Elia} {et~al.}(2017){Elia}, {Molinari}, {Schisano}, {Pestalozzi},
  {Pezzuto}, {Merello}, {Noriega-Crespo}, {Moore}, {Russeil}, {Mottram},
  {Paladini}, {Strafella}, {Benedettini}, {Bernard}, {Di Giorgio}, {Eden},
  {Fukui}, {Plume}, {Bally}, {Martin}, {Ragan}, {Jaffa}, {Motte}, {Olmi},
  {Schneider}, {Testi}, {Wyrowski}, {Zavagno}, {Calzoletti}, {Faustini},
  {Natoli}, {Palmeirim}, {Piacentini}, {Piazzo}, {Pilbratt}, {Polychroni},
  {Baldeschi}, {Beltr{\'a}n}, {Billot}, {Cambr{\'e}sy}, {Cesaroni},
  {Garc{\'\i}a-Lario}, {Hoare}, {Huang}, {Joncas}, {Liu}, {Maiolo}, {Marsh},
  {Maruccia}, {M{\`e}ge}, {Peretto}, {Rygl}, {Schilke}, {Thompson},
  {Traficante}, {Umana}, {Veneziani}, {Ward-Thompson}, {Whitworth}, {Arab},
  {Band ieramonte}, {Becciani}, {Brescia}, {Buemi}, {Bufano}, {Butora},
  {Cavuoti}, {Costa}, {Fiorellino}, {Hajnal}, {Hayakawa}, {Kacsuk}, {Leto}, {Li
  Causi}, {Marchili}, {Martinavarro-Armengol}, {Mercurio}, {Molinaro},
  {Riccio}, {Sano}, {Sciacca}, {Tachihara}, {Torii}, {Trigilio}, {Vitello}, \&
  {Yamamoto}}]{elia2017}
{Elia}, D., {Molinari}, S., {Schisano}, E., {et~al.} 2017, \mnras, 471, 100,
  \dodoi{10.1093/mnras/stx1357}

\bibitem[{{Evans} {et~al.}(2009){Evans}, {Dunham}, {J{\o}rgensen}, {Enoch},
  {Mer{\'\i}n}, {van Dishoeck}, {Alcal{\'a}}, {Myers}, {Stapelfeldt}, {Huard},
  {Allen}, {Harvey}, {van Kempen}, {Blake}, {Koerner}, {Mundy}, {Padgett}, \&
  {Sargent}}]{evans2009}
{Evans}, Neal~J., I., {Dunham}, M.~M., {J{\o}rgensen}, J.~K., {et~al.} 2009,
  \apjs, 181, 321, \dodoi{10.1088/0067-0049/181/2/321}

\bibitem[{{Fischer} {et~al.}(2019){Fischer}, {Safron}, \&
  {Megeath}}]{fischer2019}
{Fischer}, W.~J., {Safron}, E., \& {Megeath}, S.~T. 2019, \apj, 872, 183,
  \dodoi{10.3847/1538-4357/ab01dc}

\bibitem[{{Frimann} {et~al.}(2016){Frimann}, {J{\o}rgensen}, {Padoan}, \&
  {Haugb{\o}lle}}]{frimann2016}
{Frimann}, S., {J{\o}rgensen}, J.~K., {Padoan}, P., \& {Haugb{\o}lle}, T. 2016,
  \aap, 587, A60, \dodoi{10.1051/0004-6361/201527622}

\bibitem[{{Fromang} {et~al.}(2006){Fromang}, {Hennebelle}, \&
  {Teyssier}}]{Fromang06}
{Fromang}, S., {Hennebelle}, P., \& {Teyssier}, R. 2006, \aap, 457, 371,
  \dodoi{10.1051/0004-6361:20065371}

\bibitem[{{Giannetti} {et~al.}(2017){Giannetti}, {Leurini}, {Wyrowski},
  {Urquhart}, {Csengeri}, {Menten}, {K{\"o}nig}, \&
  {G{\"u}sten}}]{giannetti2017}
{Giannetti}, A., {Leurini}, S., {Wyrowski}, F., {et~al.} 2017, \aap, 603, A33,
  \dodoi{10.1051/0004-6361/201630048}

\bibitem[{{Ginsburg} {et~al.}(2017){Ginsburg}, {Goddi}, {Kruijssen}, {Bally},
  {Smith}, {Galv{\'a}n-Madrid}, {Mills}, {Wang}, {Dale}, {Darling},
  {Rosolowsky}, {Loughnane}, {Testi}, \& {Bastian}}]{ginsburg2017}
{Ginsburg}, A., {Goddi}, C., {Kruijssen}, J.~M.~D., {et~al.} 2017, \apj, 842,
  92, \dodoi{10.3847/1538-4357/aa6bfa}

\bibitem[{{Girichidis} {et~al.}(2011){Girichidis}, {Federrath}, {Banerjee}, \&
  {Klessen}}]{Girichidis11}
{Girichidis}, P., {Federrath}, C., {Banerjee}, R., \& {Klessen}, R.~S. 2011,
  \mnras, 413, 2741, \dodoi{10.1111/j.1365-2966.2011.18348.x}

\bibitem[{{Goddi} {et~al.}(2018){Goddi}, {Ginsburg}, {Maud}, {Zhang}, \&
  {Zapata}}]{goddi2018}
{Goddi}, C., {Ginsburg}, A., {Maud}, L., {Zhang}, Q., \& {Zapata}, L. 2018,
  arXiv e-prints, arXiv:1805.05364.
\newblock \doarXiv{1805.05364}

\bibitem[{{Guszejnov} {et~al.}(2020){Guszejnov}, {Grudi{\'c}}, {Hopkins},
  {Offner}, \& {Faucher-Gigu{\`e}re}}]{guszejnov2020}
{Guszejnov}, D., {Grudi{\'c}}, M.~Y., {Hopkins}, P.~F., {Offner}, S. S.~R., \&
  {Faucher-Gigu{\`e}re}, C.-A. 2020, \mnras, 496, 5072,
  \dodoi{10.1093/mnras/staa1883}

\bibitem[{{Guszejnov} {et~al.}(2016){Guszejnov}, {Krumholz}, \&
  {Hopkins}}]{guszejnov2016}
{Guszejnov}, D., {Krumholz}, M.~R., \& {Hopkins}, P.~F. 2016, \mnras, 458, 673,
  \dodoi{10.1093/mnras/stw315}

\bibitem[{{Hennebelle}(2012)}]{Hennebelle2012}
{Hennebelle}, P. 2012, \aap, 545, A147, \dodoi{10.1051/0004-6361/201219440}

\bibitem[{{Hennebelle} \& {Chabrier}(2008)}]{HC08}
{Hennebelle}, P., \& {Chabrier}, G. 2008, \apj, 684, 395,
  \dodoi{10.1086/589916}

\bibitem[{{Hennebelle} {et~al.}(2020){Hennebelle}, {Commer{\c{c}}on}, {Lee}, \&
  {Charnoz}}]{hetal2020}
{Hennebelle}, P., {Commer{\c{c}}on}, B., {Lee}, Y.-N., \& {Charnoz}, S. 2020,
  \aap, 635, A67, \dodoi{10.1051/0004-6361/201936714}

\bibitem[{{Hennebelle} {et~al.}(2019){Hennebelle}, {Lee}, \&
  {Chabrier}}]{hetal2019}
{Hennebelle}, P., {Lee}, Y.-N., \& {Chabrier}, G. 2019, \apj, 883, 140,
  \dodoi{10.3847/1538-4357/ab3d46}

\bibitem[{{Hopkins}(2013)}]{hopkins2013}
{Hopkins}, P.~F. 2013, \mnras, 433, 170, \dodoi{10.1093/mnras/stt713}

\bibitem[{{Hosokawa} \& {Omukai}(2009)}]{hosokawa2009}
{Hosokawa}, T., \& {Omukai}, K. 2009, \apj, 691, 823,
  \dodoi{10.1088/0004-637X/691/1/823}

\bibitem[{{Hsieh} {et~al.}(2018){Hsieh}, {Murillo}, {Belloche}, {Hirano},
  {Walsh}, {van Dishoeck}, \& {Lai}}]{hsieh2018}
{Hsieh}, T.-H., {Murillo}, N.~M., {Belloche}, A., {et~al.} 2018, \apj, 854, 15,
  \dodoi{10.3847/1538-4357/aaa7f6}

\bibitem[{{Inutsuka}(2001)}]{inutsuka2001}
{Inutsuka}, S.-i. 2001, \apj, 559, L149, \dodoi{10.1086/323786}

\bibitem[{{Jappsen} {et~al.}(2005){Jappsen}, {Klessen}, {Larson}, {Li}, \& {Mac
  Low}}]{jappsen2005}
{Jappsen}, A.-K., {Klessen}, R.~S., {Larson}, R.~B., {Li}, Y., \& {Mac Low},
  M.-M. 2005, \aap, 435, 611, \dodoi{10.1051/0004-6361:20042178}

\bibitem[{{Jones} \& {Bate}(2018)}]{jones2018}
{Jones}, M.~O., \& {Bate}, M.~R. 2018, \mnras, 478, 2650,
  \dodoi{10.1093/mnras/sty1250}

\bibitem[{{Kenyon} \& {Hartmann}(1995)}]{kenyon1995}
{Kenyon}, S.~J., \& {Hartmann}, L. 1995, \apjs, 101, 117,
  \dodoi{10.1086/192235}

\bibitem[{{Kroupa}(2001)}]{kroupa2001}
{Kroupa}, P. 2001, \mnras, 322, 231, \dodoi{10.1046/j.1365-8711.2001.04022.x}

\bibitem[{{Krumholz} {et~al.}(2007){Krumholz}, {Klein}, \&
  {McKee}}]{krumholz2007}
{Krumholz}, M.~R., {Klein}, R.~I., \& {McKee}, C.~F. 2007, \apj, 656, 959,
  \dodoi{10.1086/510664}

\bibitem[{{Krumholz} {et~al.}(2012){Krumholz}, {Klein}, \&
  {McKee}}]{krumholz2012}
---. 2012, \apj, 754, 71, \dodoi{10.1088/0004-637X/754/1/71}

\bibitem[{{Krumholz} {et~al.}(2016){Krumholz}, {Myers}, {Klein}, \&
  {McKee}}]{krumholz2016}
{Krumholz}, M.~R., {Myers}, A.~T., {Klein}, R.~I., \& {McKee}, C.~F. 2016,
  \mnras, 460, 3272, \dodoi{10.1093/mnras/stw1236}

\bibitem[{{Kuiper} \& {Yorke}(2013)}]{kuiper2013}
{Kuiper}, R., \& {Yorke}, H.~W. 2013, \apj, 772, 61,
  \dodoi{10.1088/0004-637X/772/1/61}

\bibitem[{{Larson}(1969)}]{larson69}
{Larson}, R.~B. 1969, \mnras, 145, 271

\bibitem[{{Lee} \& {Hennebelle}(2016{\natexlab{a}})}]{lee2016b}
{Lee}, Y.-N., \& {Hennebelle}, P. 2016{\natexlab{a}}, \aap, 591, A31,
  \dodoi{10.1051/0004-6361/201527982}

\bibitem[{{Lee} \& {Hennebelle}(2016{\natexlab{b}})}]{lee2016a}
---. 2016{\natexlab{b}}, \aap, 591, A30, \dodoi{10.1051/0004-6361/201527981}

\bibitem[{{Lee} \& {Hennebelle}(2018{\natexlab{a}})}]{leeh2018b}
---. 2018{\natexlab{a}}, \aap, 611, A89, \dodoi{10.1051/0004-6361/201731523}

\bibitem[{{Lee} \& {Hennebelle}(2018{\natexlab{b}})}]{leeh2018a}
---. 2018{\natexlab{b}}, \aap, 611, A88, \dodoi{10.1051/0004-6361/201731522}

\bibitem[{{Lee} \& {Hennebelle}(2019)}]{lee2019}
---. 2019, \aap, 622, A125, \dodoi{10.1051/0004-6361/201834428}

\bibitem[{{Lee} {et~al.}(2017){Lee}, {Hennebelle}, \& {Chabrier}}]{lhc2017}
{Lee}, Y.-N., {Hennebelle}, P., \& {Chabrier}, G. 2017, ArXiv e-prints.
\newblock \doarXiv{1709.01446}

\bibitem[{{Lee} {et~al.}(2020){Lee}, {Offner}, {Hennebelle}, {Andr{\'e}},
  {Zinnecker}, {Ballesteros-Paredes}, {Inutsuka}, \& {Kruijssen}}]{lee2020}
{Lee}, Y.-N., {Offner}, S. S.~R., {Hennebelle}, P., {et~al.} 2020, \ssr, 216,
  70, \dodoi{10.1007/s11214-020-00699-2}

\bibitem[{{Masunaga} {et~al.}(1998){Masunaga}, {Miyama}, \&
  {Inutsuka}}]{Masunaga98}
{Masunaga}, H., {Miyama}, S.~M., \& {Inutsuka}, S.-i. 1998, \apj, 495, 346,
  \dodoi{10.1086/305281}

\bibitem[{{Mathew} \& {Federrath}(2020)}]{mathew2020}
{Mathew}, S.~S., \& {Federrath}, C. 2020, \mnras,
  \dodoi{10.1093/mnras/staa1931}

\bibitem[{{Motte} {et~al.}(2018){Motte}, {Nony}, {Louvet}, {Marsh}, {Bontemps},
  {Whitworth}, {Men'shchikov}, {Nguyen Luong}, {Csengeri}, {Maury}, {Gusdorf},
  {Chapillon}, {K{\"o}nyves}, {Schilke}, {Duarte-Cabral}, {Didelon}, \&
  {Gaudel}}]{motte2018}
{Motte}, F., {Nony}, T., {Louvet}, F., {et~al.} 2018, Nature Astronomy, 2, 478,
  \dodoi{10.1038/s41550-018-0452-x}

\bibitem[{{Murray} \& {Chang}(2015)}]{chang2015}
{Murray}, N., \& {Chang}, P. 2015, \apj, 804, 44,
  \dodoi{10.1088/0004-637X/804/1/44}

\bibitem[{{Ntormousi} \& {Hennebelle}(2019)}]{ntormousi2019}
{Ntormousi}, E., \& {Hennebelle}, P. 2019, \aap, 625, A82,
  \dodoi{10.1051/0004-6361/201834094}

\bibitem[{{Offner} {et~al.}(2014){Offner}, {Clark}, {Hennebelle}, {Bastian},
  {Bate}, {Hopkins}, {Moraux}, \& {Whitworth}}]{offner2014}
{Offner}, S.~S.~R., {Clark}, P.~C., {Hennebelle}, P., {et~al.} 2014, Protostars
  and Planets VI, 53, \dodoi{10.2458/azu_uapress_9780816531240-ch003}

\bibitem[{{Offner} {et~al.}(2009){Offner}, {Klein}, {McKee}, \&
  {Krumholz}}]{offner2009}
{Offner}, S. S.~R., {Klein}, R.~I., {McKee}, C.~F., \& {Krumholz}, M.~R. 2009,
  \apj, 703, 131, \dodoi{10.1088/0004-637X/703/1/131}

\bibitem[{{Offner} \& {McKee}(2011)}]{offner2011}
{Offner}, S. S.~R., \& {McKee}, C.~F. 2011, \apj, 736, 53,
  \dodoi{10.1088/0004-637X/736/1/53}

\bibitem[{{Padoan} {et~al.}(1997){Padoan}, {Nordlund}, \& {Jones}}]{padoan1997}
{Padoan}, P., {Nordlund}, A., \& {Jones}, B.~J.~T. 1997, \mnras, 288, 145,
  \dodoi{10.1093/mnras/288.1.145}

\bibitem[{{Padoan} {et~al.}(2020){Padoan}, {Pan}, {Juvela}, {Haugb{\o}lle}, \&
  {Nordlund}}]{padoan2020}
{Padoan}, P., {Pan}, L., {Juvela}, M., {Haugb{\o}lle}, T., \& {Nordlund},
  {\r{A}}. 2020, \apj, 900, 82, \dodoi{10.3847/1538-4357/abaa47}

\bibitem[{{Salpeter}(1955)}]{salpeter55}
{Salpeter}, E.~E. 1955, \apj, 121, 161, \dodoi{10.1086/145971}

\bibitem[{{Saumon} \& {Chabrier}(1992)}]{saumon1992}
{Saumon}, D., \& {Chabrier}, G. 1992, \pra, 46, 2084,
  \dodoi{10.1103/PhysRevA.46.2084}

\bibitem[{{Saumon} {et~al.}(1995){Saumon}, {Chabrier}, \& {van
  Horn}}]{saumon1995}
{Saumon}, D., {Chabrier}, G., \& {van Horn}, H.~M. 1995, \apjs, 99, 713,
  \dodoi{10.1086/192204}

\bibitem[{{Schneider} {et~al.}(2018){Schneider}, {Sana}, {Evans},
  {Bestenlehner}, {Castro}, {Fossati}, {Gr{\"a}fener}, {Langer},
  {Ram{\'\i}rez-Agudelo}, {Sab{\'\i}n-Sanjuli{\'a}n}, {Sim{\'o}n-D{\'\i}az},
  {Tramper}, {Crowther}, {de Koter}, {de Mink}, {Dufton}, {Garcia}, {Gieles},
  {H{\'e}nault-Brunet}, {Herrero}, {Izzard}, {Kalari}, {Lennon}, {Ma{\'\i}z
  Apell{\'a}niz}, {Markova}, {Najarro}, {Podsiadlowski}, {Puls}, {Taylor}, {van
  Loon}, {Vink}, \& {Norman}}]{schneider2018}
{Schneider}, F.~R.~N., {Sana}, H., {Evans}, C.~J., {et~al.} 2018, Science, 359,
  69, \dodoi{10.1126/science.aan0106}

\bibitem[{{Semenov} {et~al.}(2003){Semenov}, {Henning}, {Helling}, {Ilgner}, \&
  {Sedlmayr}}]{semenov2003}
{Semenov}, D., {Henning}, T., {Helling}, C., {Ilgner}, M., \& {Sedlmayr}, E.
  2003, \aap, 410, 611, \dodoi{10.1051/0004-6361:20031279}

\bibitem[{{Stamatellos} {et~al.}(2011){Stamatellos}, {Whitworth}, \&
  {Hubber}}]{stamatellos2011}
{Stamatellos}, D., {Whitworth}, A.~P., \& {Hubber}, D.~A. 2011, \apj, 730, 32,
  \dodoi{10.1088/0004-637X/730/1/32}

\bibitem[{{Teyssier}(2002)}]{Teyssier02}
{Teyssier}, R. 2002, \aap, 385, 337, \dodoi{10.1051/0004-6361:20011817}

\bibitem[{{Urban} {et~al.}(2010){Urban}, {Martel}, \& {Evans}}]{urban2010}
{Urban}, A., {Martel}, H., \& {Evans}, Neal~J., I. 2010, \apj, 710, 1343,
  \dodoi{10.1088/0004-637X/710/2/1343}

\bibitem[{{Urquhart} {et~al.}(2014){Urquhart}, {Moore}, {Csengeri}, {Wyrowski},
  {Schuller}, {Hoare}, {Lumsden}, {Mottram}, {Thompson}, {Menten}, {Walmsley},
  {Bronfman}, {Pfalzner}, {K{\"o}nig}, \& {Wienen}}]{Urquhart14}
{Urquhart}, J.~S., {Moore}, T.~J.~T., {Csengeri}, T., {et~al.} 2014, \mnras,
  443, 1555, \dodoi{10.1093/mnras/stu1207}

\bibitem[{{Vaytet} {et~al.}(2013){Vaytet}, {Chabrier}, {Audit},
  {Commer{\c{c}}on}, {Masson}, {Ferguson}, \& {Delahaye}}]{vaytet2013}
{Vaytet}, N., {Chabrier}, G., {Audit}, E., {et~al.} 2013, \aap, 557, A90,
  \dodoi{10.1051/0004-6361/201321423}

\bibitem[{{Vaytet} \& {Haugb{\o}lle}(2017)}]{Vaytet17}
{Vaytet}, N., \& {Haugb{\o}lle}, T. 2017, \aap, 598, A116,
  \dodoi{10.1051/0004-6361/201628194}

\bibitem[{{Vorobyov} \& {Basu}(2010)}]{vorobyov2010}
{Vorobyov}, E.~I., \& {Basu}, S. 2010, \apj, 719, 1896,
  \dodoi{10.1088/0004-637X/719/2/1896}

\bibitem[{{Wurster} \& {Li}(2018)}]{liwurster2019}
{Wurster}, J., \& {Li}, Z.-Y. 2018, Frontiers in Astronomy and Space Sciences,
  5, 39, \dodoi{10.3389/fspas.2018.00039}

\bibitem[{{Zhao} {et~al.}(2020){Zhao}, {Tomida}, {Hennebelle}, {Tobin},
  {Maury}, {Hirota}, {S{\'a}nchez-Monge}, {Kuiper}, {Rosen}, {Bhandare},
  {Padovani}, \& {Lee}}]{zhao2020}
{Zhao}, B., {Tomida}, K., {Hennebelle}, P., {et~al.} 2020, \ssr, 216, 43,
  \dodoi{10.1007/s11214-020-00664-z}

\end{thebibliography}
\bibliographystyle{aasjournal} % style aa.bst

\appendix

\section{Bidimensional density-temperature histograms}
\label{bidim}
To get more hint on the physical conditions within our modelled clumps, we present here 
bidimensional Temperature-density histograms for the two most extreme runs 
of the COMP type, namely COMP-ACLUM and COMP-NOFEED as well as for the most resolved one 
COMP-NOACLUMhr,
 at three different snapshots. The results are display in Fig.~\ref{Tcoldens}.
For completeness, we also show bidimensional temperature-density histograms for runs 
STAN-ACLUM, STAN-ACLUMhrhs and STAN-ACLUM-vhrhs in Fig.~\ref{Tcoldens_stan}.

Bidimensional histograms contain more information that the mean temperature as 
a function of density presented in Figs.~\ref{T_rho_comp} and Figs.~\ref{T_rho_diff}.
The bidimensional histograms reveal that there is a significant spread in temperature, even regarding the 
 gas within a narrow density range. However the bulk of the gas tends 
to lay in regions which appear better defined with significantly weaker dispersion.

By the  time of the first snapshot a few solar masses have been accreted. In the three runs the
temperature distributions are similar; as expected there is a clear transition between 
the isothermal and adiabatic regimes. The transition itself is reasonably well described by the barotropic
eos calculations. At higher density, the temperature is clearly higher than 
the eos values. However, this is  essentially a consequence of insufficient resolution (that we recall is $\simeq$2 AU 
or 1 AU for COMP-NOACLUMhr). Indeed, the temperature distribution is closer to the analytical expression 
in run COMP-NOACLUMhr, which has more resolution. Similar

%The distributions at later time ($>$0.012 Myr) are more different. This is because some more stars have formed and 
%the accretion luminosity is very significant in run COMP-ACLUM. 
% While the bulk temperature is about 20 K for run $COMP-NOFEED$, it reaches  
%values of about 100 and even 300 K for the last snapshot.

\setlength{\unitlength}{1cm}
\begin{figure*}%[h!]
%\centering
\begin{picture} (0,15)
%\put(0,15){\includegraphics[width=6.5cm]{FIG_PAPIER_RHDIMF/CLUSTER_IMF_feedback_eos_hres_modcrit_hsink2/hist2D_rho_Temp_00020.pdf}}
%\put(6.5,15){\includegraphics[width=6.5cm]{FIG_PAPIER_RHDIMF/CLUSTER_IMF_feedback_eos_hres_modcrit_hsink2/hist2D_rho_Temp_00040.pdf}}
%\put(13,15){\includegraphics[width=6.5cm]{FIG_PAPIER_RHDIMF/CLUSTER_IMF_feedback_eos_hres_modcrit_hsink2/hist2D_rho_Temp_00200.pdf}}
%\put(1.5,19){R1}
%\put(8.,19){R1}
%\put(14.5,19){R1}
\put(0,10){\includegraphics[width=6.5cm]{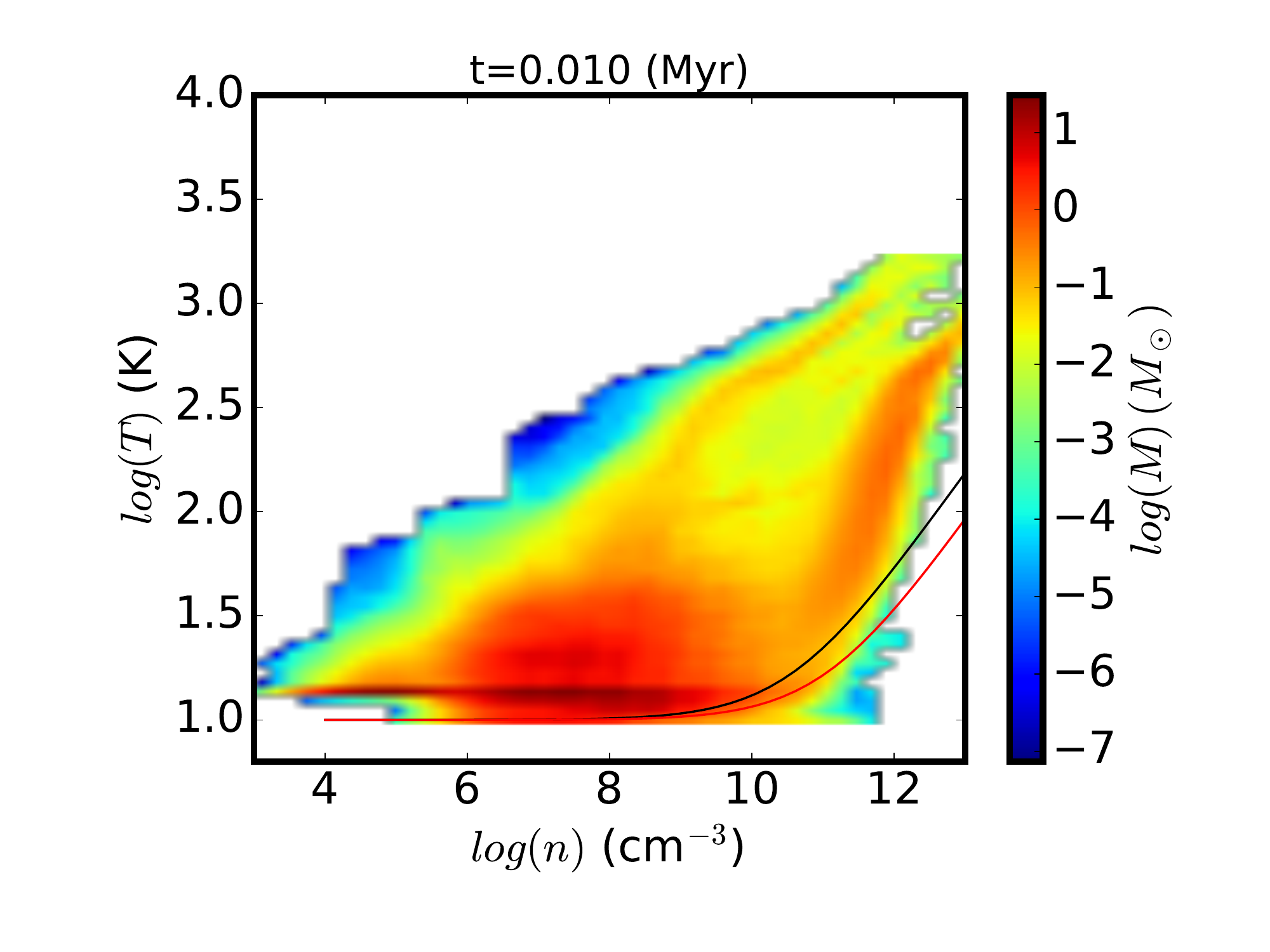}}
\put(6.5,10){\includegraphics[width=6.5cm]{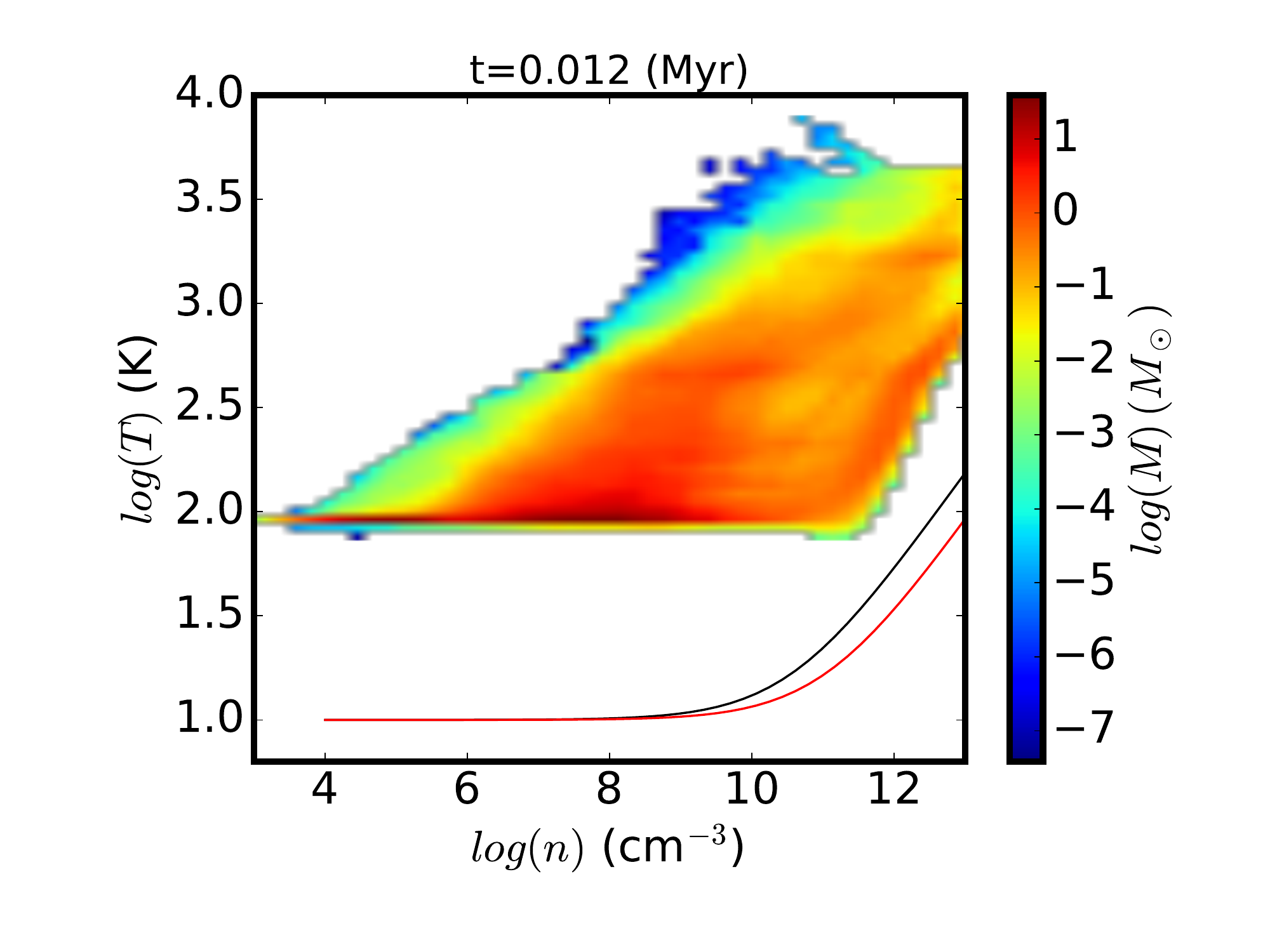}}
\put(13,10){\includegraphics[width=6.5cm]{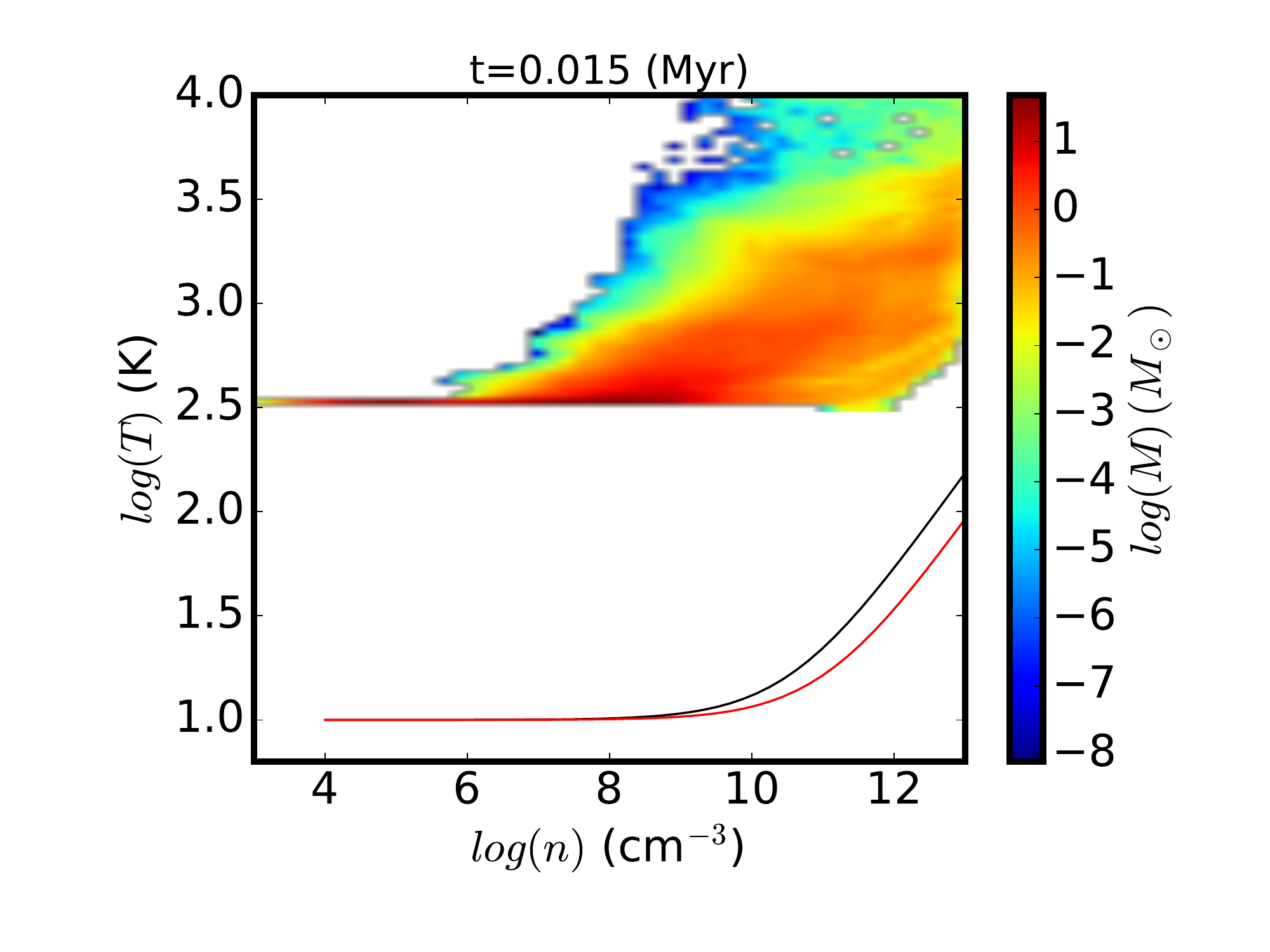}}
\put(1.5,14){COMP-ACLUM}
\put(8.,14){COMP-ACLUM}
\put(14.5,14){COMP-ACLUM}
\put(0,5){\includegraphics[width=6.5cm]{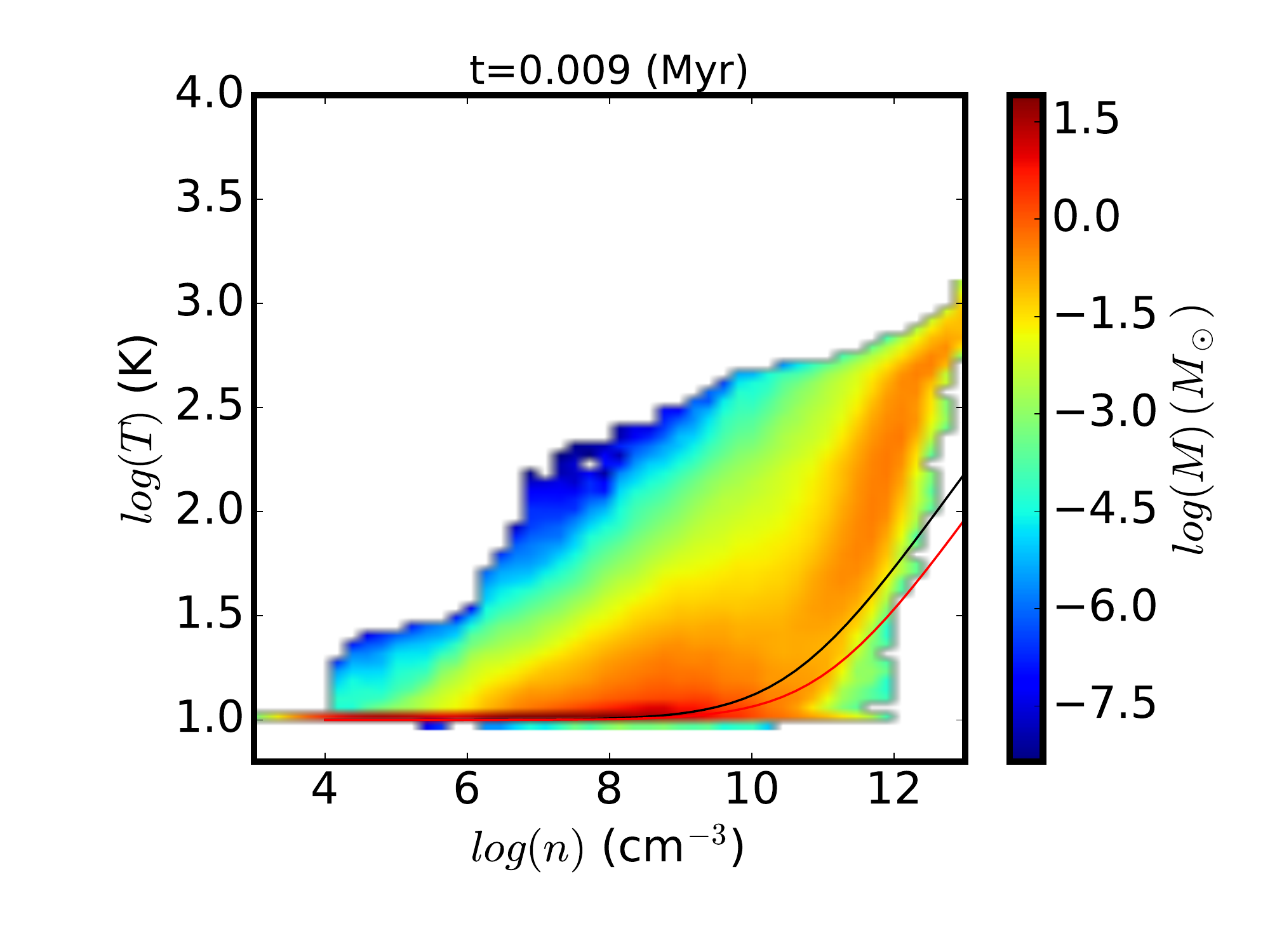}}
\put(6.5,5){\includegraphics[width=6.5cm]{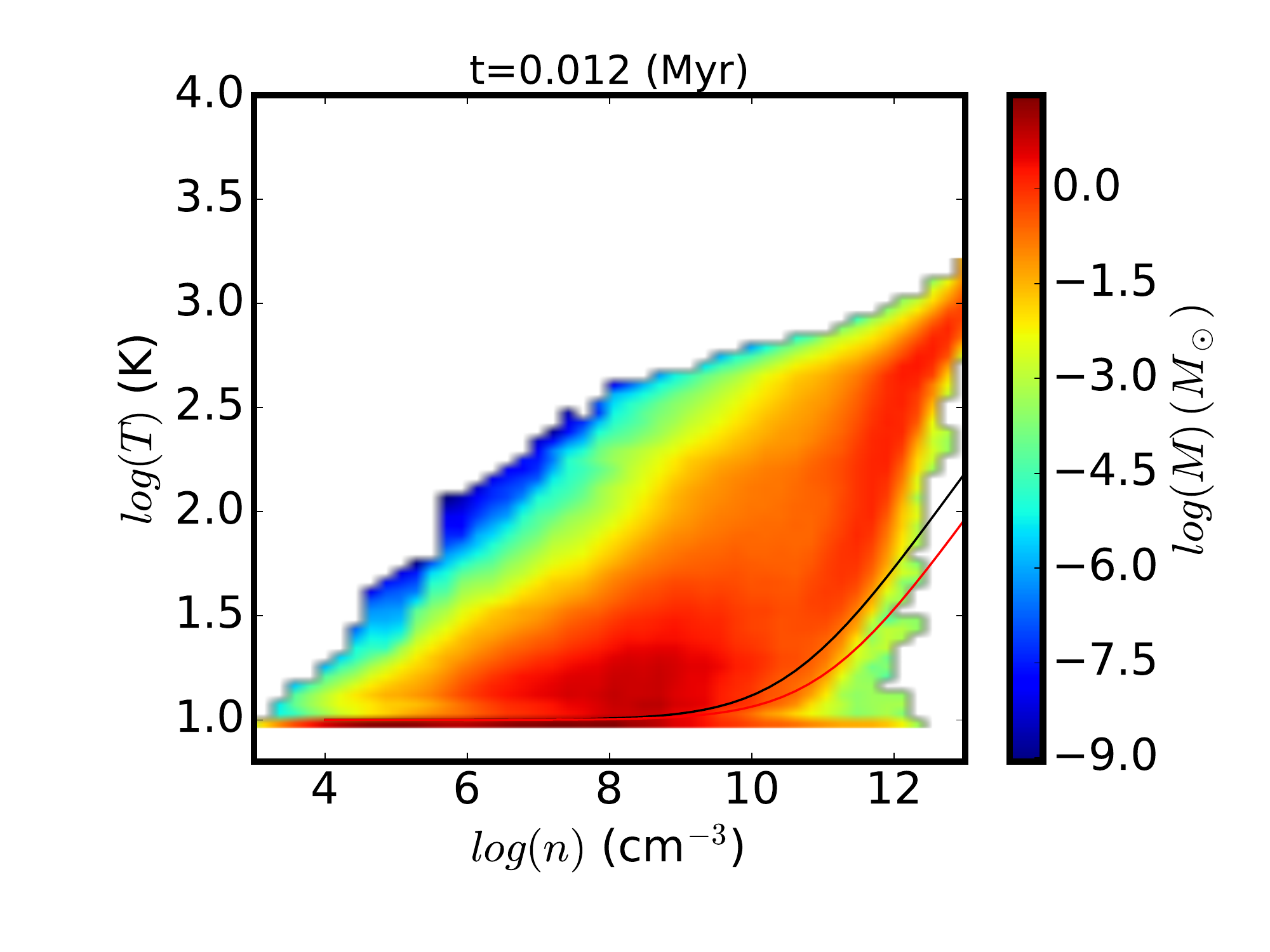}}
\put(13,5){\includegraphics[width=6.5cm]{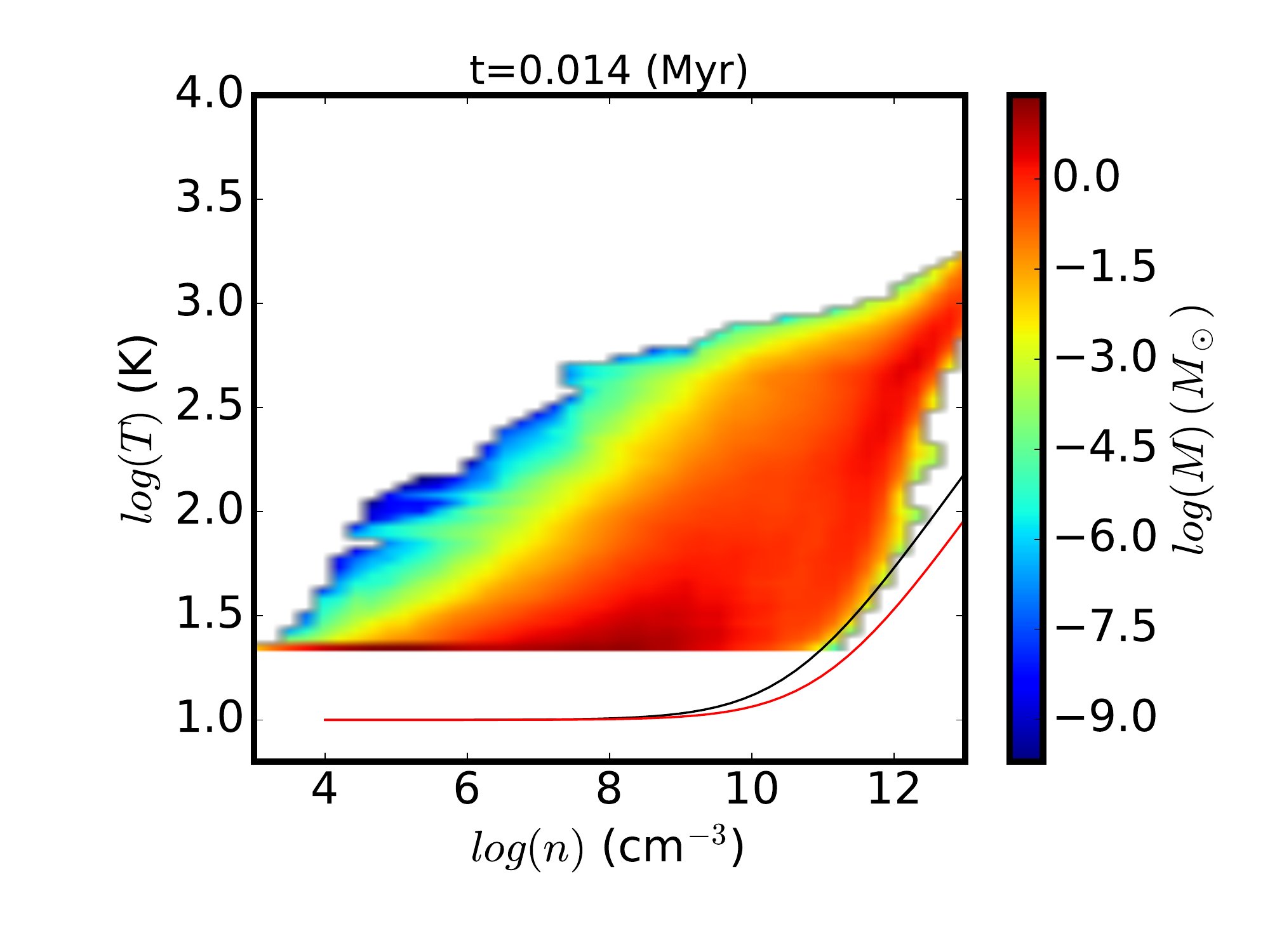}}
\put(1.5,9){COMP-NOFEED}
\put(8.,9){COMP-NOFEED}
\put(14.5,9){COMP-NOFEED}
\put(0,0){\includegraphics[width=6.5cm]{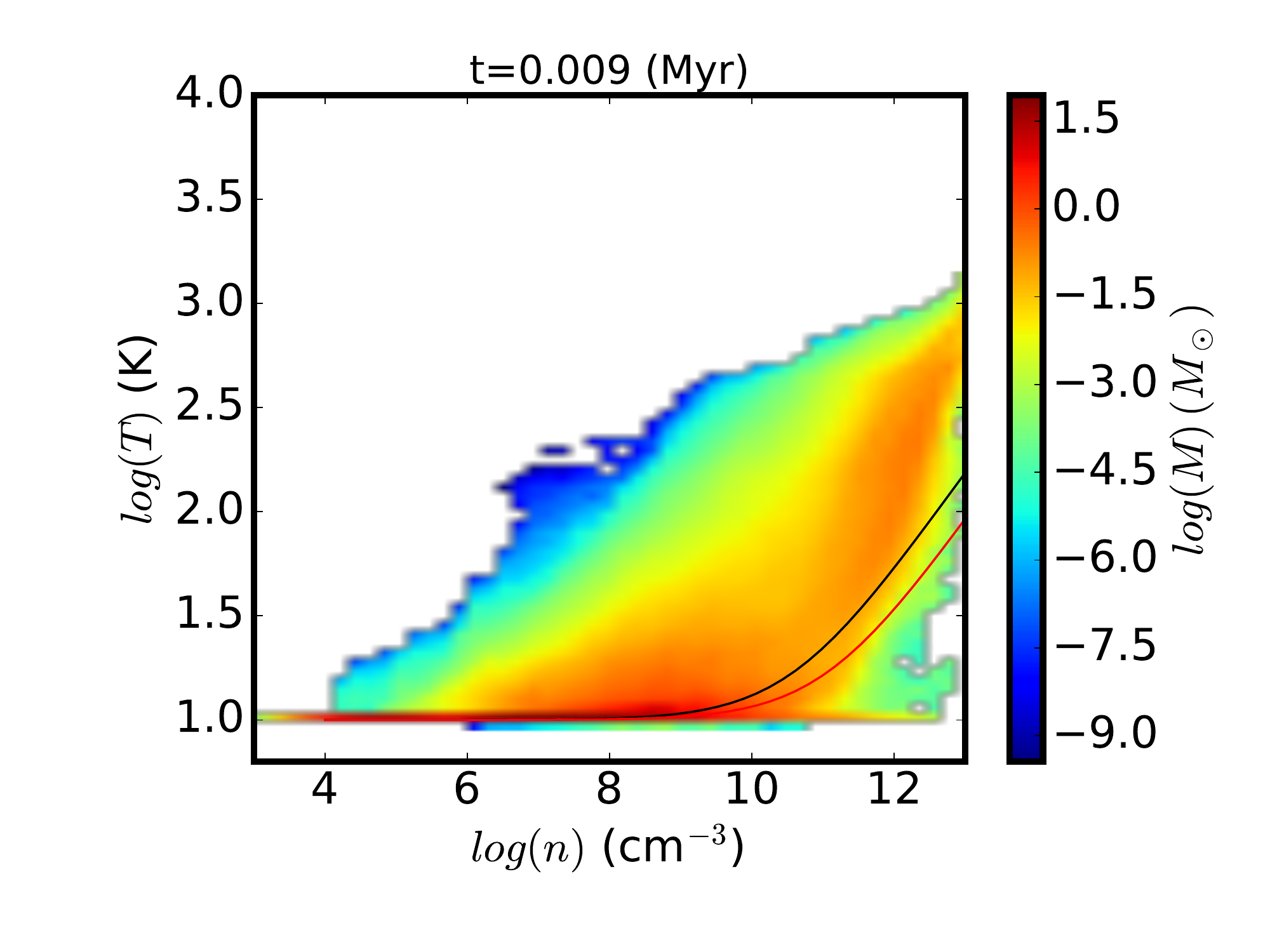}}
\put(6.5,0){\includegraphics[width=6.5cm]{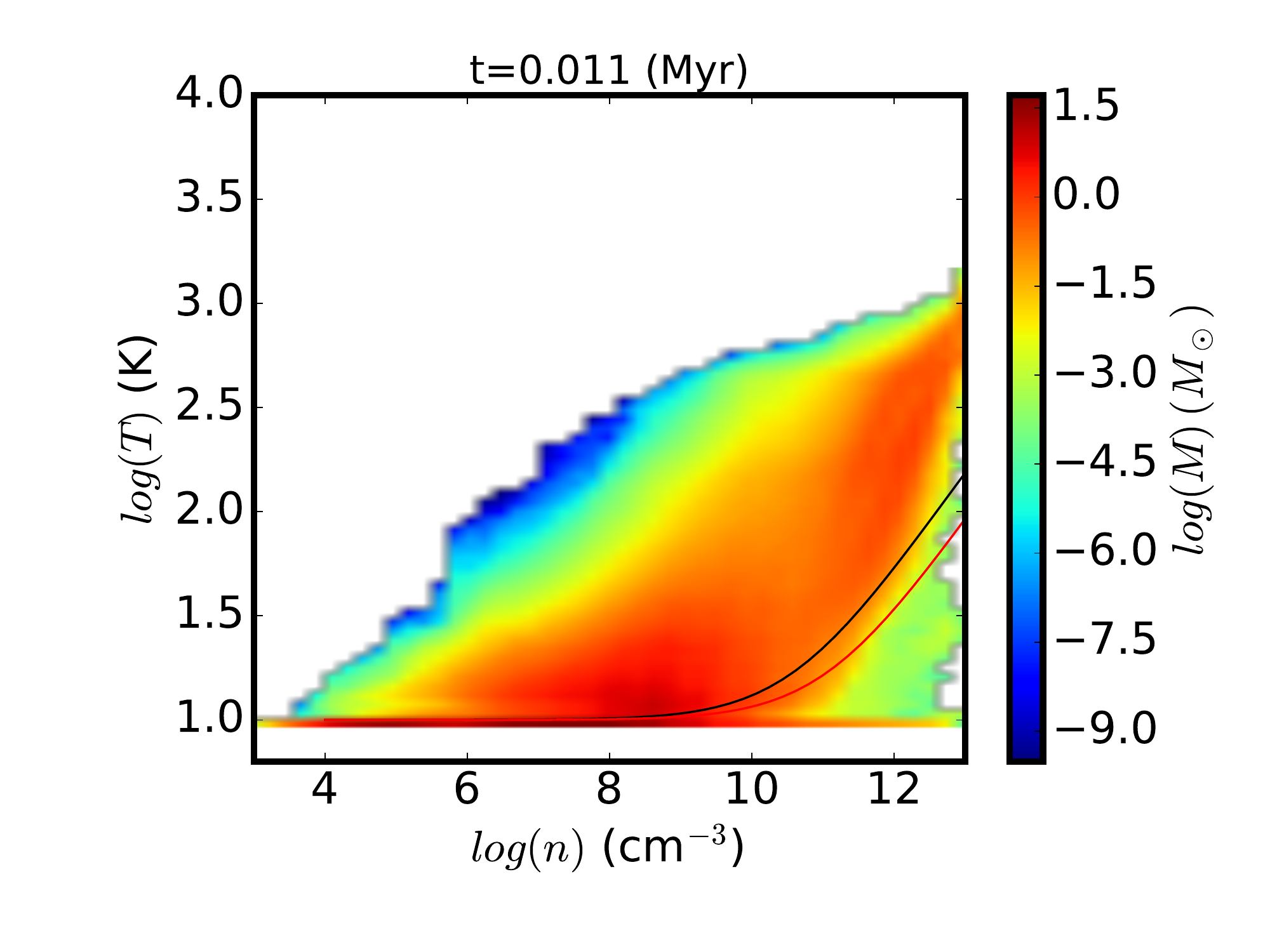}}
\put(13,0){\includegraphics[width=6.5cm]{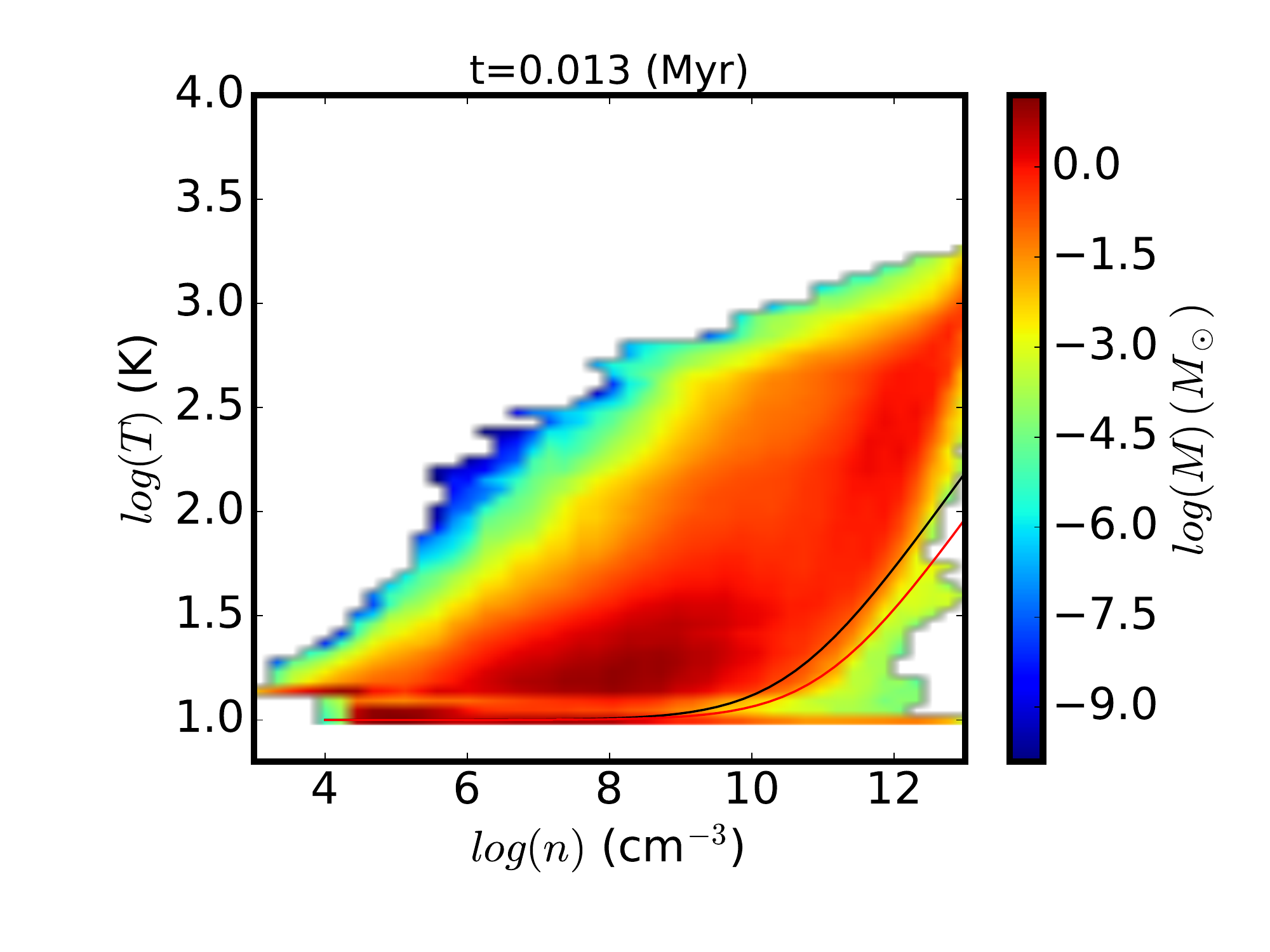}}
\put(1.5,4){COMP-NOACLUMhr}
\put(8.,4){COMP-NOACLUMhr}
\put(14.5,4){COMP-NOACLUMhr}
\end{picture}
\caption{Bidimensional Temperature-density histograms as a function at three timesteps for run COMP-ACLUM ($R=0.1$ pc initially
and accretion luminosity is taken into account) and run COMP-NOFEED (no stellar feedback).
 The left and middles column are for early time when a few tens of solar masses have been accreted. The 
right column is for later time when about 200 solar masses have been accreted. The two curves represent the 
two eos as stated by eq.~(\ref{eq_full_eos}).}
\label{Tcoldens}
\end{figure*}
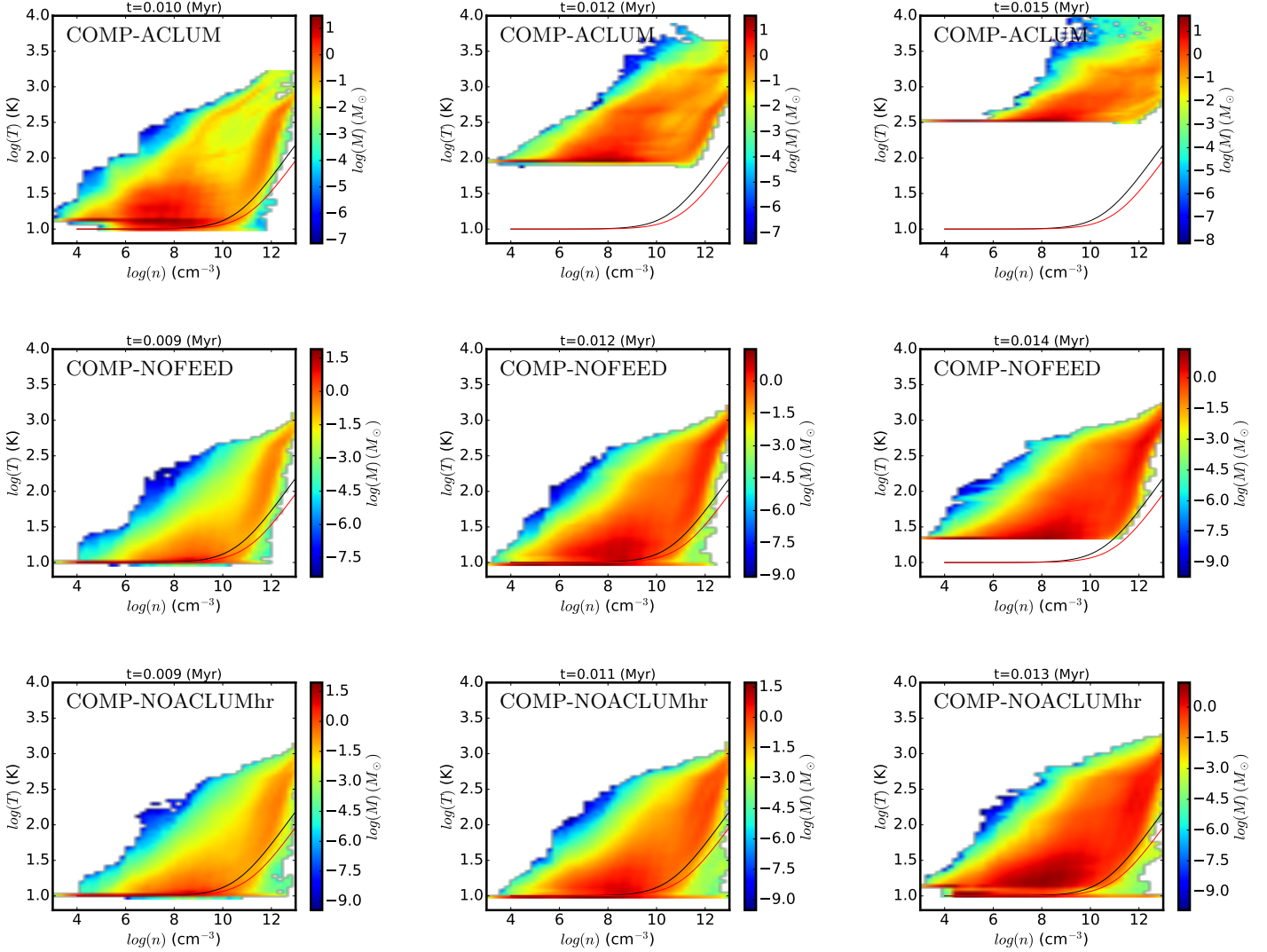

\setlength{\unitlength}{1cm}
\begin{figure*}%[h!]
%\centering
\begin{picture} (0,15)
%\put(0,15){\includegraphics[width=6.5cm]{FIG_PAPIER_RHDIMF/CLUSTER_IMF_feedback_eos_hres_modcrit_hsink2/hist2D_rho_Temp_00020.pdf}}
%\put(6.5,15){\includegraphics[width=6.5cm]{FIG_PAPIER_RHDIMF/CLUSTER_IMF_feedback_eos_hres_modcrit_hsink2/hist2D_rho_Temp_00040.pdf}}
%\put(13,15){\includegraphics[width=6.5cm]{FIG_PAPIER_RHDIMF/CLUSTER_IMF_feedback_eos_hres_modcrit_hsink2/hist2D_rho_Temp_00200.pdf}}
%\put(1.5,19){R1}
%\put(8.,19){R1}
%\put(14.5,19){R1}
\put(0,10){\includegraphics[width=6.5cm]{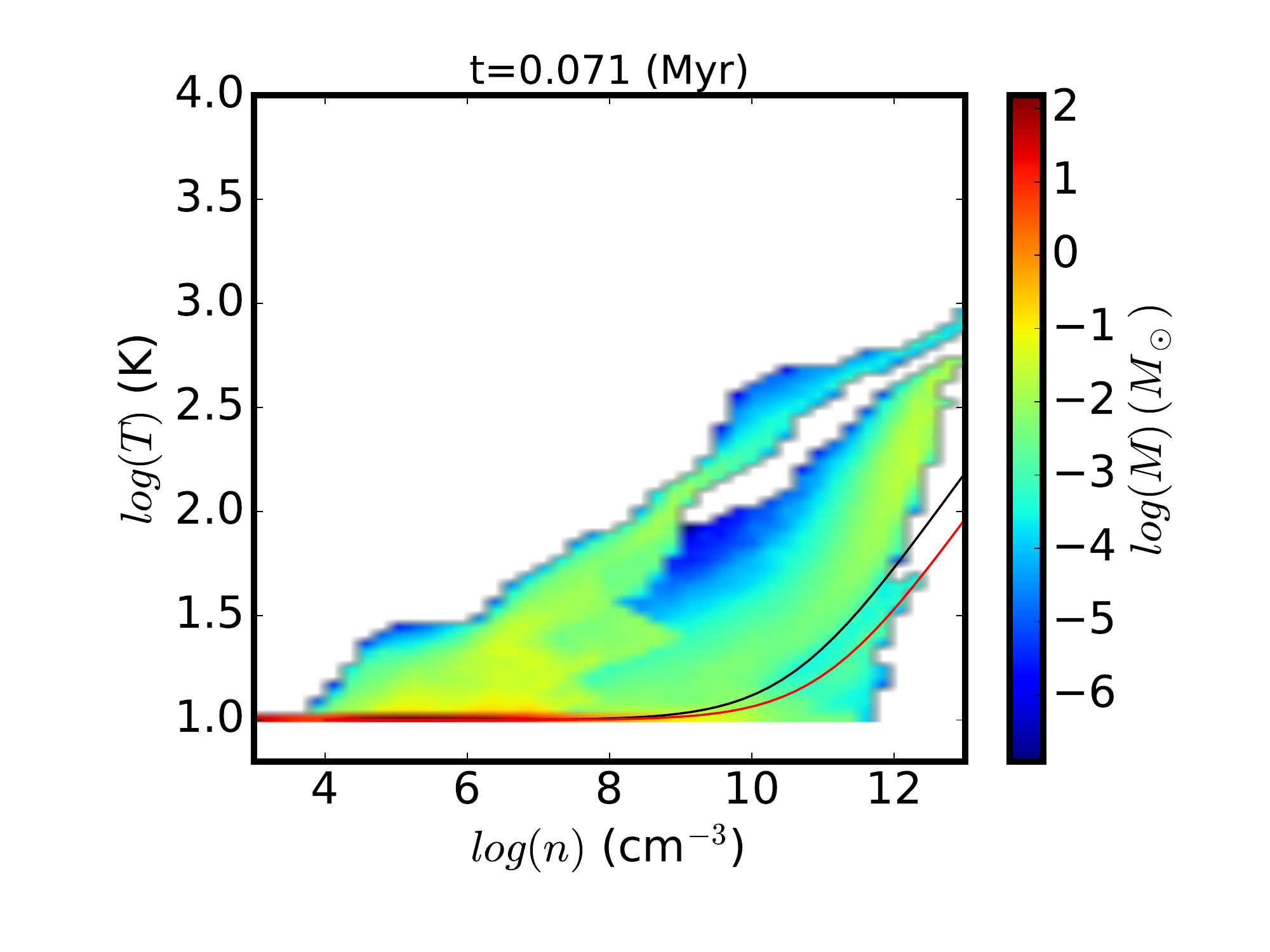}}
\put(6.5,10){\includegraphics[width=6.5cm]{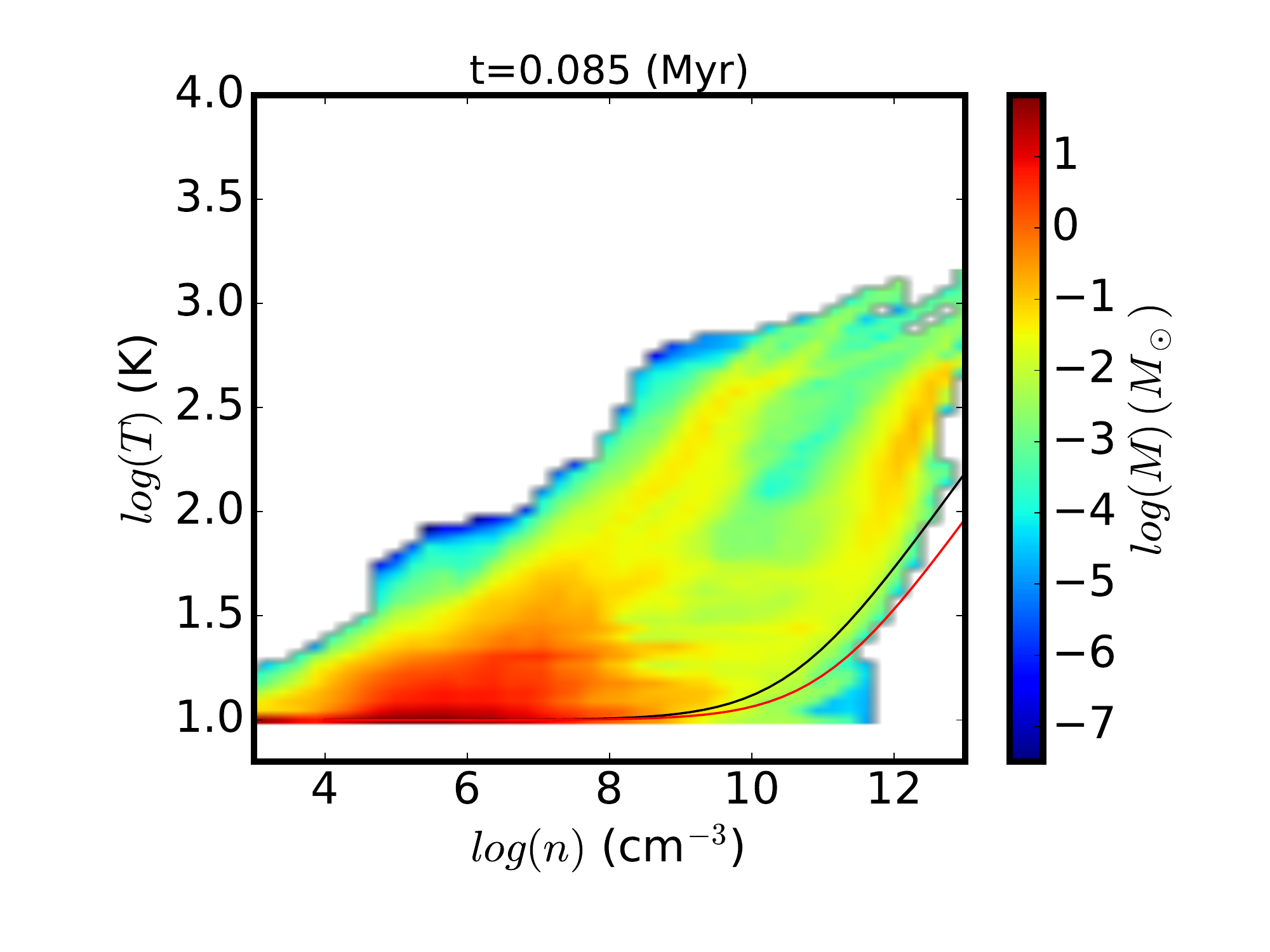}}
\put(13,10){\includegraphics[width=6.5cm]{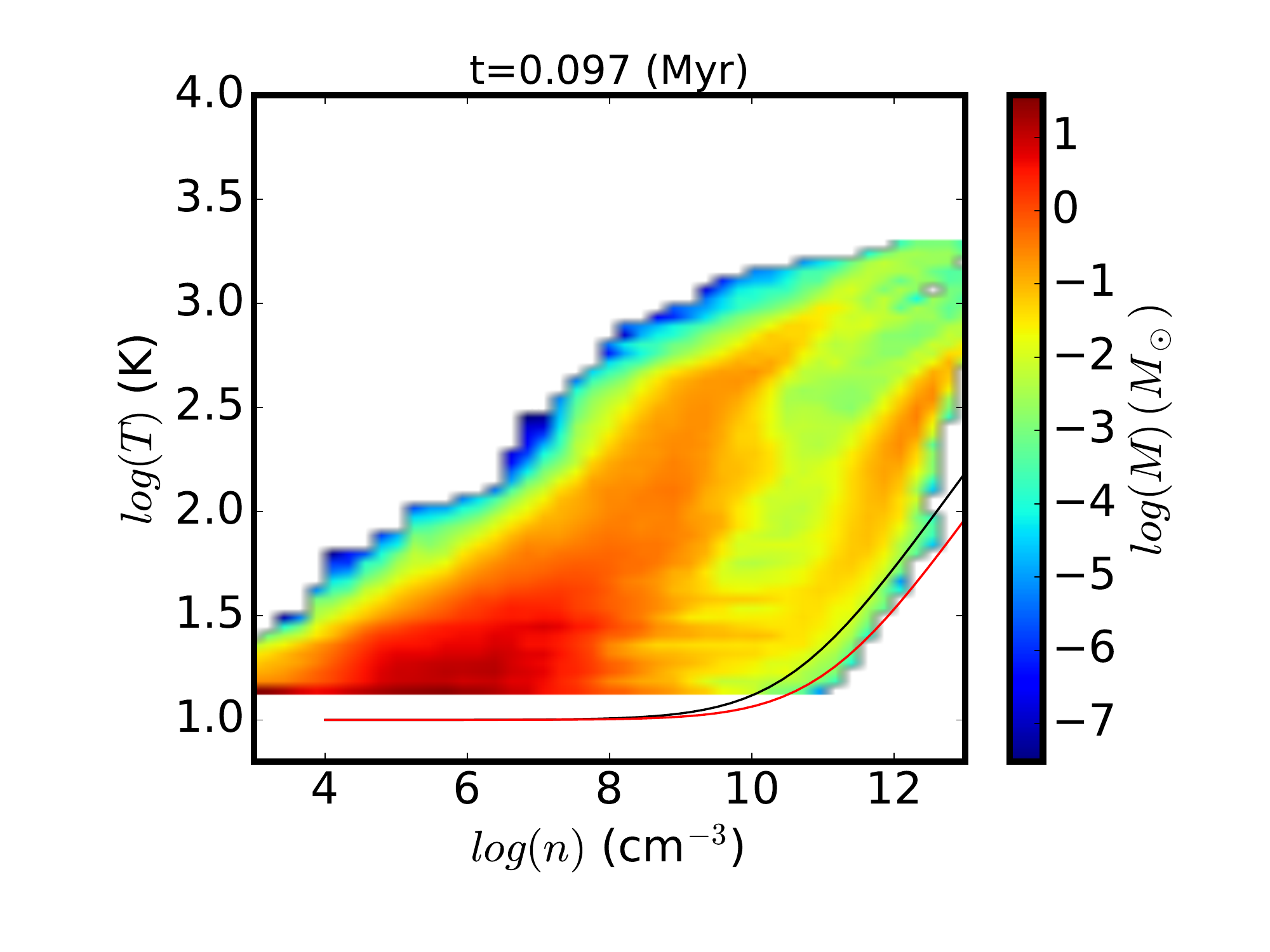}}
\put(1.5,14){STAN-ACLUMvhrhs}
\put(8.,14){STAN-ACLUMvhrhs}
\put(14.5,14){STAN-ACLUMhrhs}
\put(0,5){\includegraphics[width=6.5cm]{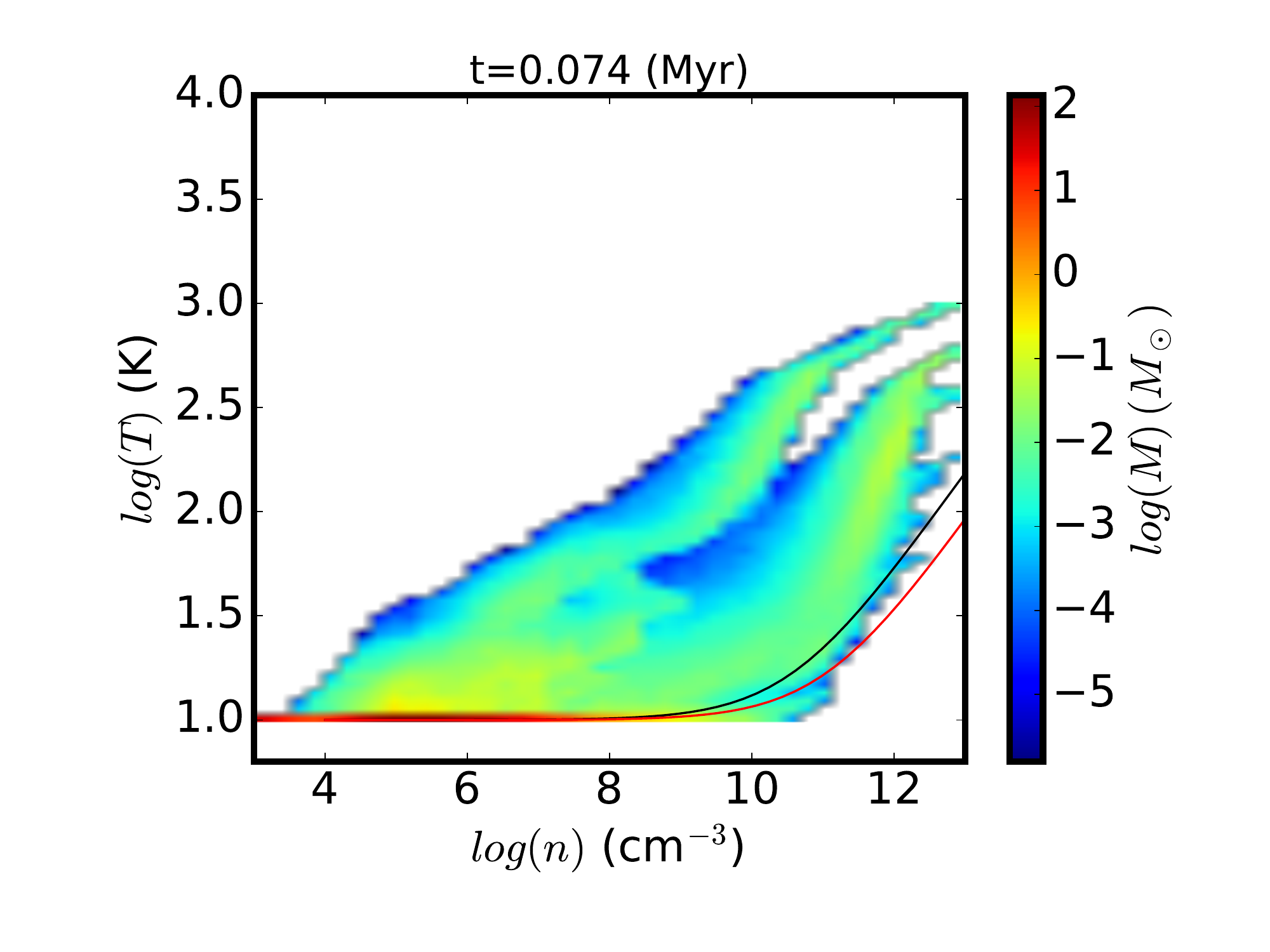}}
\put(6.5,5){\includegraphics[width=6.5cm]{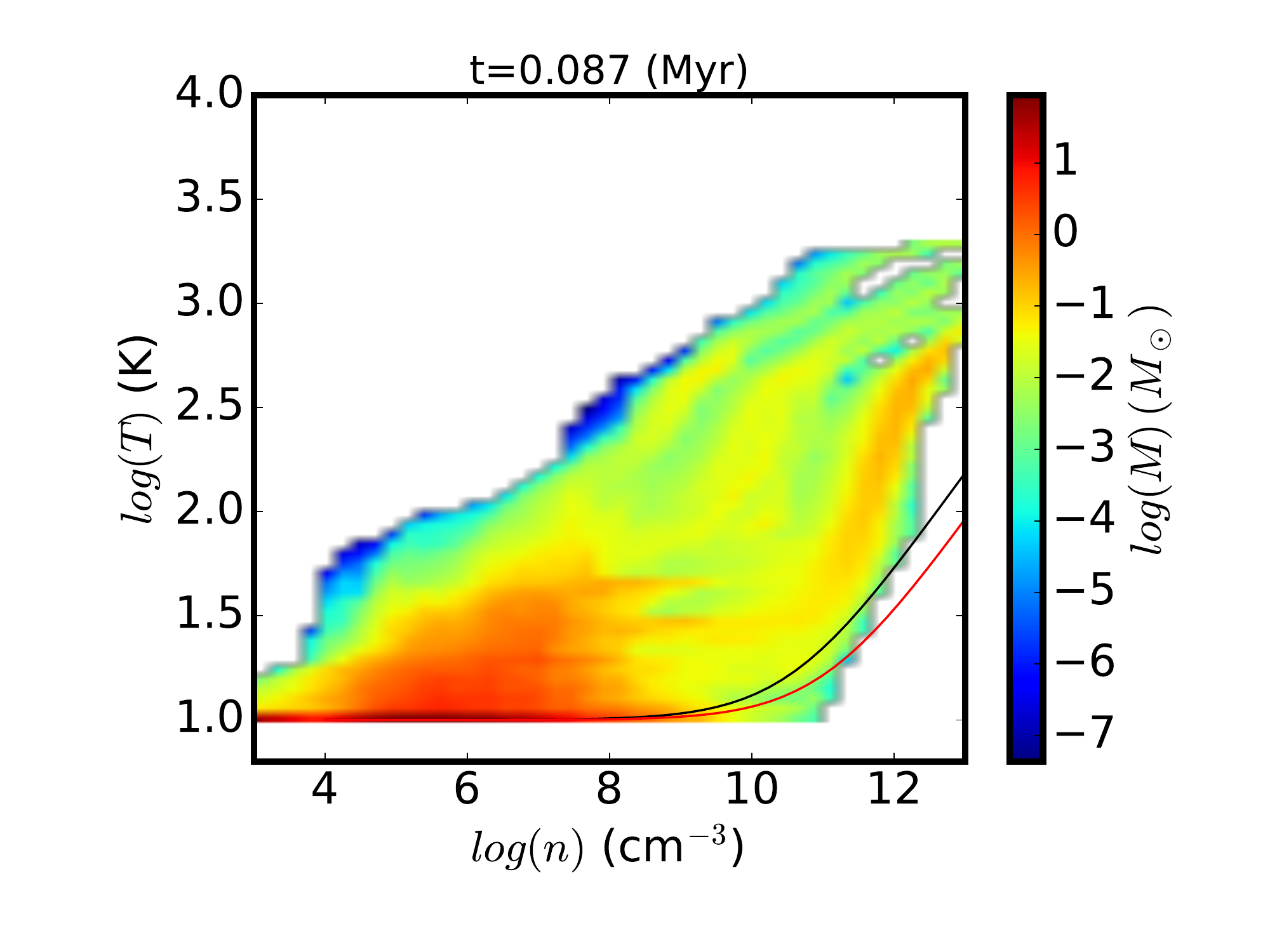}}
\put(13,5){\includegraphics[width=6.5cm]{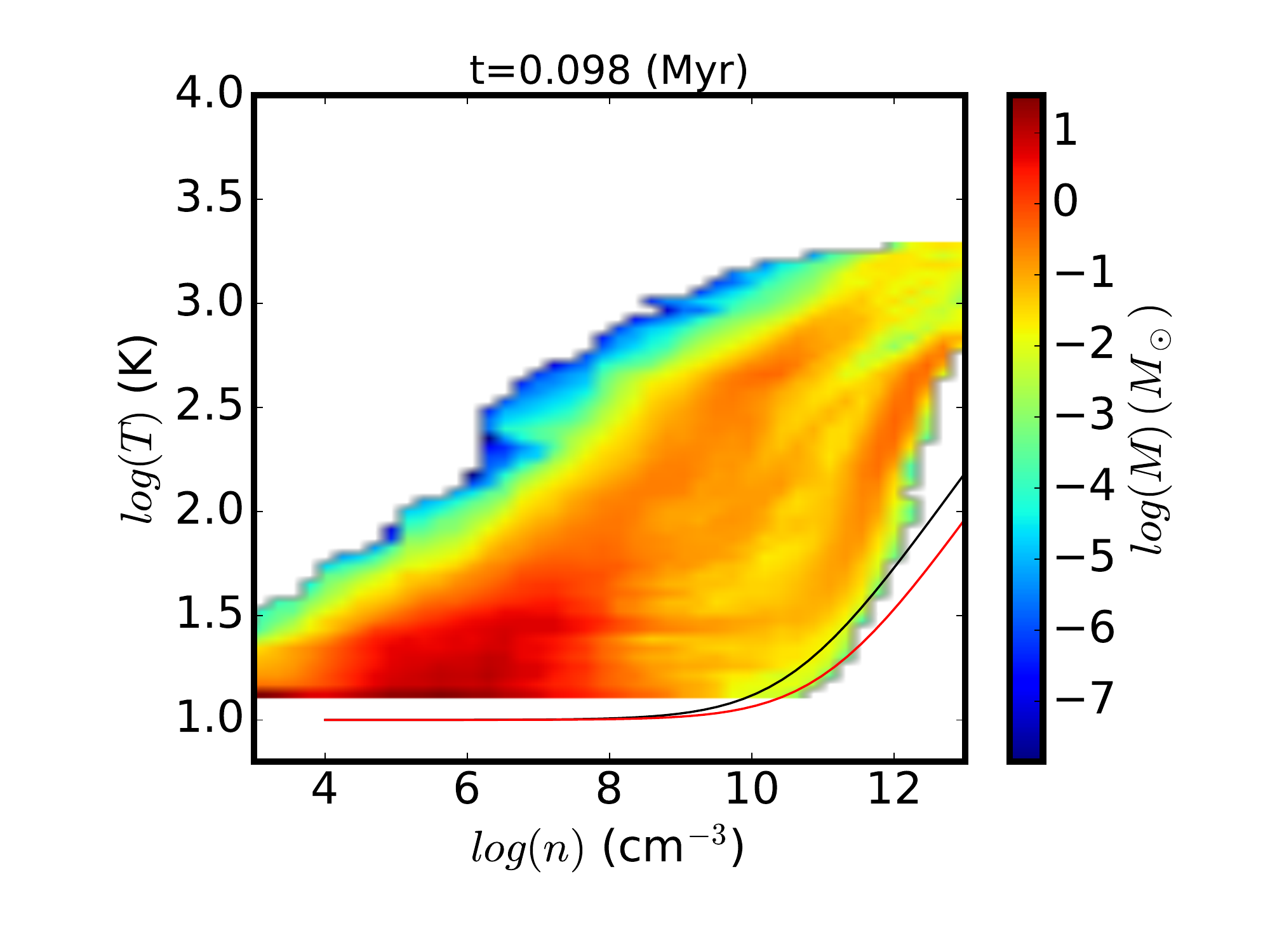}}
\put(1.5,9){STAN-ACLUMhrhs}
\put(8.,9){STAN-ACLUMhrhs}
\put(14.5,9){STAN-ACLUMhrhs}
\put(0,0){\includegraphics[width=6.5cm]{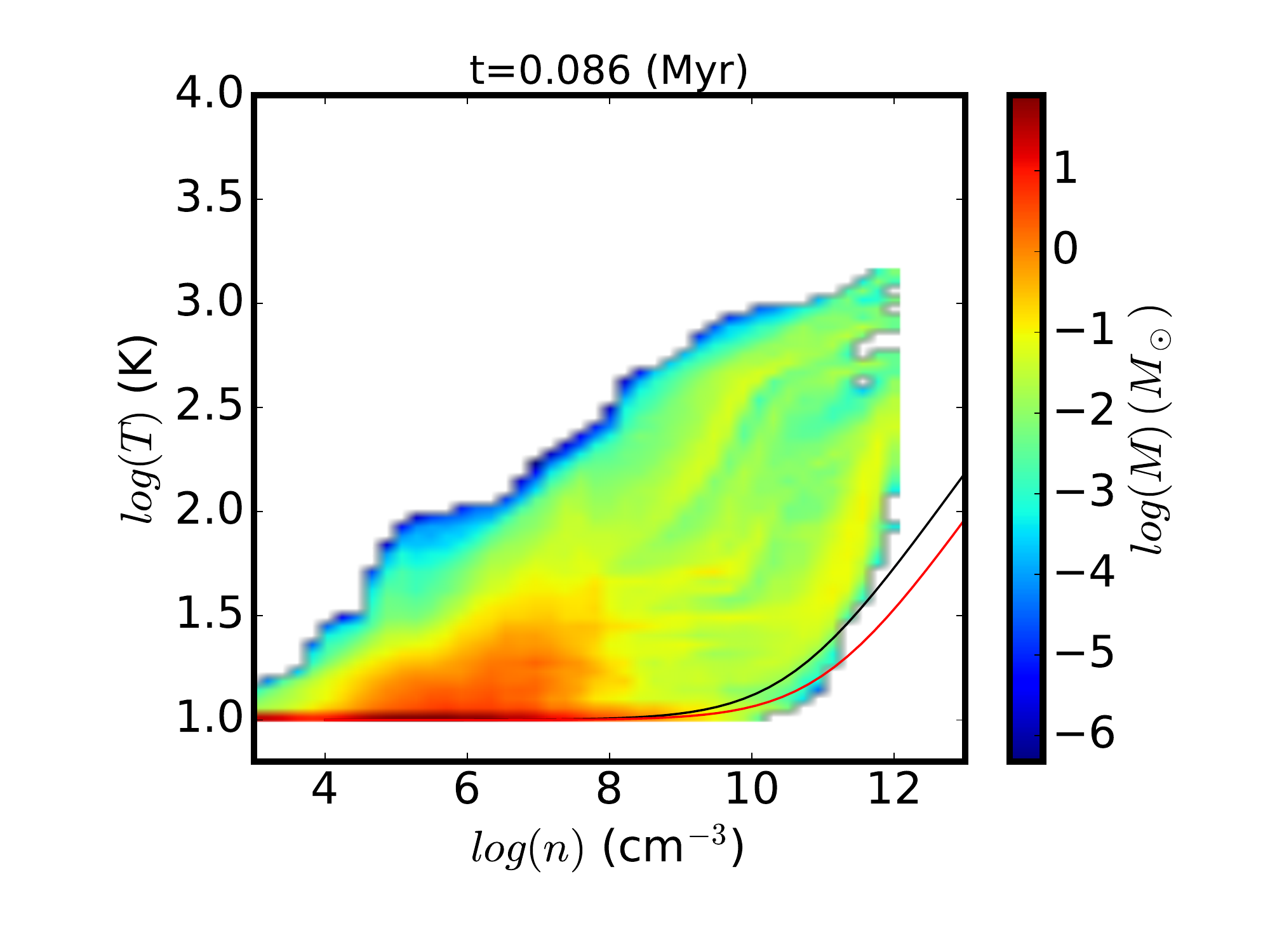}}
\put(6.5,0){\includegraphics[width=6.5cm]{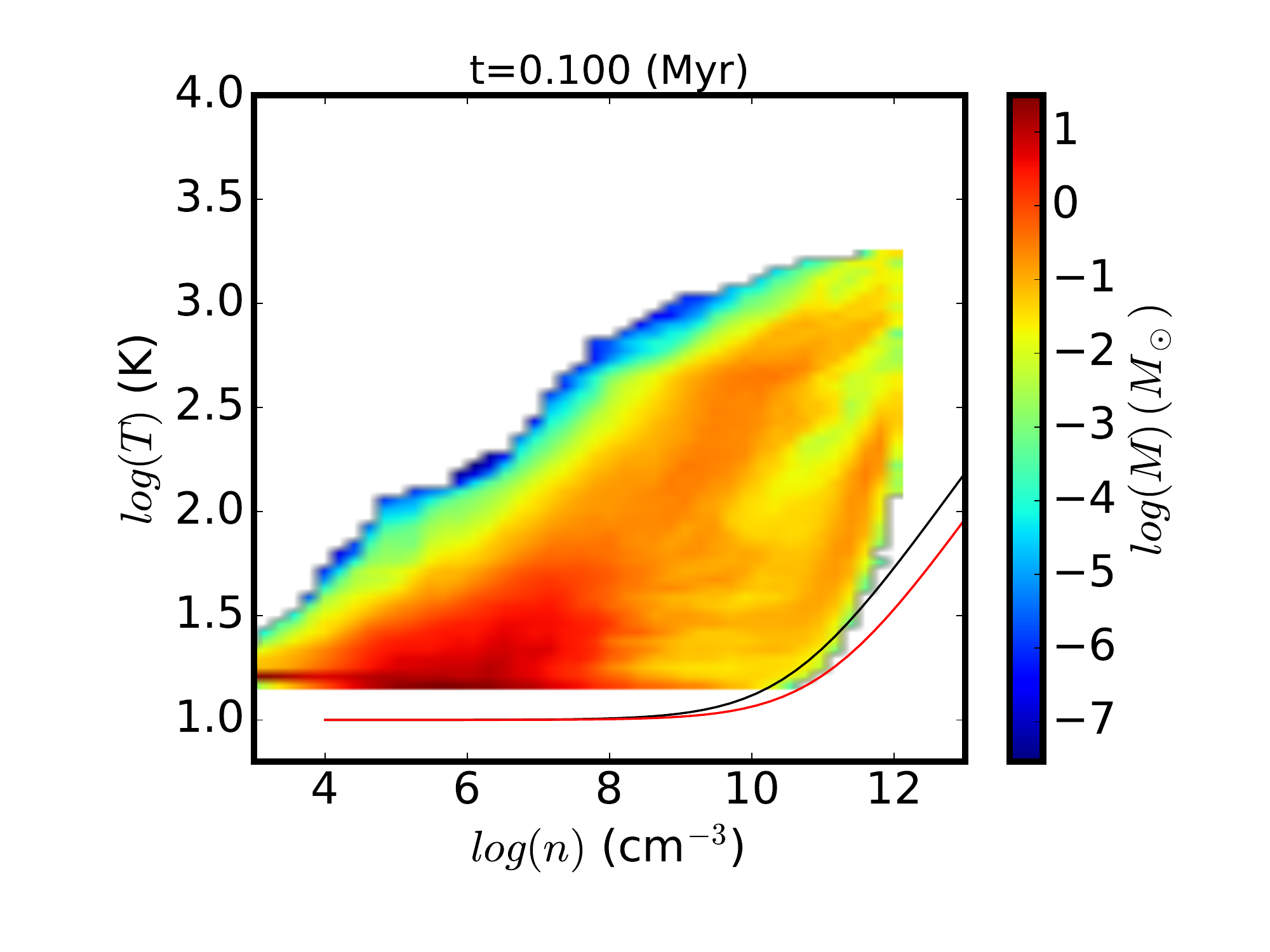}}
\put(13,0){\includegraphics[width=6.5cm]{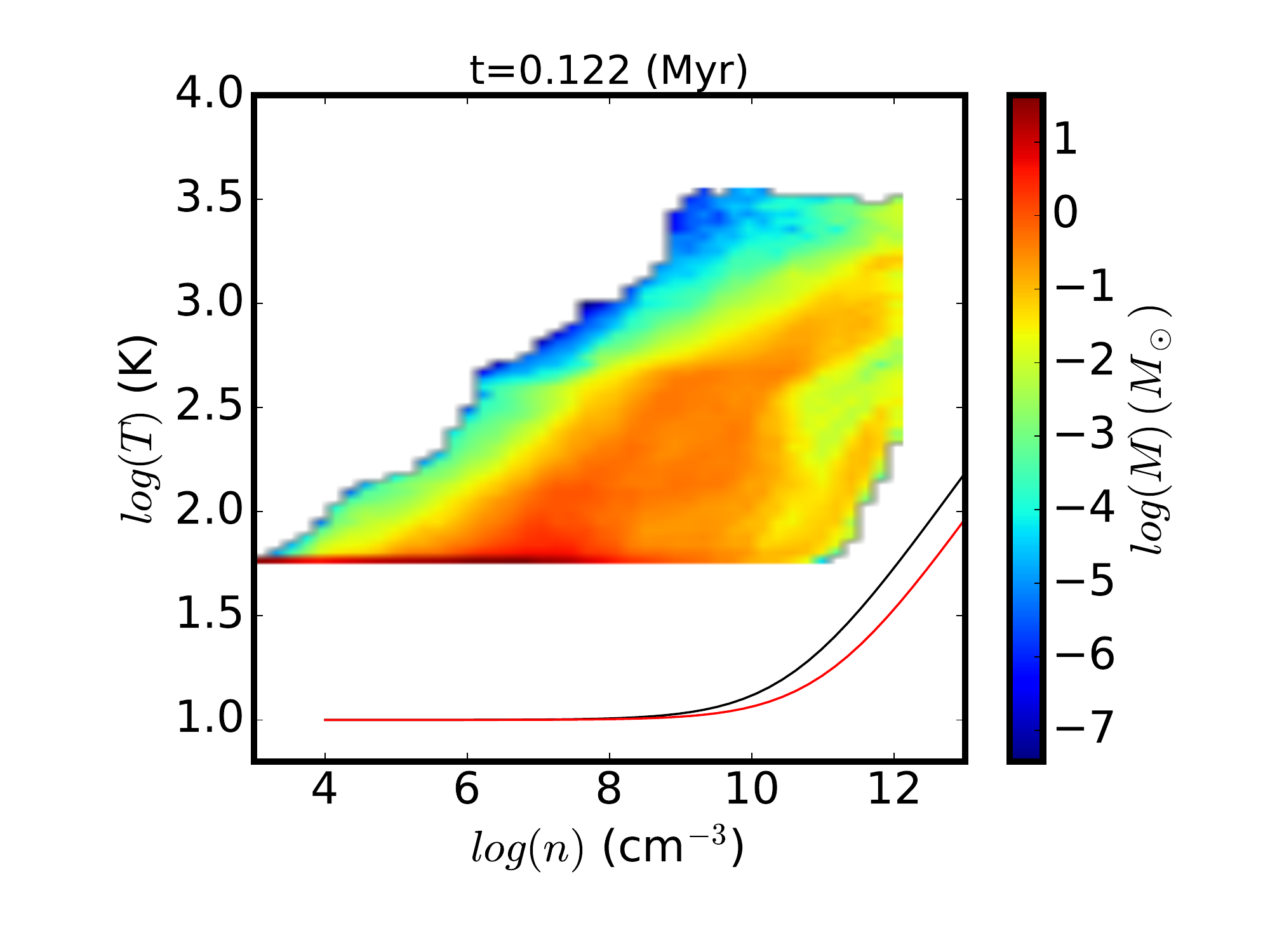}}
\put(1.5,4){STAN-ACLUM}
\put(8.,4){STAN-ACLUM}
\put(14.5,4){STAN-ACLUM}
\end{picture}
\caption{Bidimensional Temperature-density histograms as a function at three timesteps for runs STAN-ACLUM ($R=0.4$ pc initially
and accretion luminosity is taken into account), STAN-ACLUMhrhs and STAN-ACLUMvhrhs.
Comparison between runs at similar time, allows to see the influence of numerical resolution at high density.
 The two curves represent the 
two eos as stated by eq.~(\ref{eq_full_eos}).}
\label{Tcoldens_stan}
\end{figure*}
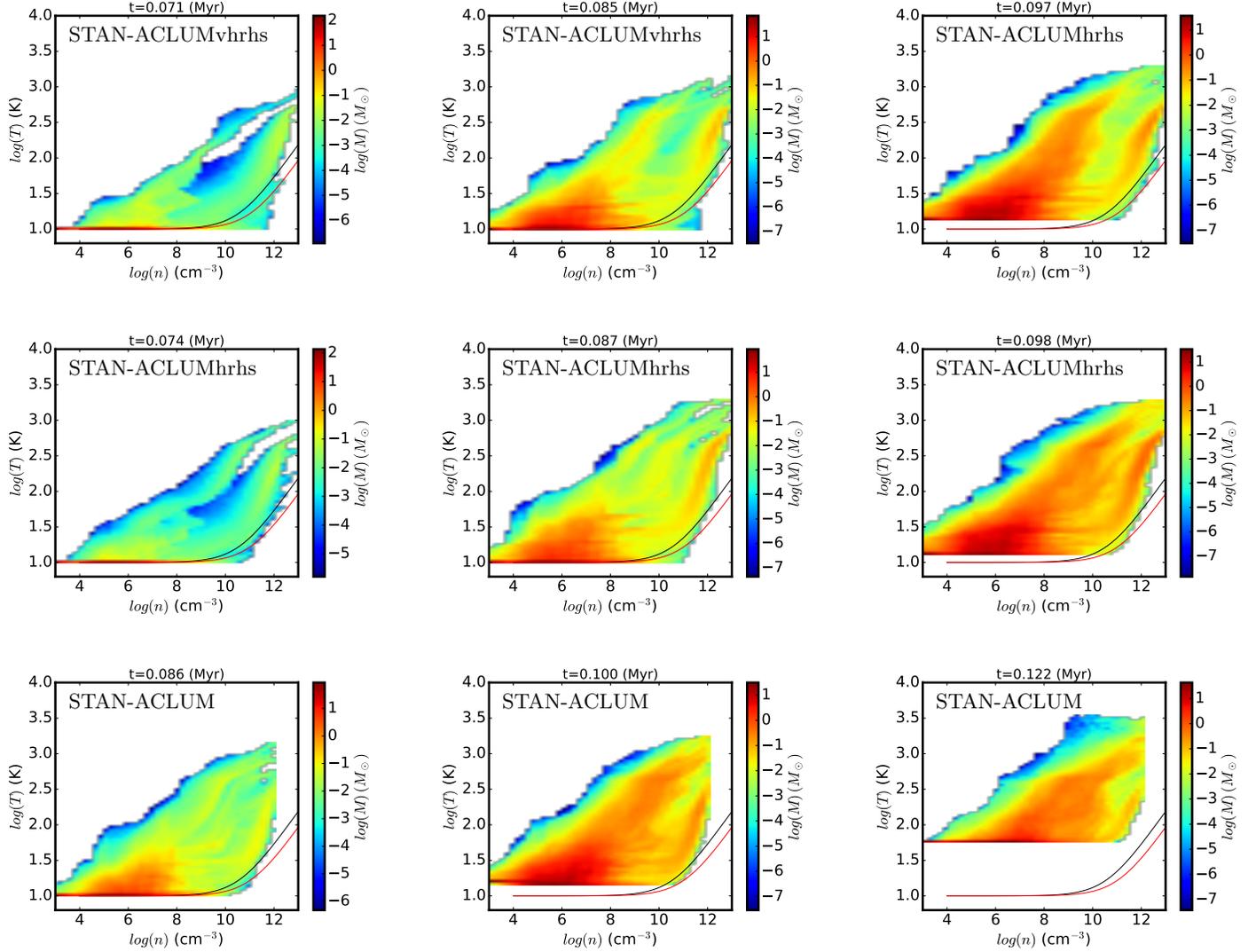

\section{Dependence on numerical parameters}
\label{num_res}

\setlength{\unitlength}{1cm}
\begin{figure*}%[h!]
%\centering
\begin{picture} (0,12)
\put(0,5.5){\includegraphics[width=8cm]{FIG_PAPIER_RHDIMF/CLUSTER_IMF_feedback_eos_hres_modcrit_hsink2/sink_hist_mass_serie.pdf}}  
\put(2,11.1){COMP-NOACLUM}
\put(8,5.5){\includegraphics[width=8cm]{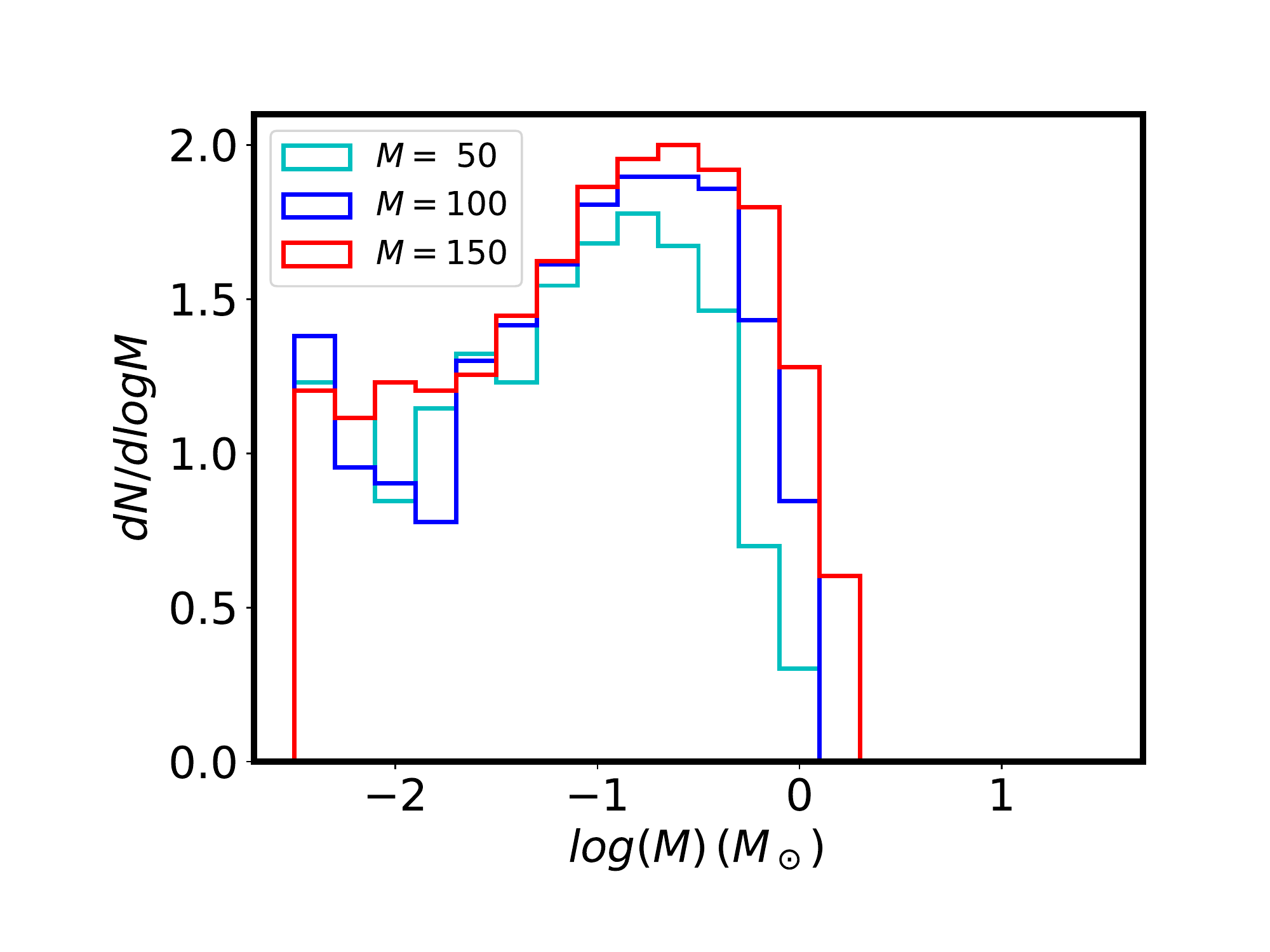}}  
\put(10,11.1){COMP-NOACLUMhr}
\put(8,0){\includegraphics[width=8cm]{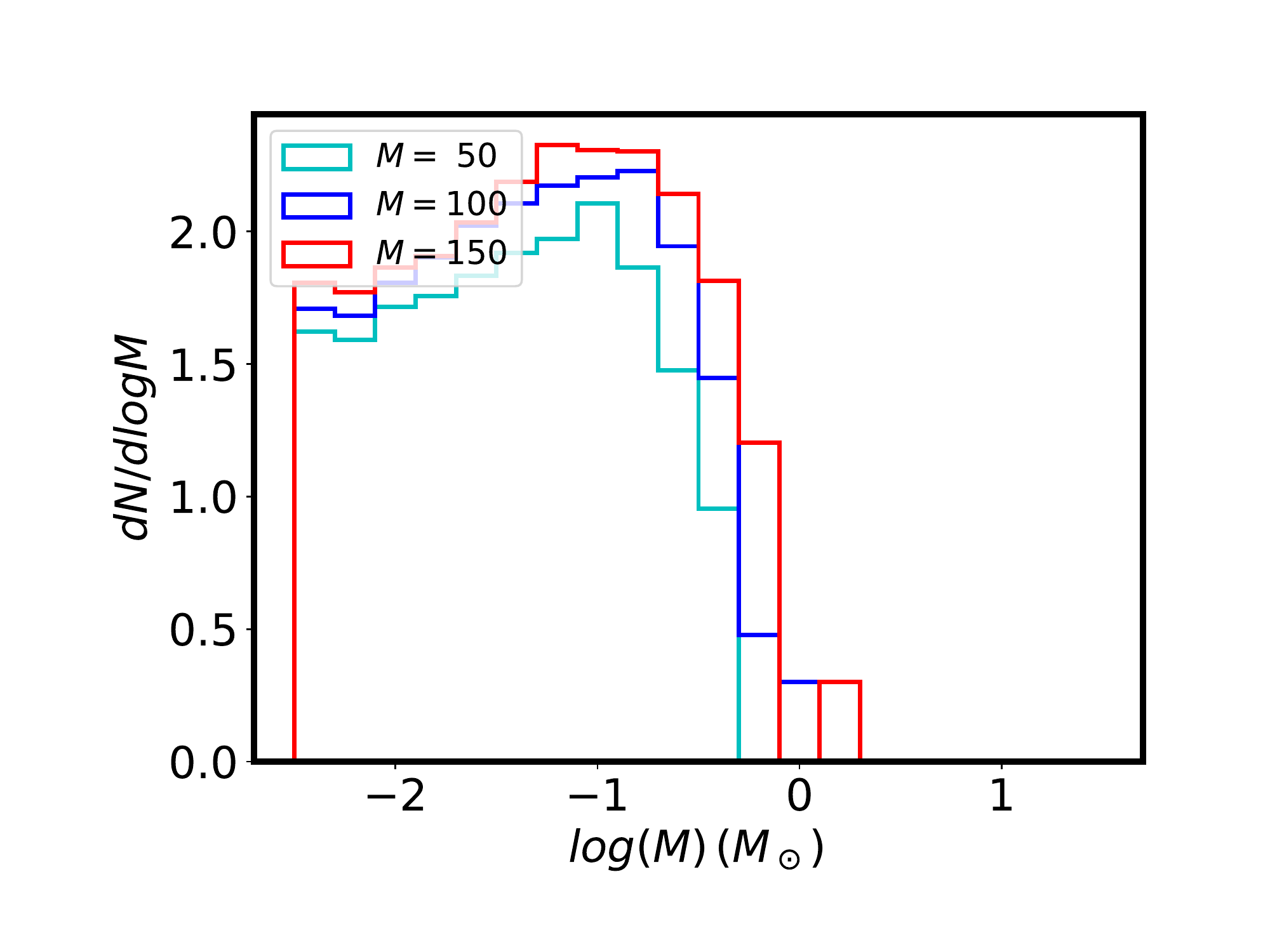}}  
\put(0,0){\includegraphics[width=8cm]{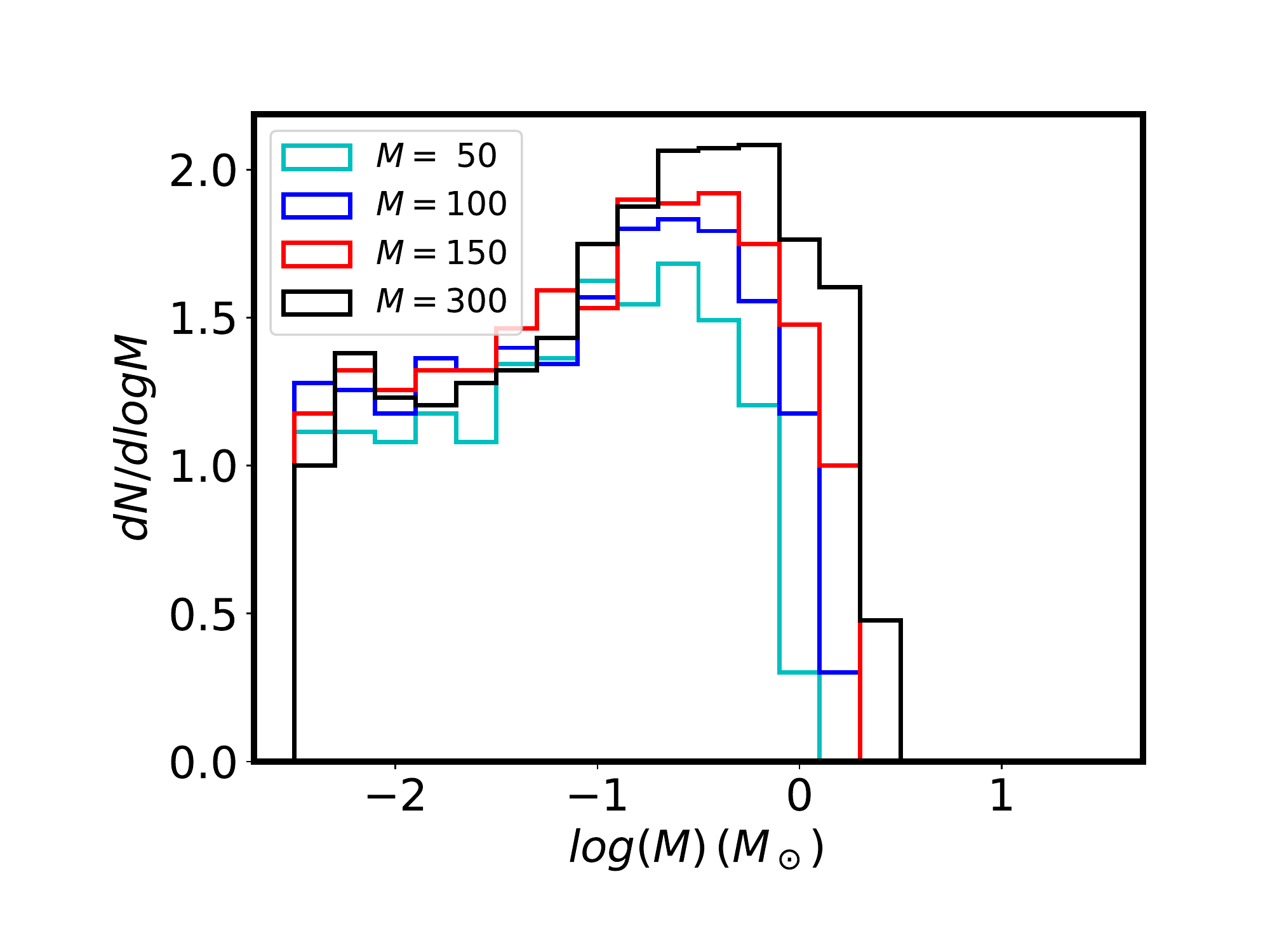}}  
\put(10,5.6){COMP-NOACLUMls}
\put(2,5.6){COMP-NOACLUMlrls}
\end{picture}
\caption{Mass spectra for the COMP-type runs and no accretion luminosity at various times and for various numerical resolution
and sink parameters (see table~\ref{table_param_num}).
}
\label{run_IMF_comp_reso}
\end{figure*}

\setlength{\unitlength}{1cm}
\begin{figure}%[h!]
%\centering
\begin{picture} (0,18)
\put(0,12){\includegraphics[width=8cm]{FIG_PAPIER_RHDIMF/CLUSTER_IMF_feedinter_lowcd_eos_hres_hsink_nobug/sink_hist_mass_serie.pdf}}  
\put(2,17.6){STAN-ACLUM}
\put(0,6){\includegraphics[width=8cm]{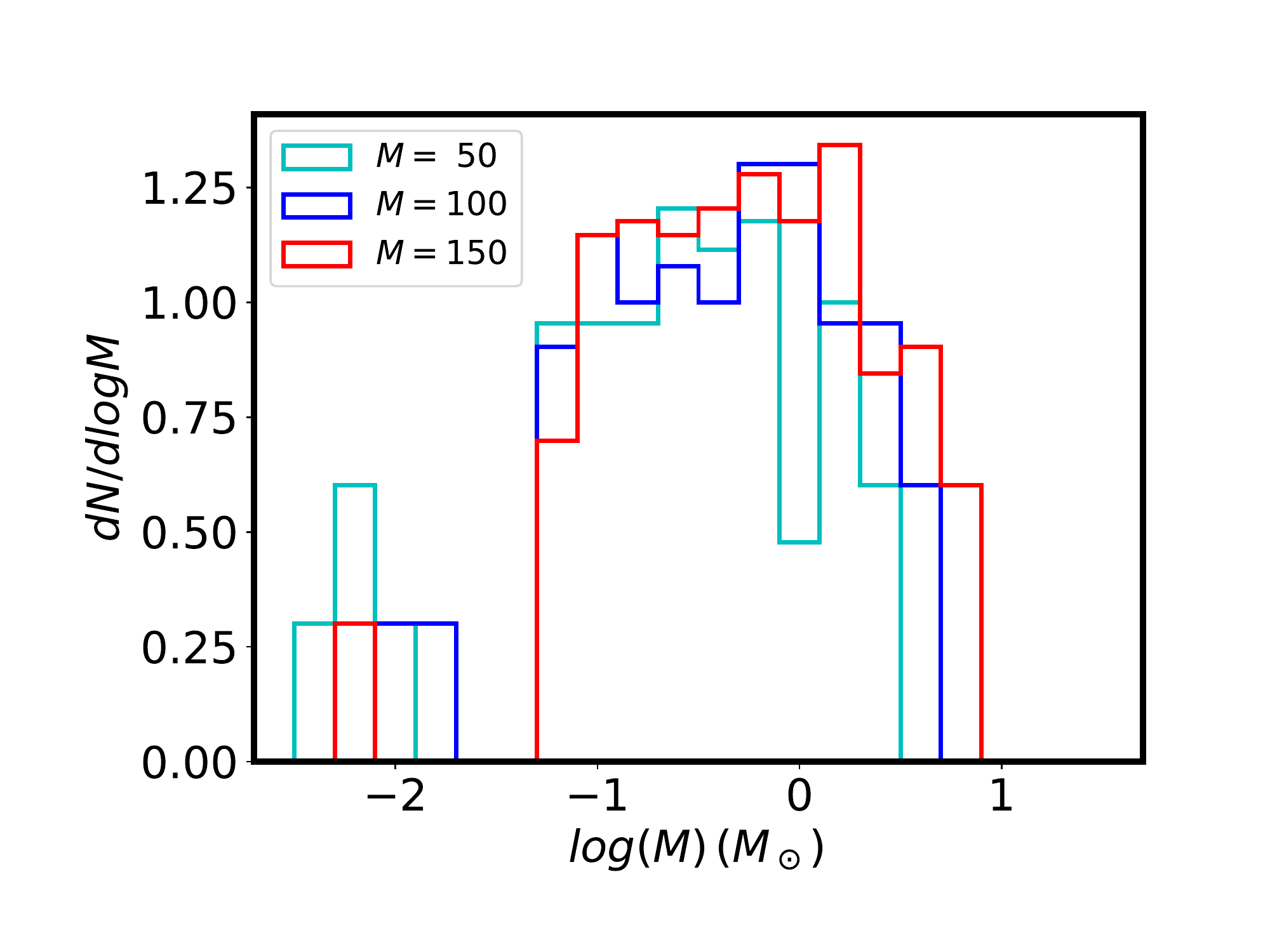}}  
\put(2,11.6){STAN-ACLUMhrhs}
\put(0,0){\includegraphics[width=8cm]{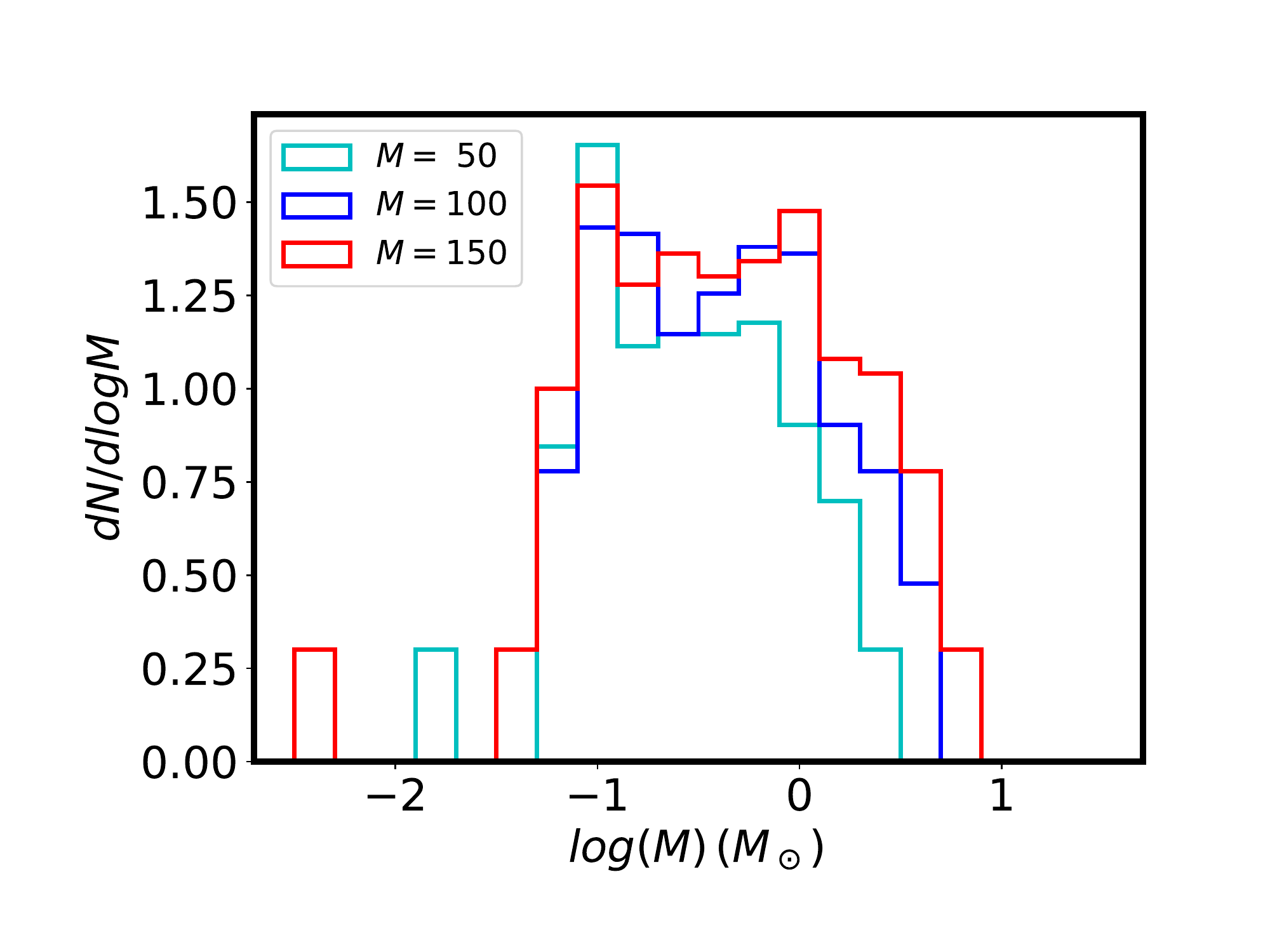}}  
\put(2,5.6){STAN-ACLUMvhrhs}
\end{picture}
\caption{Mass spectra for the STAN-type runs with accretion luminosity at various times and for three numerical resolution
and two values of $n_{\rm acc}$ (see table~\ref{table_param_num}).
}
\label{run_IMF_diff_reso}
\end{figure}

To investigate the issue of numerical convergence several runs have been performed as indicated at the end 
of Table~\ref{table_param_num}. First, we studied COMP-type runs with no accretion luminosity and 
second STAN-type ones taking the accretion luminosity into account.

The corresponding mass spectra for COMP-type runs are depicted in Fig.~\ref{run_IMF_comp_reso}.
Top-left panel reproduces for convenience the result of run COMP-NOACLUM. 
Bottom-left presents the run COMP-NOACLUMlrls which has  a resolution of 4.6 AU compared to 2.3 AU for run COMP-NOACLUM also 
sinks get introduced at $n_{\rm acc} =10^{12}$ cm$^{-3}$ instead of $n_{\rm acc} =10^{13}$ for run COMP-NOACLUM. 
Thus the mass $dx^3 n_{\rm acc}$,  i.e. the mass contained in the finest computational cells,  is nearly the same in both runs. 
Clearly the peak is located nearly at the same position. There are nevertheless some differences between the two runs.
COMP-NOACLUM has about three times more small objects ($M_* < 0.03 \, M_\odot$) and roughly 
three times  more big ones ($M_* > 3 \, M_\odot$). 
The discrepancy between runs COMP-NOACLUM  and COMP-NOACLUMls is even larger. Both runs have the same
spatial resolution, but $n_{\rm acc}$ is ten times lower in COMP-NOACLUMls, meaning that the sinks are introduced 
more easily. The peak in COMP-NOACLUMls is located at about 0.1 $M_\odot$ instead of 0.3 $M_\odot$ and the number
of small objects is even  larger. 
Therefore we see that the mass spectrum is influenced both by resolution and sink threshold. 
Run COMP-NOACLUMhr explores the influence of further numerical resolution but same 
$n_{\rm acc}$ than COMP-ACLUM. It shows that the mass 
spectrum shifts toward smaller masses but  by a factor less than 2. In particular, 
while run COMP-NOACLUMhr has the same value of $dx^3 n_{\rm acc}$ than run COMP-NOACLUMls, we find 
that the peak does not shift to much smaller value as it is the case for run COMP-NOACLUMls. We interpret
this as due to the fact that at density $n_{\rm acc} =10^{13}$ cm$^{-3}$, the gas is nearly adiabatic. 
This shows that although complete numerical convergence may have not been completely reached,
  run COMP-NOACLUMhr is probably approaching it. Indeed, contrary to the isothermal regime where 
the number of fragments increases with resolution, this is not the case in the adiabatic one. \\ \\

Figure~\ref{run_IMF_diff_reso} displays the mass spectra for STAN-type runs. 
Top panel reproduces run STAN-ACLUM to ease the comparison process.
Middle panel presents run STAN-ACLUMhrhs which has two times more resolution 
and a value of $n_{\rm acc}$ that is ten times higher leading to roughly the same 
value of $dx^3 n_{\rm acc}$
when the sink particles get introduced. As can be seen the agreement is only 
moderate. There is a tendency for run STAN-ACLUMhrhs to have more massive objects by a factor of about $\simeq 2$. 
Run STAN-ACLUMvhrhs has a resolution of 1.15 AU, which is two times higher than for run 
STAN-ACLUMhrhs. Both runs have the same $n_{\rm acc}$. We see that overall the two distributions, without 
being identical are nevertheless similar.

Let us stress that in \citet{leeh2018a} and \citet{leeh2018b}  systematic investigations of numerical convergence 
and dependency on $n_{\rm acc}$ have been performed and it has been concluded that while numerical convergence
could not be achieved when an isothermal equation was used, convergence was achieved 
when a barotropic one was used. That is to say when the 
adiabatic exponent becomes larger than 4/3 above a certain density, numerical convergence was obtained. 
Regarding the value of $n_{\rm acc}$, it has been found that if the equation of state has an exponent that at very high density 
is close enough to 4/3 (for instance 7/5 but not 5/3), then the value of $n_{\rm acc}$ is not {\it too} consequential. 
The situation in the present paper appears to be more difficult as we found dependence both on numerical resolution 
and on $n_{\rm acc}$. The reason is probably that while in barotropic calculations, the eos is imposed irrespectively 
of the resolution, the fully radiative calculation runs meant as describing self-consistently the 
thermal state of the gas. In particular the structure of the first hydrostatic core, which is argued to set 
the peak of the IMF \citep{leeh2018b,hetal2019}, is expected to be self-consistently calculated. However the size of 
the first hydrostatic cores is about 5 AU. It is therefore relatively unsurprising that simulations 
with only few AU of resolution have not reached full numerical convergence yet.  
The similarities between on one hand runs COMP-NOACLUM and COMP-NOACLUMhr, and on the other hand, 
runs STAN-ACLUM, STAN-ACLUMhrhs and STAN-ACLUMvhrhs, also suggest that our simulations are probably approaching convergence.

\section{The mean density profile}
\label{delta_sink}

\setlength{\unitlength}{1cm}
\begin{figure}%[h!]
%\centering
\begin{picture} (0,12)
\put(8,0){\includegraphics[width=8cm]{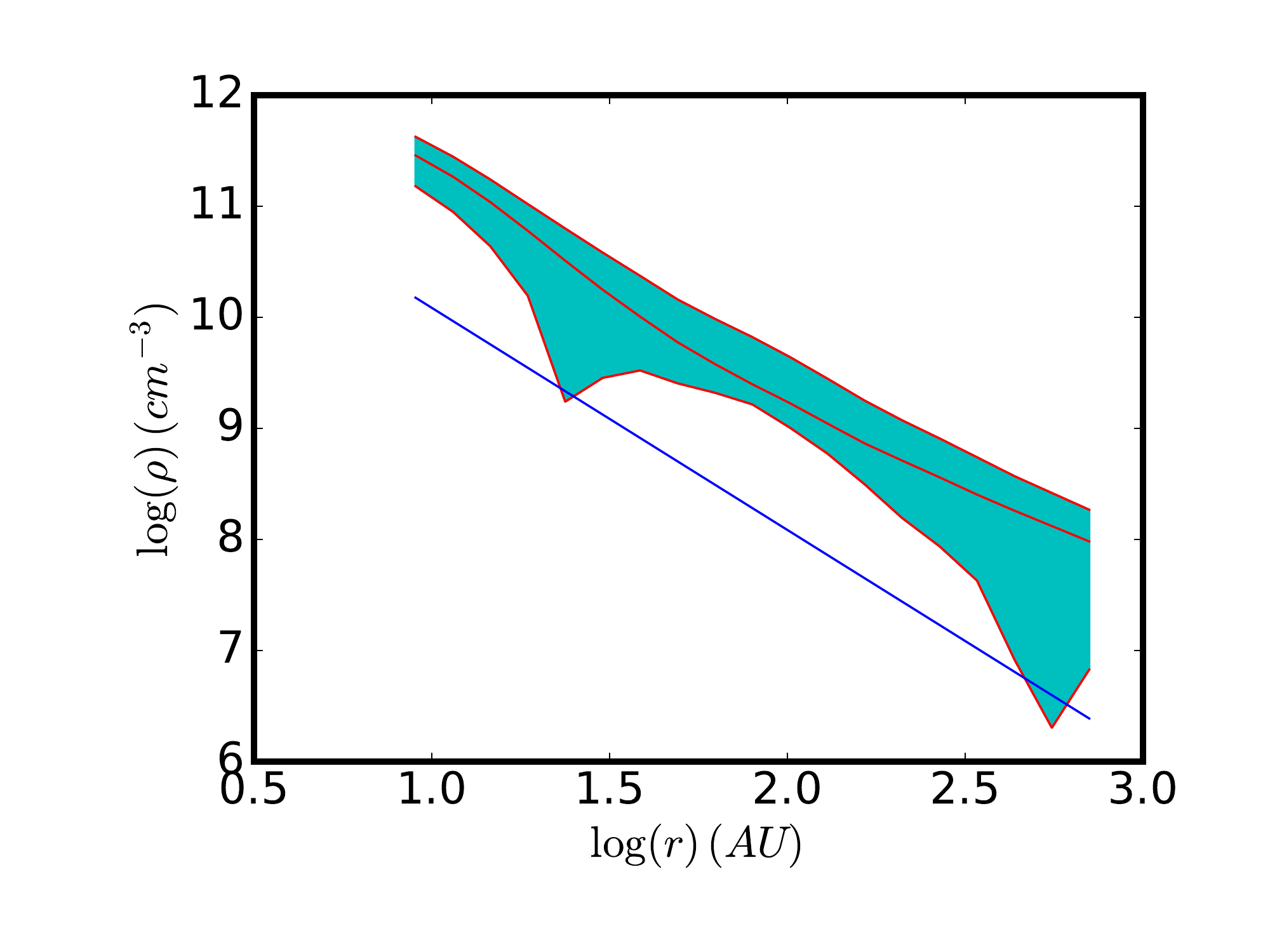}}  
\put(10,5.6){COMP-NOACLUMls}
\put(0,0){\includegraphics[width=8cm]{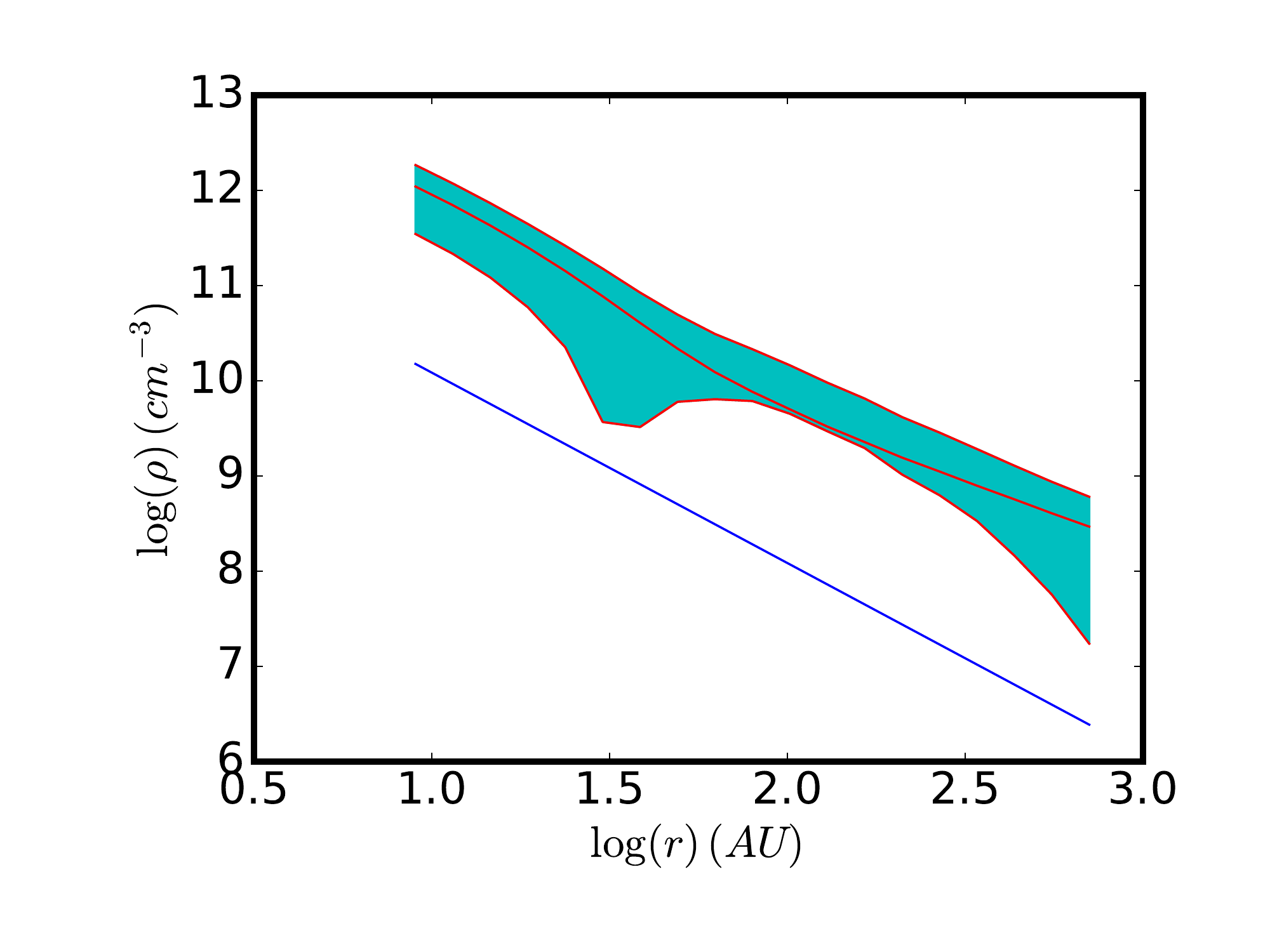}}  
\put(2,5.6){COMP-NOACLUM}
\put(0,6){\includegraphics[width=8cm]{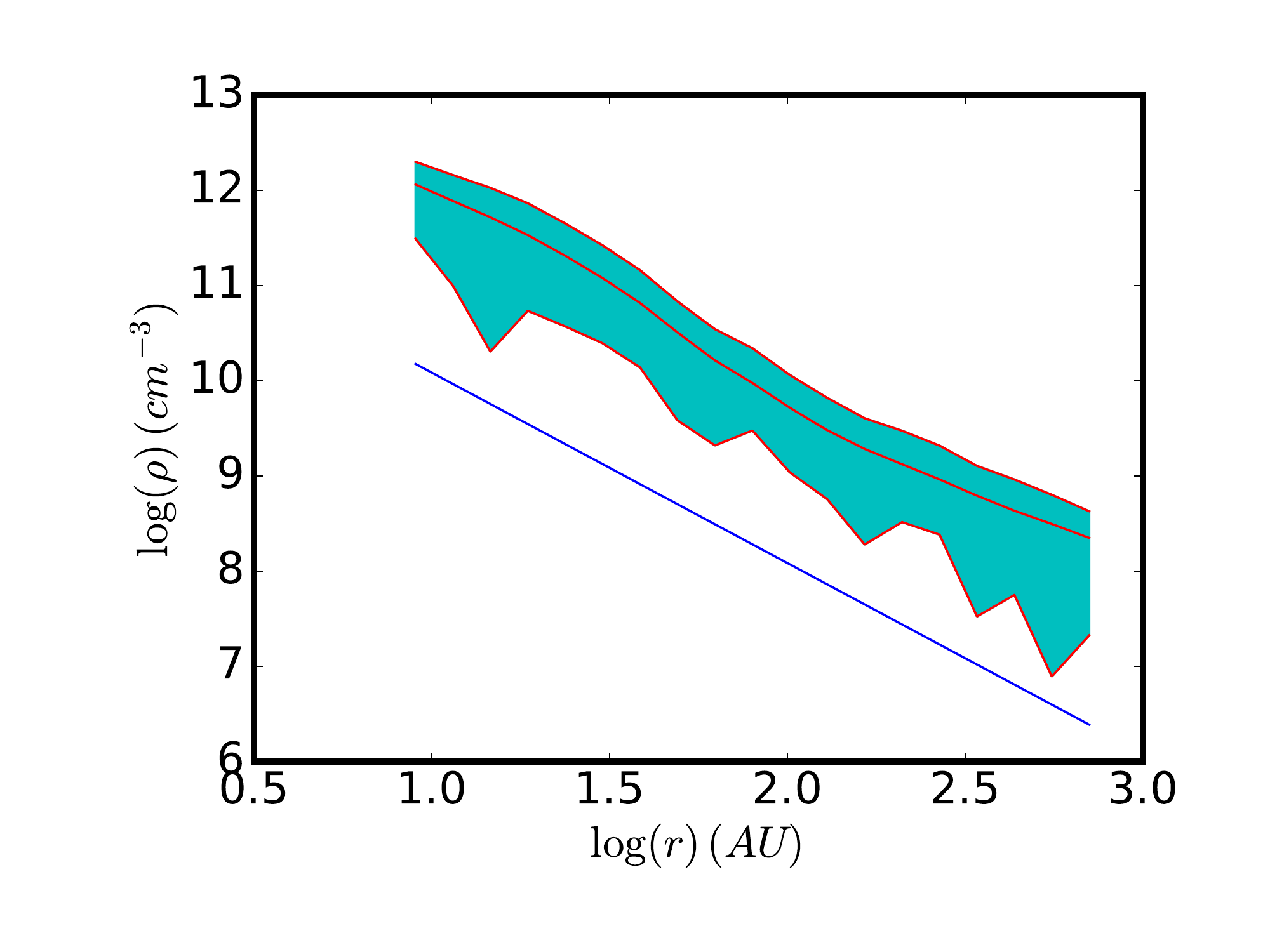}}  
\put(2,11.6){COMP-ACLUM}
\put(8,6){\includegraphics[width=8cm]{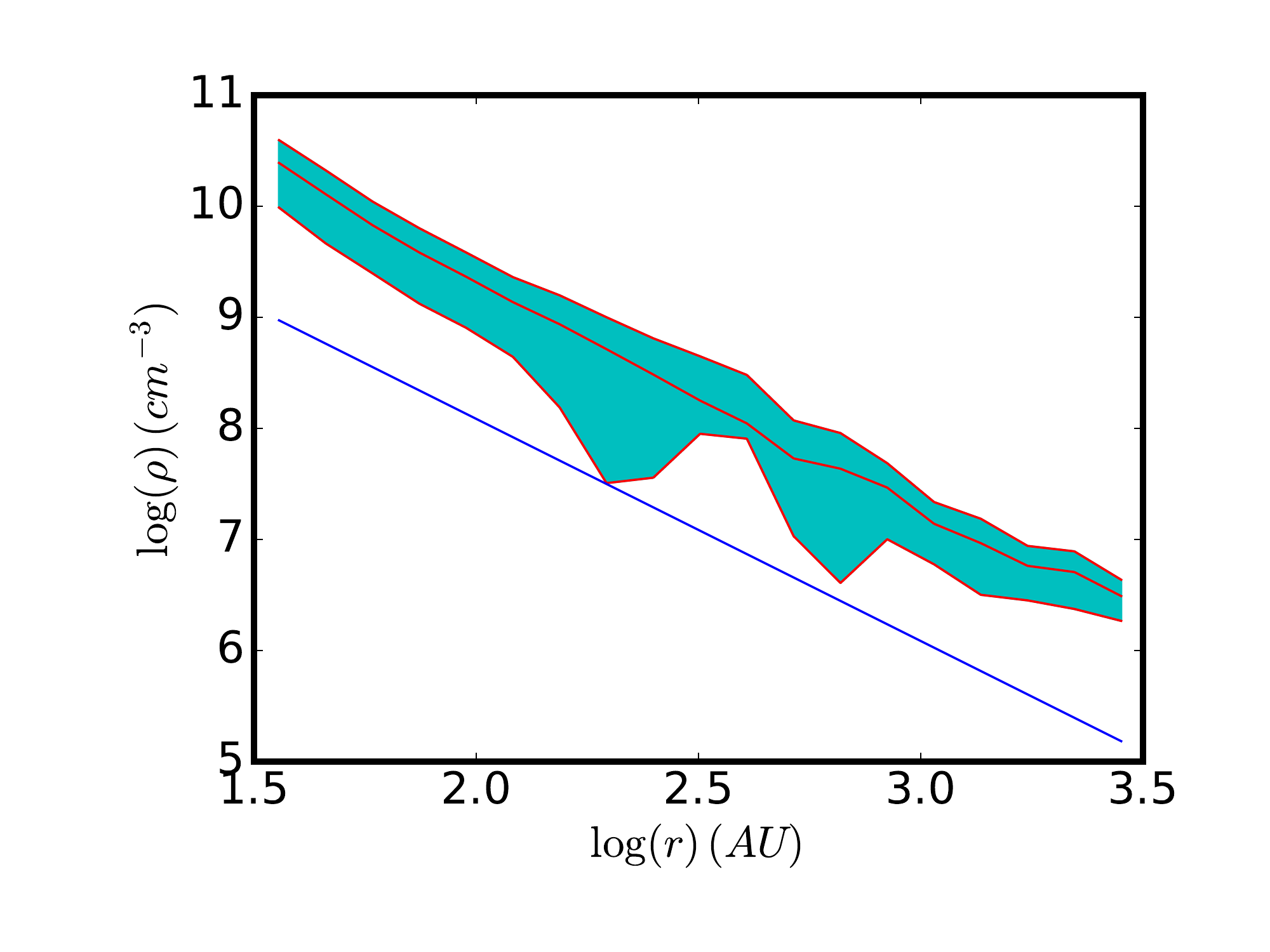}}  
\put(10,11.6){STAN-ACLUM}
\end{picture}
\caption{The mean density profile around sink particles (red lines) and standard deviation (shaded area). The blue line
is the density profile of the singular isothermal sphere.
}
\label{sink_rho}
\end{figure}

\setlength{\unitlength}{1cm}
\begin{figure}%[h!]
%\centering
\begin{picture} (0,6)
%\put(8,0){\includegraphics[width=8cm]{FIG_PAPIER_RHDIMF/CLUSTER_IMF_feedback_eos_hres_modcrit_hsink/sink_delta_mass.pdf}}  
%\put(10,5.6){COMP-NOACLUMls}
%\put(0,0){\includegraphics[width=8cm]{FIG_PAPIER_RHDIMF/CLUSTER_IMF_feedback_eos_hres_modcrit_hsink2/sink_delta_mass.pdf}}  
%\put(2,5.6){COMP-NOACLUM}
\put(0,0){\includegraphics[width=8cm]{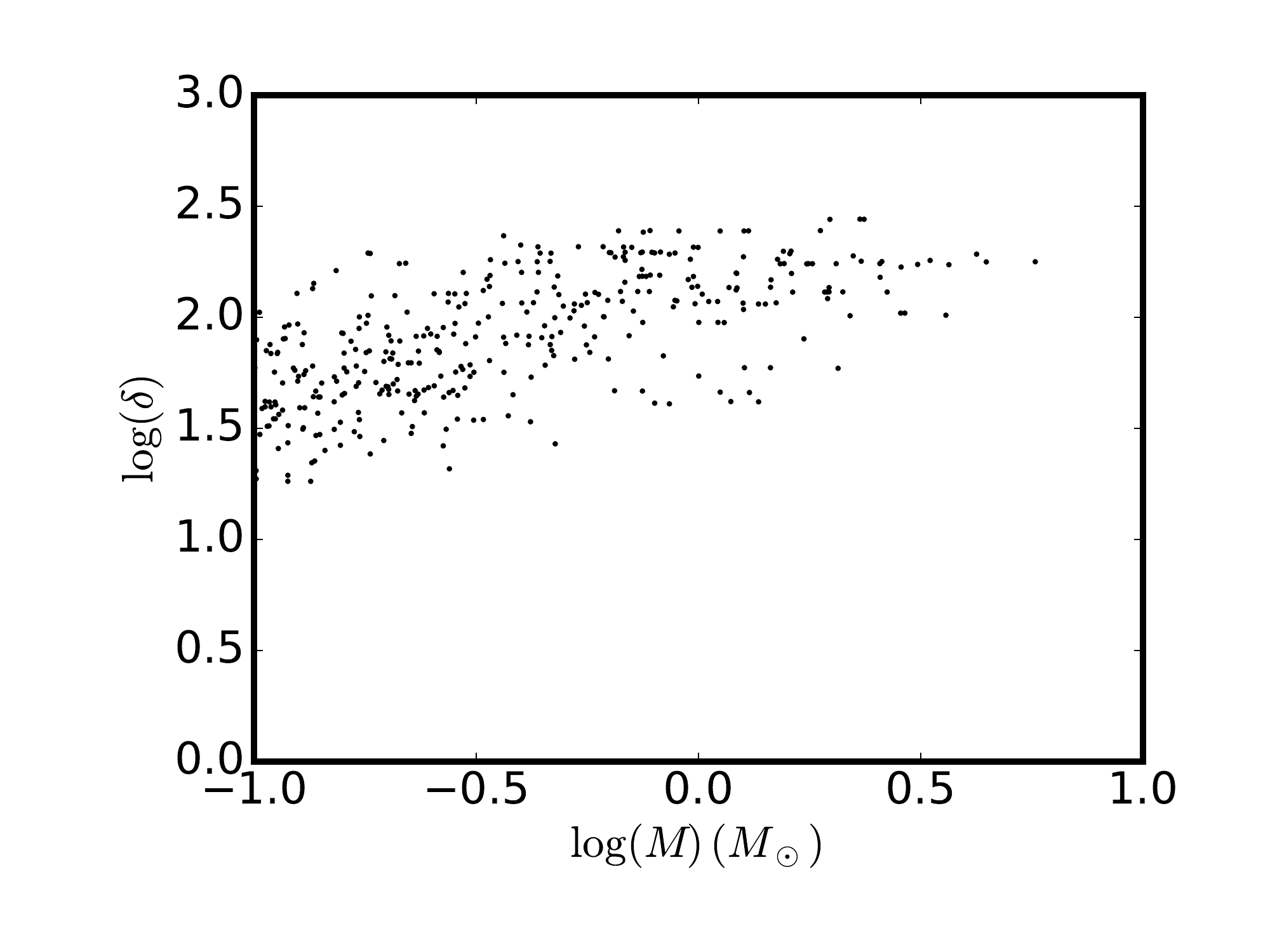}}  
\put(2,5.6){COMP-ACLUM}
\put(8,0){\includegraphics[width=8cm]{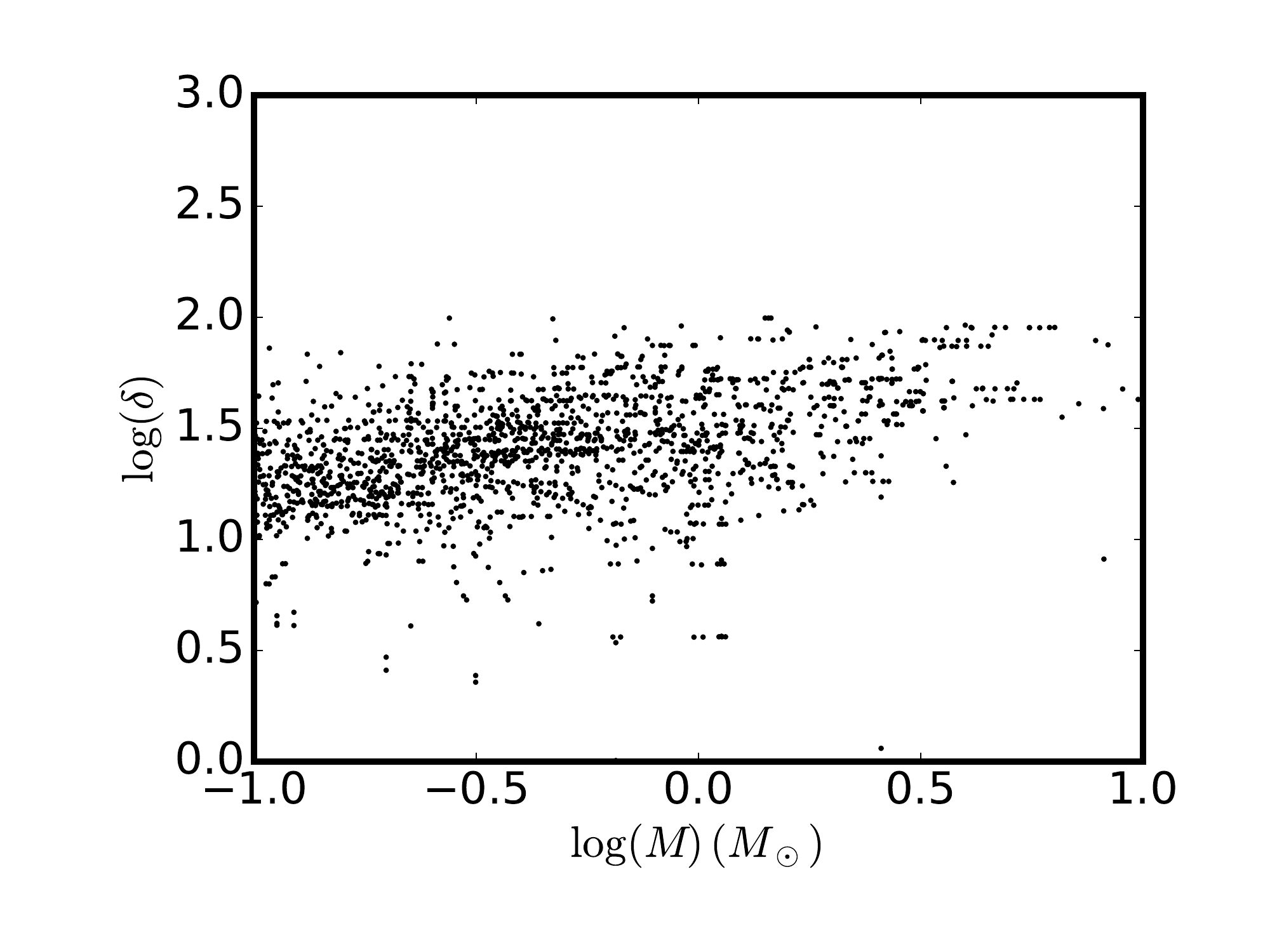}}  
\put(10,5.6){STAN-ACLUM}
\end{picture}
\caption{The distribution of $\delta_\rho$ (ratio between density and singular isothermal sphere density around sink particles).
Each point corresponds to a sink particle.
}
\label{sink_delta}
\end{figure}

To get an estimate of the $\delta$ parameter which appears in eq.~(\ref{total_approx_num}), we proceed
like in \citet{hetal2019}, i.e. we measure the mean density in concentric shells around sink particles. 
We then compute the mean density value and the standard deviation (shaded area). 
The result is displayed in Fig.~\ref{sink_rho}. 
The density field is typically $\propto r^{-2}$ and in the case of COMP-ACLUM is about 100 times above the 
density of the singular isothermal sphere (blue line). For run STAN-ACLUM, it is more on the order of 20-30, 
which is roughly 4 times lower than for COMP-ACLUM. This is likely a direct consequence of their 
respective initial radii, which precisely differ by a factor of four. 

As the singular isothermal sphere density is $\propto C_s^2$, it is worth investigating the density profile
of run COMP-NOACLUM since its temperature are factors 3-5 lower than the ones of COMP-ACLUM. 
The result is displayed in left-bottom panel, which reveals that the density distribution of run 
COMP-NOACLUM is very comparable to the one of COMP-ACLUM. The reason is that this is likely the turbulence which is 
playing here the role of an effective sound speed \citep{chang2015}.

Finally, it is also worth to display the density distribution of run COMP-ACLUMls as we saw in \S~\ref{num_res}
that this run (that we remind uses $n_{\rm acc} =10^{12}$ cm$^{-3}$) has much more sink particles than 
run COMP-ACLUM. The density field is typically a factor of nearly 3-5 below the one of run COMP-ACLUM. 
This is probably a consequence of the fact that there are more numerous sinks, meaning that the 
density field around a given objects is partially accreted by the numerous neighbours.
Also  the sinks  are about three times 
less massive on average and as seen in Fig.~\ref{sink_delta}, the density around an object increases with its 
mass.

To get a better estimate of the parameter $\delta _\rho$, we have plotted for each sink and at various 
timesteps, the mean value of $\delta _\rho$, obtained by taking the mean of the  density divided by the 
singular isothermal sphere density, as a function of the sink mass. The results are displayed in Fig.~\ref{sink_delta}
for runs COMP-ACLUM and STAN-ACLUM. The two distributions span almost an order of magnitude ranging 
from respectively 20 to 300 and 10 to 100. There is a trend for $\delta _\rho$ to increase with the sink mass.

%\begin{thebibliography}{55}
%\end{thebibliography}{55}

\end{document}